\documentclass[preprint,12pt]{elsarticle}

\usepackage{makeidx}
\usepackage{multirow}
\usepackage{multicol}
\usepackage[dvipsnames,svgnames,table]{xcolor}
\usepackage{graphicx}
\usepackage{ulem}
\usepackage{amssymb,amsmath}
\usepackage[modulo]{lineno}
\usepackage{epstopdf}
\usepackage{hyperref}

\makeatletter
	{\par\setlength{\parindent}{#3}
	\setlength{\leftmargin}{#1}       \setlength{\rightmargin}{#2}%
	\advance\linewidth -\leftmargin       \advance\linewidth -\rightmargin%
	\advance\@totalleftmargin\leftmargin  \@setpar{{\@@par}}%
	\parshape 1\@totalleftmargin \linewidth\ignorespaces}{\par}%
\makeatother

\begin{document}

\begin{frontmatter}

\title{The building blocks of magnonics}



\author{B. Lenk}
\author[mue]{H. Ulrichs}
\author{F. Garbs}
\author{M. M\"{u}nzenberg}

\address{I. Physikalisches Institut, Georg-August Universit\"{a}t G\"{o}ttingen, Germany}
\address[mue]{current address: Institut f\"{u}r Angewandte Physik, Westf\"{a}lische Wilhelms-Universit\"{a}t M\"{u}nster, Germany}

\begin{abstract}
Novel material properties can be realized by designing waves' dispersion relations in artificial crystals. The crystal's structural length scales may range from nano- (light) up to centimeters (sound waves). Because of their emergent properties these materials are called metamaterials.
Different to photonics, where the dielectric constant dominantly determines the index of refraction, in a ferromagnet the spin-wave index of refraction can be dramatically changed already by the magnetization direction. This allows a different flexibility in realizing dynamic wave guides or spin-wave switches.
The present review will give an introduction into the novel functionalities of spin-wave devices, concepts for spin-wave based computing and magnonic crystals. The parameters of the magnetic metamaterials are adjusted to the spin-wave $k$-vector such that the magnonic band structure is designed.
However, already the elementary building block of an antidot lattice, the singular hole, owns a strongly varying internal potential determined by its magnetic dipole field and a localization of spin-wave modes.
Photo-magnonics reveal a way to investigate the control over the interplay between localization and delocalization of the spin-wave modes using femtosecond lasers, which is a major focus of this review. We will discuss the crucial parameters to realize free Bloch states and how, by contrast, a controlled localization might allow to gradually turn on and manipulate spin-wave interactions in spin-wave based devices in the future.
\end{abstract}

\end{frontmatter}

\linenumbers

\tableofcontents

\section{Introduction}
Magnonics is a young and evolving research field. Its aim is to control and manipulate spin waves in ferromagnetic materials~\cite{ref1}. In analogy to photonics taking control of the flow of light, it allows for the design of material properties for spin waves.
The experimental realization of computing with spin waves will be discussed first. These are logic gates (NOR, XNOR, etc.)~--~novel devices that benefit from the particular properties of spin waves. To connect and to guide information, spin-wave frequency filters and wave guides are under research. In such, novel computing concepts have been described. One of these are reconfigurable mesh structures using spin waves for parallel graphics computing.
This review will introduce how periodic structures can be realized in different dimensions, which will determine the line, plane or volume in which the artificial material properties are tailored and affect propagating spin waves.
A large part of the present review will map how they can be investigated using laser light --~hence the term photo-magnonics~-- a technique developed in the last years. In particular, two-dimensional structures will be considered.
Generally, the response function to a dynamic excitation propagating through a periodic crystal determines its propagation defined by the dispersion~$E(\vec{k})$, which in turn determines the available density of states.
For weak potentials, the concept of constructing Bloch states using plane waves and calculating their dispersion in the reduced zone scheme is straight-forward. Applied to photons in periodic dielectrics, we show how band gaps at the Brillouin zone boundary tailor novel functionalities.

In the case of spin waves the dispersion can be quite complex in the unstructured materials already; we give a short introduction into the peculiarities of the spin-wave dispersion, namely the anisotropy of the dispersion with respect to the magnetization direction.
The same concepts as for electrons and photons can be applied to magnons to form Bloch states in a magnonic crystal. In the uniform mode approximation simple band structures can be directly calculated.
However, these ignore the detailed shape of the potential and are a good approximation for a weakly varying potential only. A more realistic potential landscape will be discussed in the case of the periodic dipolar potential for a simple antidot lattice. The individual shape of each spin-wave scatterer, the building block of the magnonic crystal, is determined by its detailed variation of the dipole field.
Spin-wave modes in periodically structured materials have been studied experimentally by different techniques: microwave-based techniques (ferromagnetic resonance (FMR), vector network analyzer FMR or pulse-inductive microwave magnetometer (PIMM)), in combination with spatial-resolved Kerr microscopy. A somewhat larger part will be devoted to optical pump-probe techniques with femtosecond lasers in this review.
The interplay between localization and delocalization of the dynamic modes in the magnonic crystal is shown to be crucial to understand the observed spin-wave modes.
The deep distortion of the internal field by the dipole field at an antidot site results in a localization of the spin waves in many cases, prohibiting the observation of delocalized modes. However, in order to realize magnonic wave guides and active spin-wave materials, it is crucial to control the interplay between localization and delocalization of the dynamic modes in the magnonic crystal.
For a high filling fraction of the antidot lattice (i.e.\ amount of material removed), spin waves are trapped over a large field range at places with reduced internal field in the magnonic crystal. Generally, only for small filling fractions delocalized free Bloch modes are observed, whose $k$-vector is determined by the crystal's unit cell. We will discuss how these Bloch modes can be identified in photo-magnonics.
At the end we will give a short outlook on future possibilities, perspectives and developments in the field, such as guiding spin waves in magnonic wave guides, spin-wave resonators or tuning the degree of localization of a spin wave and their interaction.

\section{Computing with spin waves}
On the semiconductor roadmap of 2009, possible emerging research devices and emerging materials for future electronics and logic are presented~\cite{ref2}. On the materials side, graphene, carbon nanotubes, and nanowires are discussed as possible candidates to develop materials to meet requirements for future semiconductor devices.
On the device side, the use of the electron spin for information storage, and the implementation of spin waves for performing non-volatile logic functions on a CMOS chip are discussed. The idea to use waves for signal processing is not new: light in glass fibers is used as an interconnect between chips in powerful computers.
The IBM blue gene super computer link modules have optical printed circuits on each board. However, their function is to convert electrical signals into optical signals that can be directed to the next step of data processing. As long as the processing relies on electrons the extra effort of transforming the data into light pulses for effective and fast throughput to other regions of the processor is unavoidable. Building a computer that uses solely photons for information processing and its non-volatile logic functions is not that easy a task.
It is difficult to realize a light switch that can be operated by light; to switch light, one would need to modify the index of refraction, which is related to the electron density of the material. Consequently, a high density of photons is needed to have a significant effect on the dielectric properties of a material. To realize a switch that should turn on and off the total number of photons, certain tricks have to be played. We will discuss in section~\ref{sec:slow-photons} the progress in this area briefly.

How is it possible to take advantage of the properties of spin waves, the intrinsic strong non-linearities to realize switches for spin waves and to realize non-charge based persistent devices for spin-wave based computing? In recent years a large portion of the effort of the magnetism community was devoted to the field of spin electronics, to develop spin-based semiconductor devices for spin-based transistors and spin-based logic.
Magnetic random access memories (MRAMs) have been developed that exploit the advantage of non-volatile memory storage, while reducing power consumption. Power consumption is one of the biggest problems of high integration circuits today, because of the heating of the chip, reduced battery lifetimes and environmental harmfulness.
The idea of looking at spin waves stands to reason, which has, however, not been done until recently. Even though spin waves have been studied as narrow frequency filters in the 60s to 80s in YIG-based high frequency devices~\cite{IEEE-stuff, sykes1976}, the idea to use spin waves for data transmission and processing is very new.
It can be realized with control on the nanometer scale by now and will be discussed in section~\ref{sec:experiments}. What makes spin waves favorable for technological applications is the corresponding index of refraction, which can easily be manipulated.
\linelabel{line-index-refr}The term 'index of refraction' should here be understood in analogy to photonic crystals where optical resonances can be shifted by a periodic modulation of the dielectric constant.
In the ferromagnetic case, the modulation may be in the parameters exchange constant ($A$) and/or saturation magnetization ($M_\mathrm{S}$) while an additional directional asymmetry is introduced by the applied magnetic field. Corresponding Bloch-like resonances of the spin-wave spectrum will be discussed in section~\ref{sec:bloch-modes}.
In general, spin-wave frequency and wave length can both be tuned from MHz to THz on micron to sub-nm scales, respectively. Therefore, the use of spin waves for logic devices is the natural extension of magnetic non-volatile elements for storage.
The flexibility of spin-wave based materials opens up the possibility of a spin-wave based computing architecture. This is the underlying idea of magnonics: to use spin waves for power-saving computing.
Nevertheless, certain prerequisites in manipulating and guiding spin waves have to be demonstrated a priori \cite{ref3,ref4}: (i)~the controlled excitation of spin waves at defined frequencies, (ii)~guiding spin waves in magnetic wave guides and (iii)~active spin-wave devices for spin-wave manipulation and information processing.
\begin{figure}[h!]
\centering
\includegraphics[width=350pt]{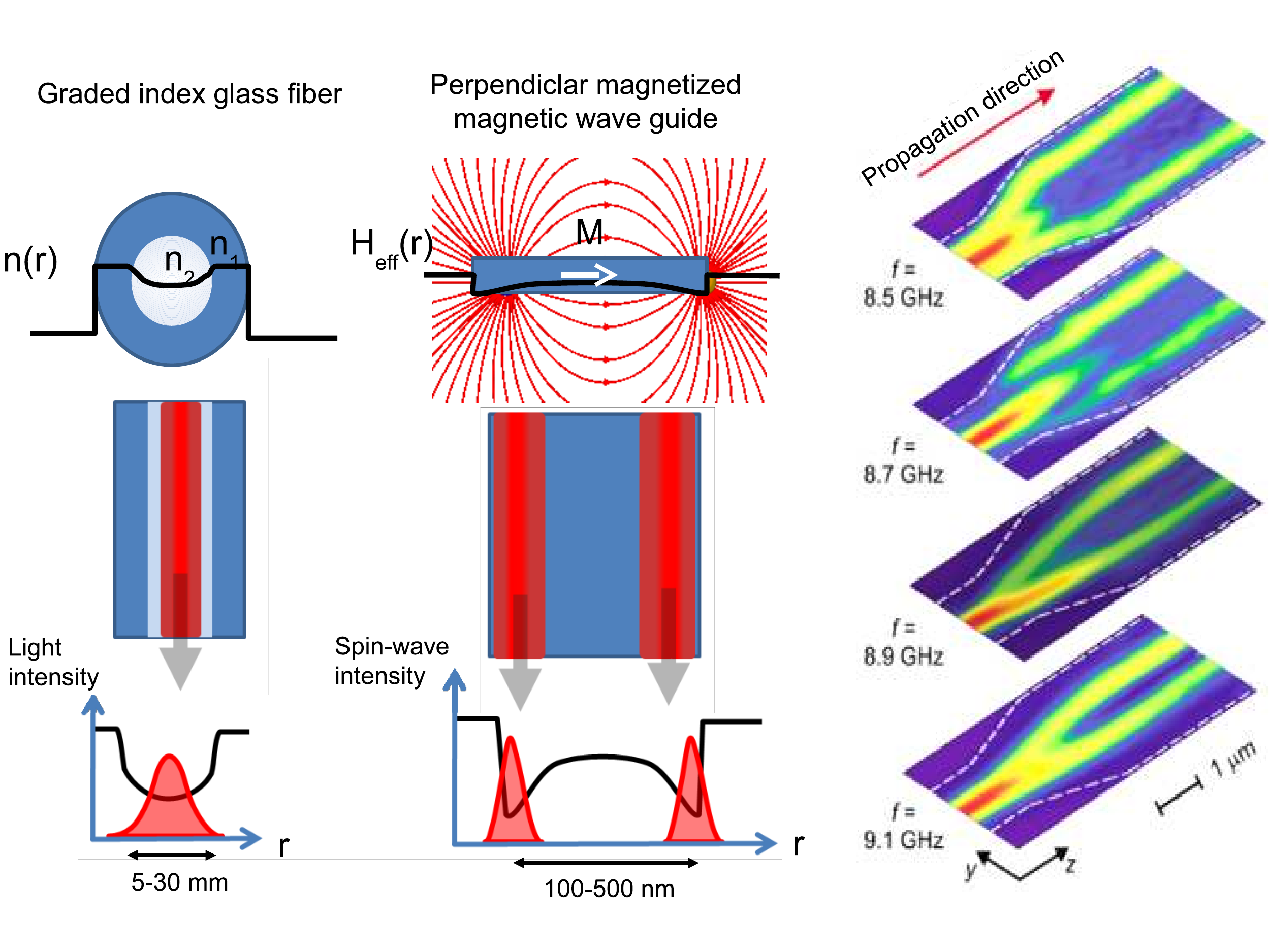}
\caption{A graded index glass fiber (left column) concentrates the intensity of the light modes in the area of low index of refraction, i.e.\ small dielectric constant (depicted by a solid black line). In a simple, homogeneously magnetized plane magnetic wire (center column), the internal effective field (solid black line) determines the localization of the spin-wave modes. Localization of spin waves along a spin-wave guide (right column) measured with micro-BLS (Brillouin Light Scattering) (adapted from~\cite{ref3}). The internal field changes as the width of the wire changes, resulting in different localization axes of the beam path.}
\label{fig:guides}
\end{figure}

\subsection{What are spin-wave guides?}
An object often referred to as a spin-wave guide is a (quasi) one-dimensional stripe of ferromagnetic material. Through its physical boundaries, it restricts the region of existence for spin waves. In optics, an analogue object is called a fiber.
Different from glass fibers in optics, a spin-wave guide is a homogeneously magnetized wire that can channelize, split, and manipulate submicrometer-width spin-wave beams, as was shown by Demidov and Demokritov~\cite{ref3}. The differences are depicted schematically in Fig.~\ref{fig:guides} where the properties of a graded index glass fiber are compared to a magnetic wave guide. In the graded index glass fiber, the index of refraction gradually increases to the outer shell of the glass fiber having a higher index.
The light is totally reflected and conducted in the core of the fiber. The gradual transition leads to a smooth curvature of the wave fronts so that the Gaussian shape of the intensity profile and temporal coherence of the signal pulse is only slightly distorted.
The index of refraction, arising from the dielectric properties of the materials, serves as a potential landscape for the light wave. For a spin wave, the situation is sketched in the middle of Fig.~\ref{fig:guides}. A flat, structured stripe with a width in the micron range serves as a spin-wave guide.
The details of propagation depend on the magnetization direction: magnetized in-plane perpendicular to the wire axis, Damon-Eshbach modes that are a species of dipolar spin waves (see section~\ref{sec:spin-wave-ranges}) can travel along the wire. Magnetic charges at the boundaries lead to a decrease of the internal magnetic field in the wire at both sides; i.e., due to the demagnetization field of the wire magnetized perpendicular to the wire axis, the effective field is reduced at the edges of the spin-wave guide (see sketch in Fig.~\ref{fig:guides}). The well structure of the internal field shows minima at both sides of the wave guide.
Depending on the width of the stripe, the internal magnetic field in the center of the stripe is also diminished more or less in strength for smaller or wider wave guides. One can excite modes, which are localized in these wells -- so-called `edge modes', as well as modes in the middle of the wave guide -- so-called `center modes'. In this context, it is important to notice that edge-modes always have lower frequencies, as compared to center modes.

The right panel in Fig.~\ref{fig:guides} shows micro-Brillouin Light Scattering (micro-BLS) maps of such modes. Spin waves are excited in the region with smaller width, and thus stronger reduced internal field at the center. They propagate towards the region with increased width and thus increased internal field in the center. A transformation from a center into an edge-mode takes places in the transition zone.
Due to the upward shift of the spectrum in the wider stripe, the excited wave can no longer exist in the center.
This localization process depends on the frequency of the initially excited mode: for higher frequencies, the edge-modes move toward the center. This example nicely shows that the understanding of and control over the spin-wave dispersion, localization, and delocalization with the internal field distribution is one of the key aspects to progress in this novel sub-field.

\subsection{Spin waves on a chip: Reconfigurable mesh design}
\begin{figure}
\centering
\includegraphics[width=350pt]{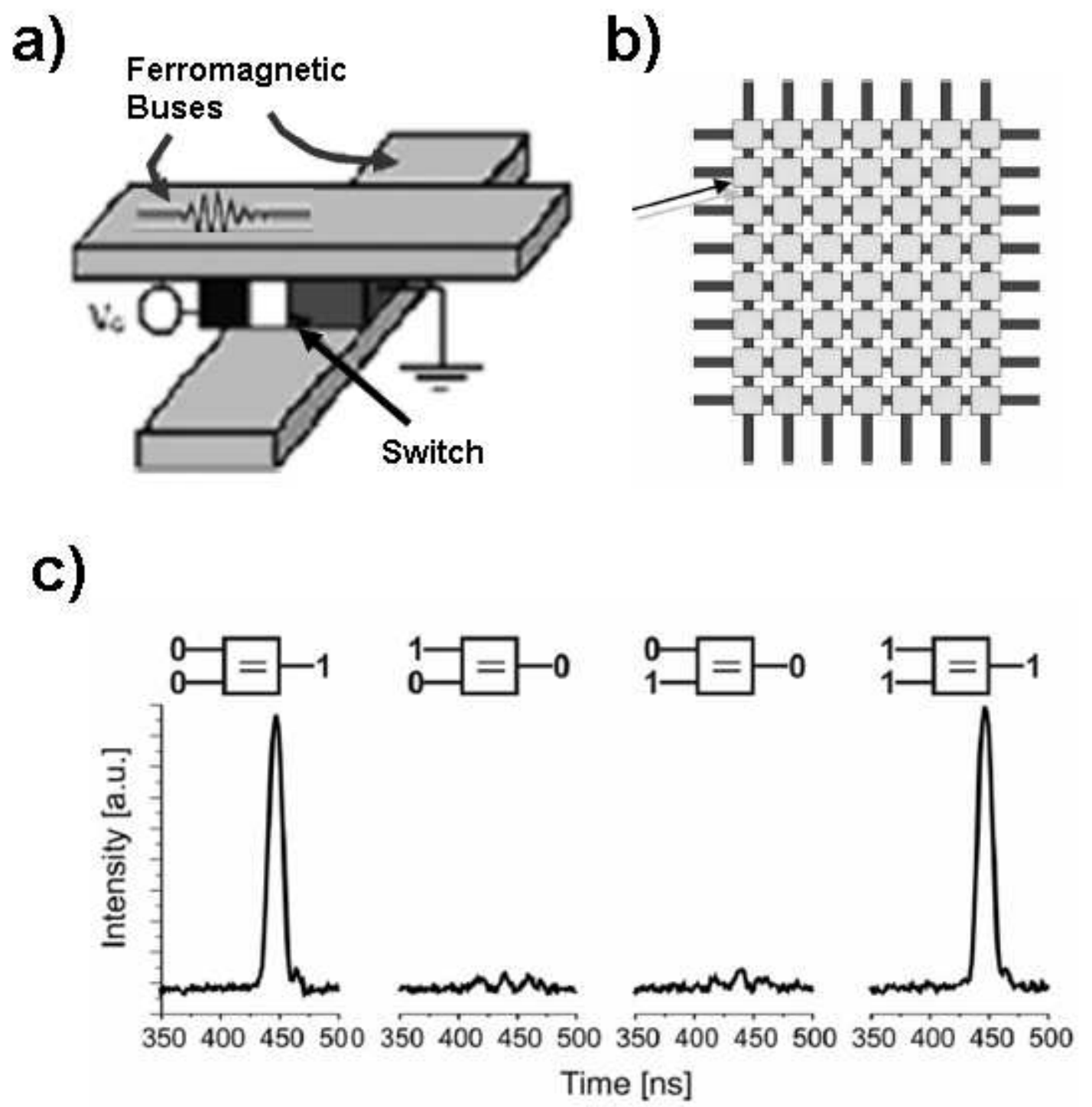}
\caption{Concepts and first realization using spin-wave based computation: a)~Schematics of a spin-wave bus, and b)~building blocks of a reconfigurable mesh structure (from~\cite{ref5}). c)~Experimental realization of a XNOR functionality (logic equality) using a spin-wave interferometer with phase shifter as reproduced from~\cite{ref6}.}
\label{fig:02}
\end{figure}

Concepts to use spin waves for data processing in a chip have been developed by Kang Wang et al., a short review on spin-wave based computing has been published in the series of the first ``International Seminar and Workshop Magnonics: From Fundamentals to Applications'' held August 2009 in Dresden~\cite{ref4} based on their earlier work published in~\cite{ref5}.
Many current applications require a vast amount of data to be processed in parallel. A typical example is a graphic chip, with its algorithms for parallel image processing. Their architecture and computation power has been increased enormously in the last decade. From the scientific point of view, numerical simulation programs, which can be parallelized, can take advantage of the significant progress in that field, so that when put together, an array of graphic chips makes a new super computer.
The idea of the spin-wave based computing concept to parallelize computation uses a reconfigurable mesh architecture. Spin-wave guides transmit the signal at each line. The chip consists of a mesh of $N \times N$ spin-wave switches interconnected by ferromagnetic spin-wave guides. Each node is realized by a ferromagnetic switch. If the switch is ``on'', the spin wave is guided into the crossed line to the spin-wave buses output.
The switching frequency is in the order of GHz and transmission speed is $10^4\, \mathrm{m\, s^{-1}}$, allowing fast data processing. In the architecture presented in Refs.~\cite{ref4,ref5}, excitation is realized by a strip-line; detection will be realized by inductive detection at a second strip-line.
The spin-wave switch at each crossing could be realized by a diluted magnetic semiconductor which can be switched from a ferromagnetic to a paramagnetic state by applying positive or negative voltages.
It should be remarked that as of today, this concept is lacking a realistic practical implementation. Ferromagnetic semiconductors like Mn-doped GaAs have both a very strong spin-orbit interaction and a critical temperature that is below room temperature, and hence will not work for actual devices.
However, using a mesh structure allows for large image data to be processed in the reconfigurable nodes. This is a typical application needed for example to process image data for compression in a television graphics chip.
Additionally, different to a standard reconfigurable mesh design, spin-wave buses could in principle address different frequencies. Therefore, parallel operation seems possible as well.
The layout is shown in Fig.~\ref{fig:02}. In a), the schematic layout of one node is shown, which can be addressed by the voltage turning on/off the magnetization, which then guides the spin-wave package to the lower magnetic wave guide. In b), the full layout of the mesh structure is depicted.

\subsection{Elements of spin-wave computing and logic}
At the same time as concepts that take advantage of spin-wave computing have been put forward, spin-wave based logic elements have been demonstrated. Their principle relies on a spin-wave beam splitter. Two spin-wave packets of same amplitude are split into different arms of a spin-wave guide. In one arm, the propagating spin wave receives a phase shift by~180$^\circ$ ($\pi$).
The arms are united and both spin waves interfere destructively. This can be compared to a spin wave Mach-Zehnder-type interferometer. Different realizations for phase shifters have been suggested; for example, in micromagnetic simulations, it was shown that a domain wall could be utilized as a phase shifter in one of the interferometer arms~\cite{ref7}.
More generally, the phase shift, slowing down or acceleration in one interferometer arm relative to the other, depends on the dispersion of the spin wave. The dispersion is naturally modified by the (sometimes complicated) magnetic structure of the domain walls, thus the average internal field in the domain wall.
Another possibility is to shift the frequency by locally applying an external magnetic field. In the yttrium iron garnet (YIG) wave guides, which have the lowest damping of any magnetic material, millimeter propagation length of spin waves in the microwave frequency range can be realized. In these systems, spin-wave propagation, spin-wave amplification, spin-wave pumping, phase shifters, interferometers and filters were demonstrated~\cite{ref6,ref8}. Schneider and coworkers developed in these wave guides a spin-wave logic realizing exclusive-not-OR and not-AND gates based on the Mach-Zehnder-type interferometer, which will be presented in more detail. Input and output into the YIG wave guides is implemented by microwave antennas; the signal is propagated in the YIG film and split into the two arms of the interferometer.
The phase shifter is realized by applying an Oersted field on top of the YIG, which changes the carrier wave number of the spin-wave packet within that region. By changing the amplitude of the Oersted field, the phase shift can be chosen to sum up to one half of a wavelength. The output signal is shown in Fig.~\ref{fig:02}~c). If no field is applied at either arm, both spin-wave packets are equal in phase and the full signal is detected. If for any of the arms the relative phase is shifted (input 0,1; 1,0), the signals detected at the microwave detection antenna have opposite signs and cancel out. The output signal is zero.
If both arms are subject to an Oersted field, both spin waves are shifted by the same phase, and again, the full signal is measured at the output. This gives the functionality of an XNOR logic device (logic equality). A second logic device functionality is implemented by using the Oersted field on top of the YIG arm to completely suppress the propagation (spin-wave switch), which gives a zero transmission (output 0) if both inputs are on (input 1,1), else (input 0,0; 0,1; 1,0) a signal is detected (output 1), which is the NAND functionality.
The disadvantage of YIG-based magnonic devices is that YIG cannot be integrated into standard semiconductor technology because high quality YIG films cannot be grown on silicon at the moment, but special substrates are needed. The industrial demand of miniaturization and integration into semiconductor technology is much better met by Permalloy (Ni$_{80}$Fe$_{20}$).
This is why the aforementioned concepts need to be transferred to smaller structures.
\begin{figure}[!ht]
\centering
\includegraphics[width=0.5\columnwidth]{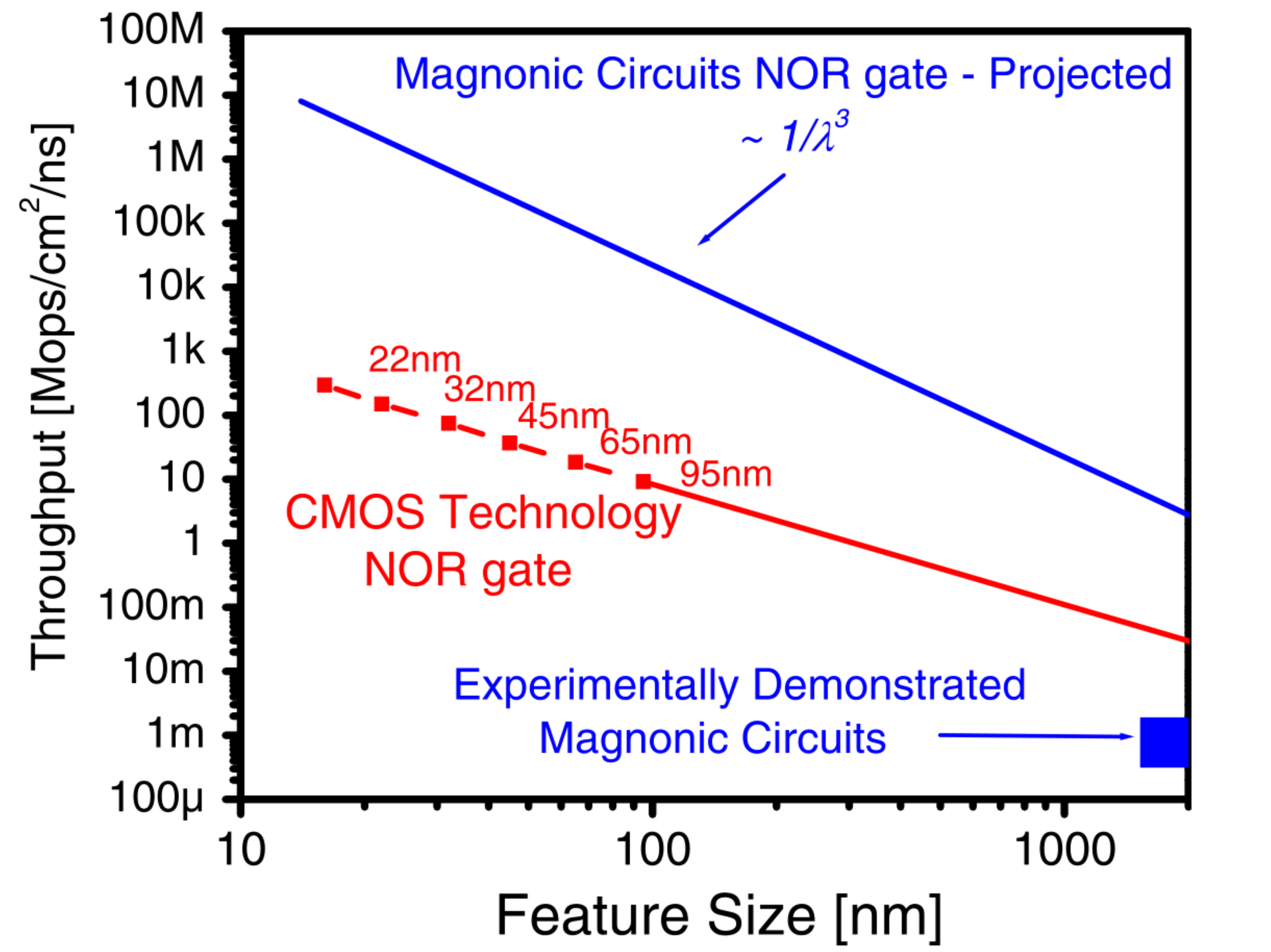}
\caption{ CMOS technology scaling versus spin-wave based magnetic circuits, from~\cite{ref4}. For spin-wave circuits, the number of operations per area per time (throughput) scales inversely proportional with the minimum feature size (wave length)~$\lambda$.}
\label{fig:03}
\end{figure}

In Fig.~\ref{fig:03} Wang et al.\ compare MOS logic and magnonic logic in terms of throughput (the number of operations per area per time) as a function of the minimum feature size~$\lambda$, which is the gate length for CMOS or the wavelength for a spin-wave circuit, respectively.
The throughput in devices realized so far is quite small and feature sizes lie in the micrometer range.
However, according to the projected estimates, spin logic may provide a throughput advantage of more than three orders of magnitude over CMOS in the future. Magnonic logics have several scaling advantages, one being the inversely proportional throughput of the spin circuit to the wavelength.
However, scaling down to spin-wave length in the nanometer range is in first order a material problem: one has to find materials where the damping of spin waves is low~\cite{ref9}. Half metals are an ideal candidate here, since electron and spin scattering channels can be decoupled~\cite{ref10}.
\begin{figure}[!ht]
\centering
\includegraphics[width=448pt]{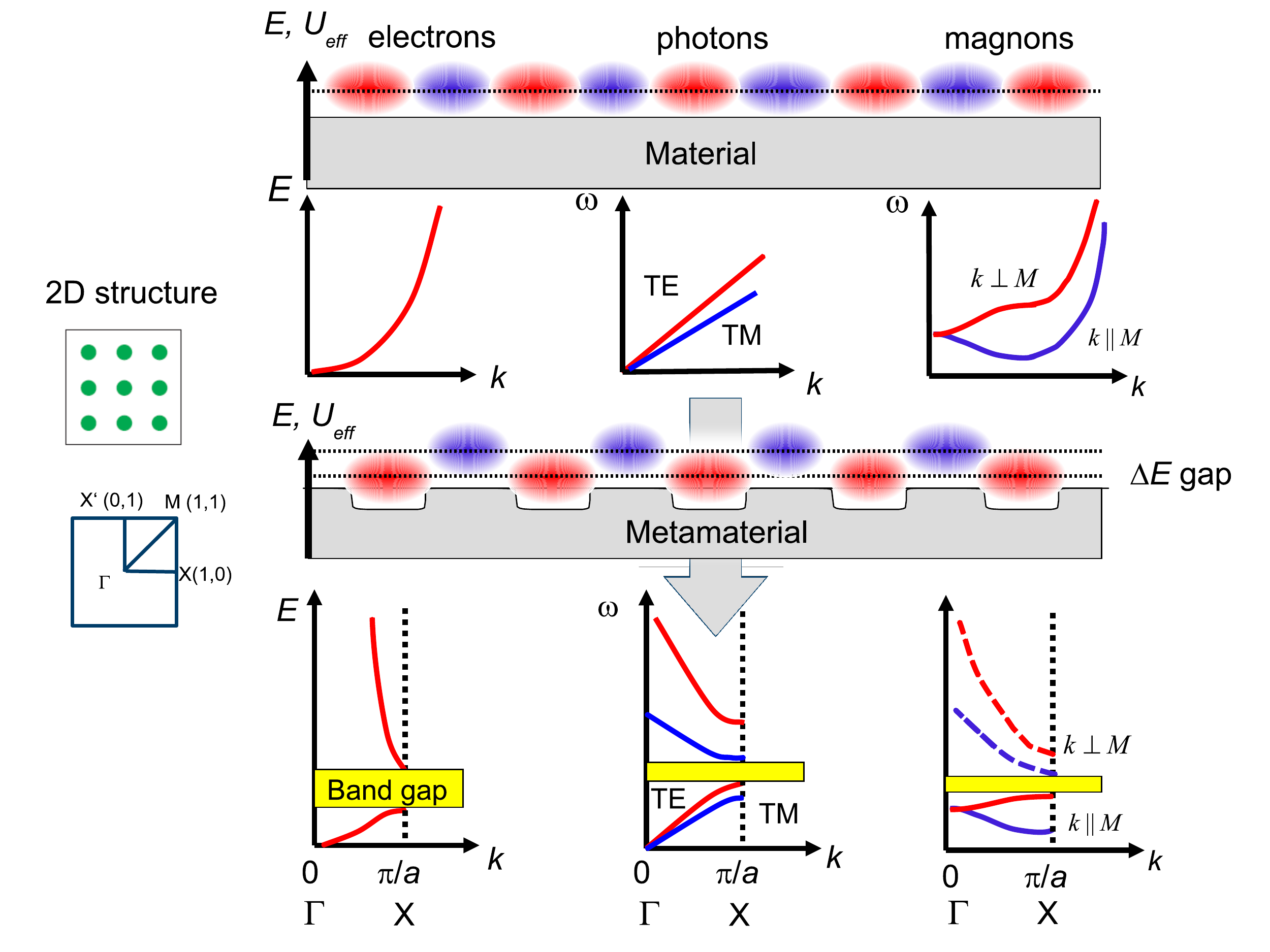}
\caption{Schematic analogy of electrons, photons, and magnons in periodic potentials. The dispersion is shown in the reduced zone scheme. Different dispersions result in characteristic differences in the band structure. A common feature is the possibility of the formation of band gaps. The polarization (transverse electric (TE) or transverse magnetic (TM)) for photons and the propagation direction~$k$, with respect to the applied magnetic field for magnons (parallel or perpendicular), are additional free parameters determining the band formation for the latter. ($k$-scale is logarithmic for magnons)}
\label{fig:04}
\end{figure}

\section{Tailoring artificial materials}
To realize devices described in the previous section, it is most convenient to microscopically manipulate the properties of a single material. This is the origin of the success of the semiconducting materials as silicon, which has a band gap at the Fermi level. Thus, no states are available for conduction inside the gap region.
To design the electronic properties, doping is necessary. The potential landscape is then formed by a combination of differently doped regions and various devices such as transistors can be realized to switch current flow through a device on and off. It is the flexibility of the material that allows a change in the conductivity by orders of magnitude and allows the shaping of the density of states at well-defined energies with a spatial precision down to nanometers. The band gap and a controlled introduction of defect states defines this flexibility.
In a similar manner, photonic materials can be designed. Spectacular examples are materials with negative index of refraction~\cite{ref11}, photonic band gap materials already existing in nature giving it its colorful life, and controlled ``defects'' realizing high quality resonators for light waves. These materials own novel properties that are very different from those of their constituents.
Any wave propagating probes a material with effective parameters designed by structure and composition. Therefore metamaterials are per definition artificial materials where the periodicity of the structure is smaller or equal than the wavelength.
This can be centimeter or millimeter in size: first metamaterials have been realized for microwave radiation~\cite{ref12}. Artificial materials also include those acting on water waves~\cite{ref13}, which allow the formation of refractive elements like lenses and sonic waves, to realize sonic crystals with negative refraction~\cite{ref14}. These examples from other areas utilize rather macroscopic waves transmitted through the metamaterial.
In the following, we will remind the reader of the basics of band structure formation in metals and covalent bonded materials. As a next step we will apply this formalism to photons in a periodically structured material with alternating index of refraction before we expand the formalism to magnons.

\subsection{Bloch conditions and band structure: weak periodic potential}
\begin{figure}
\centering
\includegraphics[width=430pt]{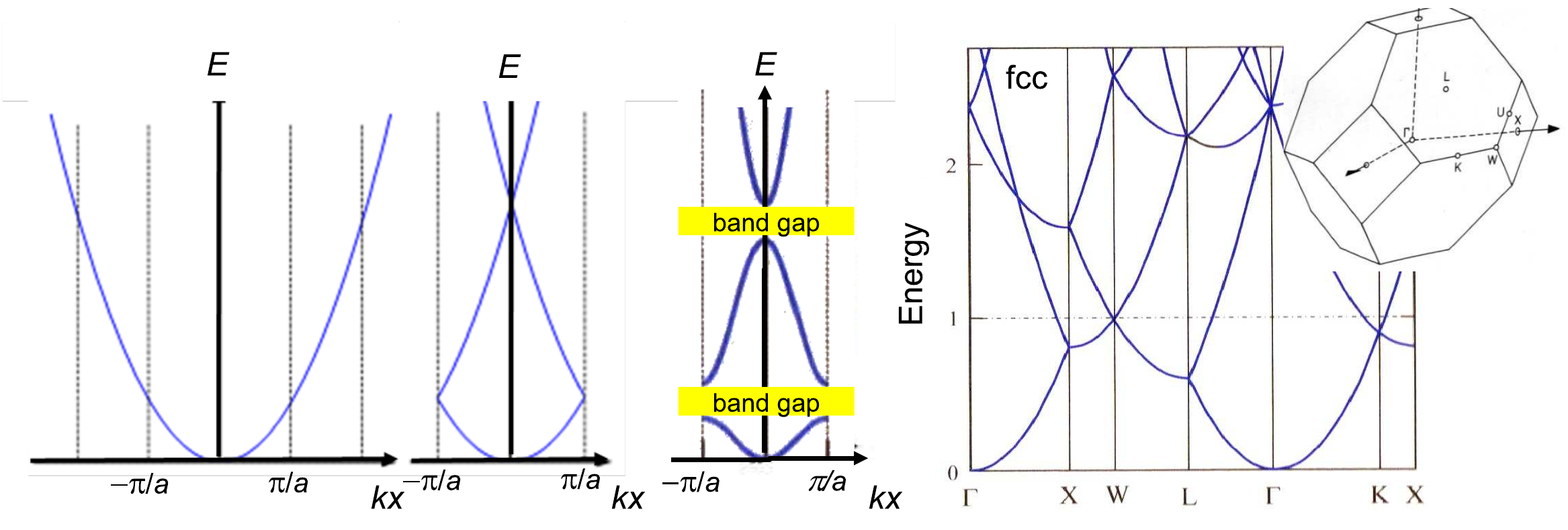}
\caption{Free electron band structure (left) in the periodic boundary scheme and reduced zone scheme. In a weak periodic potential a band gap opens up (middle). It is given by $2|V_G|$ where $V_G$ is the Fourier component of the periodic potential. Free electron band structure (right) for a fcc-lattice along the main symmetry directions (adapted from~\cite{ref15}).}
\label{fig:05}
\end{figure}

In a metal, the free electron ansatz is very successful, leading to Sommerfeld equations describing transport in an electron gas. The kinetic energy dominates the properties of the electrons. The energy levels are filled homogeneously up to the Fermi level leading to a Fermi  sphere in momentum space, which behaves isotropic for the different directions. The Hamilton operator is given by
\begin{equation*}\label{eq:hamilton}%
H = H_\mathrm{kin} + V(r).
\end{equation*}%
The solution for the free electrons where $V(r)$ is neglected, is simply given by a quadratic dispersion mirroring the increase of kinetic energy
\begin{equation*}\label{eq:wave-function-energy}%
\psi_k = e^{ikr},\quad E = \frac{ (\hbar k)^2 }{ 2m }.
\end{equation*}%
Turning on a periodic potential, the Bloch theorem allows to exploit the periodicity. The solution can be separated into two parts: a free electron propagating part with wave number~$k$ and a function that is periodic with the lattice~$u_{n,k}$. The new quantum state is described by a wave number~$k$ and a band index~$n$,
\begin{align*}%
V &= V(r+R),\\
\psi_{n,k} &= e^{ikr} u_{n,k}(r).
\end{align*}%
The periodic function~$u_{n,k}$ describes the electron distribution within one cell around the crystal atom. Without making any calculations of the exact band structure, already some general conclusions regarding the electron states in the resulting band structure can be drawn. As a consequence of the periodic potential, the electron dispersion mirrors the symmetry of the lattice.
Gaps open up especially around the zone boundary in the high symmetry directions. This can be seen in Fig.~\ref{fig:05} (middle) where, for a weak potential, the opening of the gap at the zone boundary is given.
The periodic functions~$u_{n,k}$ and their eigen energy~$E_{n,k}$ have to be calculated. To do this for a given potential, the plane wave ansatz~$\exp(ikr)$ is used to make a Fourier expansion for the wave function. It is most convenient to solve the set of Bloch states in $k$-space via the Fourier transform of the Schr\"odinger equation.
Then the Fourier coefficients of the periodic potential $V$ determine the coefficients of the Fourier expansion of the wave function.
This directly allows the calculation of the band gap: it is given by $2|V_G|$ where $V_G$ is the Fourier component of the periodic potential at the zone boundary~$k=G$.
In fact, we will see in Chapter 4 that for spin-wave states, similar approaches to calculate their band structure can be used, however, also with distinct differences arising form the different equations of motion including the symmetry of dipolar interactions mirrored by the vector nature of the magnetization~$M$.

\subsection{Electrons in a strong periodic potential}
\begin{figure}[!ht]
\centering
\includegraphics[width=430pt]{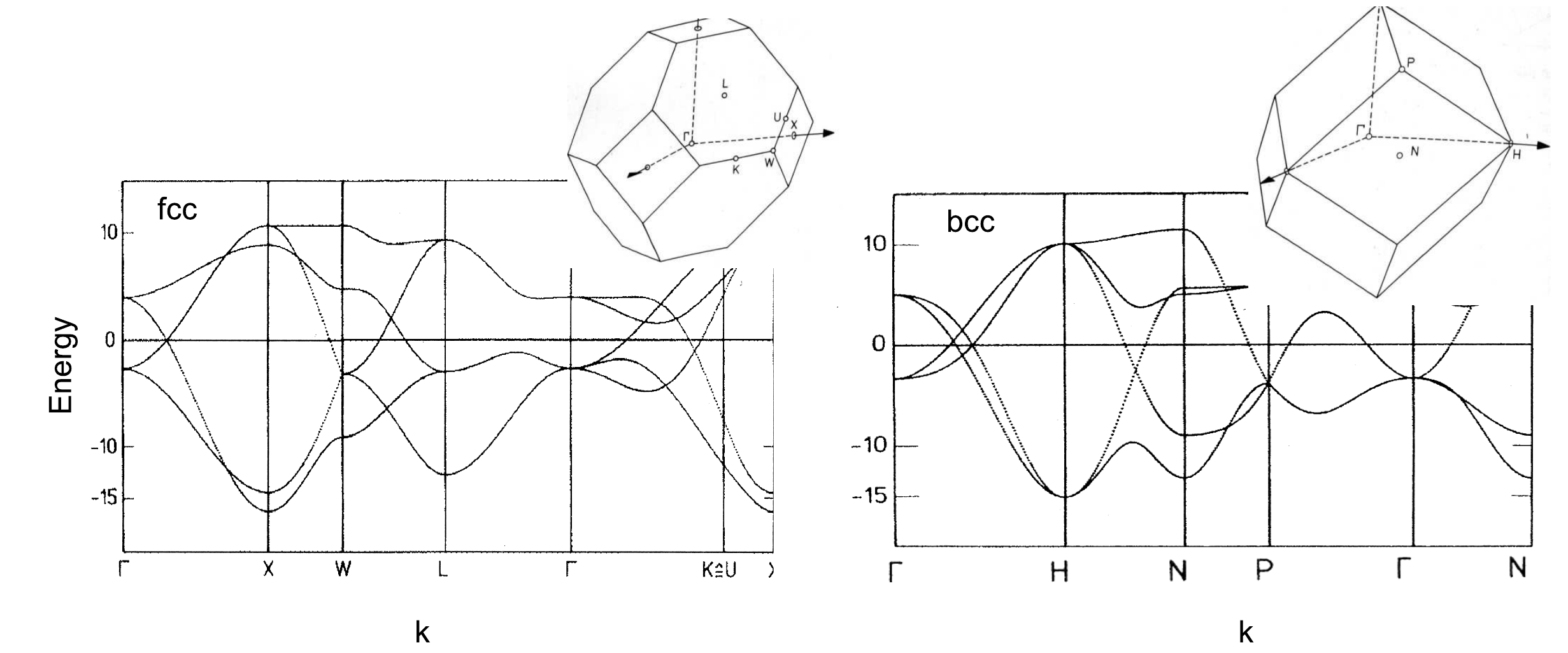}
\caption{Canonical tight binding d-band structure for fcc and bcc lattice along the main symmetry directions. The band width~$\omega_d$ of the cosine-like bands is given by the overlap~$t_d$ of the atomic orbital functions (here d-orbitals) into the different crystal directions $\sim 2 \sqrt{N_{nn}} t_d$. The overlap~$t_d$ determines the delocalization of the electrons within the band. The symmetry is determined by the underling atomic wave functions admixture (adapted from~\cite{ref15,ref16}).}
\label{fig:06}
\end{figure}

If the electron wave function possesses a localized nature, the free electron approach is not successful. The electronic band structure of such a material mirrors more strongly the orbital nature of the atomic wave functions. The s,p,d-like symmetry determines many of the fundamental properties of the material.
Examples are band ferromagnets, where the flat d-like electron bands are responsible for the formation of the ferromagnetic order~\cite{ref16}. Other examples include carbon, silicon and germanium of the group 4 elements in the diamond structure. The local electron density is described by the linear combination of s and p wave functions forming the $\mathrm{sp}^3$ tetraeder structure.
In that case, another possibility is to take the atomic wave function of the single atom as a basis to construct the bands. Typically, a Wannier function localized on an atom, but with decaying density of states to the neighboring atom, is used. The band width is given by the overlap to the next atom.
A large overlap means a large hopping rate and a strong delocalization and consequently a large bandwidth. The more localized the electrons are on the atomic-like orbitals, the smaller is the overlap and the smaller is the band width in energy. One can derive these tight binding bands for different crystal symmetries (Fig.~\ref{fig:06} for bcc and fcc). For example, in the bcc case, the splitting into the $d_\mathrm{xy}$, $d_\mathrm{xz}$, $d_\mathrm{yz}$ symmetry and  $d_\mathrm{3z^2-r^2}$, $d_\mathrm{x^2-y^2}$ symmetry are observed varying from a different distance to the next neighbor atom along the axes of the cube and along the diagonal.
The band width of the $d_\mathrm{3z^2-r^2}$, $d_\mathrm{x^2-y^2}$ bands is increased by a factor of $1.5$ along the $\Gamma$-X symmetry direction. The symmetry of the atomic wave function in connection with the lattice symmetry determines the electron states observed in the crystal; even though the states are strongly determined by their atomic wave functions and show a weaker dispersion, the next neighbor distance significantly influences the properties of the electron in the band such as localization. We will see here the similarities to the magnonic systems, where spin waves can also be of localized nature later and its symmetry is mirrored by a short range interaction.

\subsection{Photonic crystals -- photons in periodic potentials}
\begin{figure}[!ht]
\centering
\includegraphics[width=400pt]{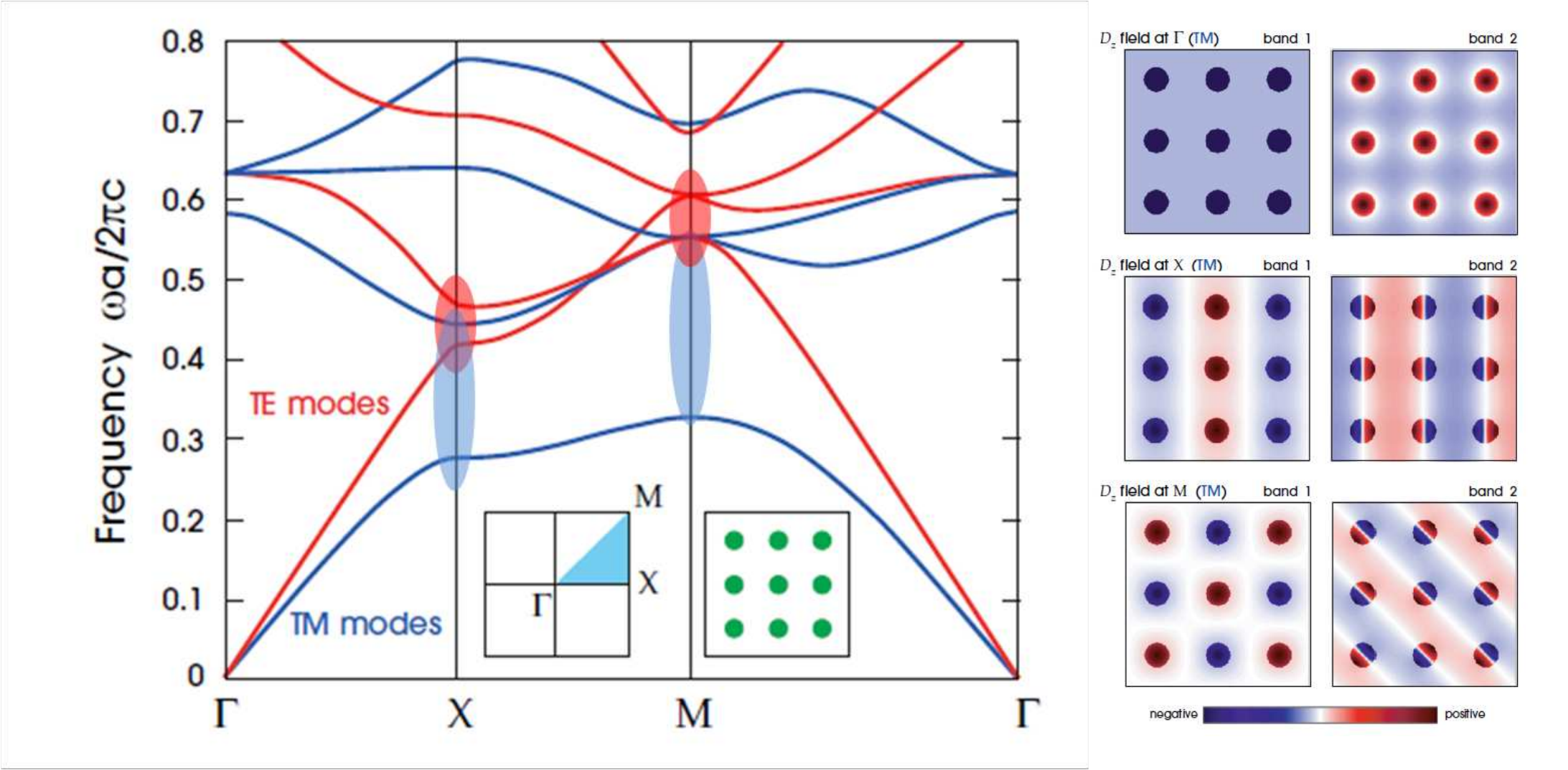}
\caption{Photonic band structure calculated for an array of two dimensional Al columns. Transversal electric (TE) and transverse magnetic (TM) modes show a different slope at the $\Gamma$-point. The TM-mode shows a lower velocity because it effectively sees the higher dielectric constant of the Al columns. Shaded areas mark the splitted bands at the high-symmetry points point. Right: At the zone boundary for the different bands, the standing waves with high and low energy are plotted (Al ($\varepsilon=8.9$) in air ($\varepsilon=1$)) (adapted from~\cite{ref17}).}
\label{fig:07}
\end{figure}

Similar to the periodic potential experienced by electrons around ion cores in a lattice, in a photonic crystal the periodic dielectric properties determine the periodic confinement for the photons. In general, if the potential varies periodically for a wave traveling through a two-dimensional artificial crystal, one can obtain a solution by labeling the modes with their band index~$n$ and wave vector~$k$ using Bloch's theorem, as described for the electrons before.
The same concepts can be applied despite the equation of motion being determined by the wave equation derived from the Maxwell equations. Equal to the periodic potential of the positive ion cores~$V_{U}$ in the solid state for electrons, for a photonic material the dielectric properties~$\varepsilon$ determine the refractive index, which determines the periodic confinement for the photons~$V_\varepsilon$~\cite{ref17}. To compare: in a magnonic material, the dispersion will be dependent on the local magnetization direction~$V_{M}$ and the induced internal field. In many cases photonic structures are two dimensional.
The mathematical ansatz to describe the electro-magnetic modes of a photonic crystal is similar to the description of delocalized electronic states in solids. The wave traveling through a two-dimensional artificial photonic crystal experiences a periodically varying potential arising from the dielectric constant~$\varepsilon$. In two dimensions the potential~$V_\varepsilon$ varies as
\begin{equation*}\label{eq:periodic-potential}%
V_{\varepsilon} = V_{\varepsilon}(r+R),
\end{equation*}%
where $R$ is a linear combination of the primitive lattice vectors for a two dimensional array in the x-y plane. Using Bloch's theorem and using $\rho$ as the projection of $r$ in the x-y plane, one obtains a solution labeling the modes with their band index~$n$ and wave vector~$k_\parallel$
\begin{equation*}\label{eq:photonics-Bloch}%
\psi_{n, k_z, k_\parallel} = e^{i k_\parallel \rho} e^{i k_z z} u_{n, k_z, k_\parallel} (\rho).
\end{equation*}%
One example taken from \cite{ref17} is given in Fig.~\ref{fig:07}. It has been calculated for a two-dimensional array of aluminum columns with the dielectric constants of Al ($\varepsilon=8$) and air ($\varepsilon=1$). In the case of an electro-magnetic wave, one has to distinguish two solutions in the two dimensional crystal. For a photon propagating through the crystal, the electric field can be polarized in the plane or out of the plane of the two dimensional structure.
Therefore, different dispersions are found for transverse electric (TE) and transverse magnetic (TM) polarization to the plane of the crystal: TE and TM modes show a different slope at the $\Gamma$-point.
The TM-mode shows a lower velocity because it effectively sees the higher dielectric constant of the Al columns. The corresponding modes at the zone boundary along the different symmetry direction at the X and M point are plotted to the right: they form standing waves which are localized either at the Al columns or in between.
The one with lowered energy is localized in the areas with high $\varepsilon$ and the one with increased energy in the areas with low $\varepsilon$, leading to distinct gaps in the photonic band structure in this direction. It can be easily understood that the gap is larger for the TM modes for the same argument that this mode, since its $E$-vector is in the plane, is strongly affected by the alternating dielectric constants.

\subsection{Novel functionalities of photonic crystals: slow photons}\label{sec:slow-photons}
A designed photonic band structure can be used to achieve novel optical properties. Quite generally, close to the zone boundary, the band flattens out. Approaching the zone boundary then means in terms of group velocity given by $d\omega / dk$, that the photon is ``slowed down''~\cite{ref18}.
Usually switching light by light is not possible: the change in the index of refraction is only small. However, when using slow light, a trick can be applied.
Exploiting the sensible reaction to the refractive index in the region where the band flattens out, it was experimentally demonstrated that it's feasible to guide light from one wave guide to another output, which is the realization of an ultrafast nanophotonic switch using slow light~\cite{ref19}. In detail, the spatial beating due to odd and even super-modes in two parallel wave guides is used for rerouting triggered by a femtosecond laser pulse.
This shows that a wave guide in a photonic material is not only a straight line of material conducting light as a glass fiber: the peculiar properties can be designed by the photonic structure around the wave guide giving the light completely new properties in the artificially structured material. We suggest it should be one aim in magnonics to put forward similar concepts to exploit the full possibilities of artificially structured materials for spin waves.
\begin{figure}
\centering
\includegraphics[width=0.7\columnwidth]{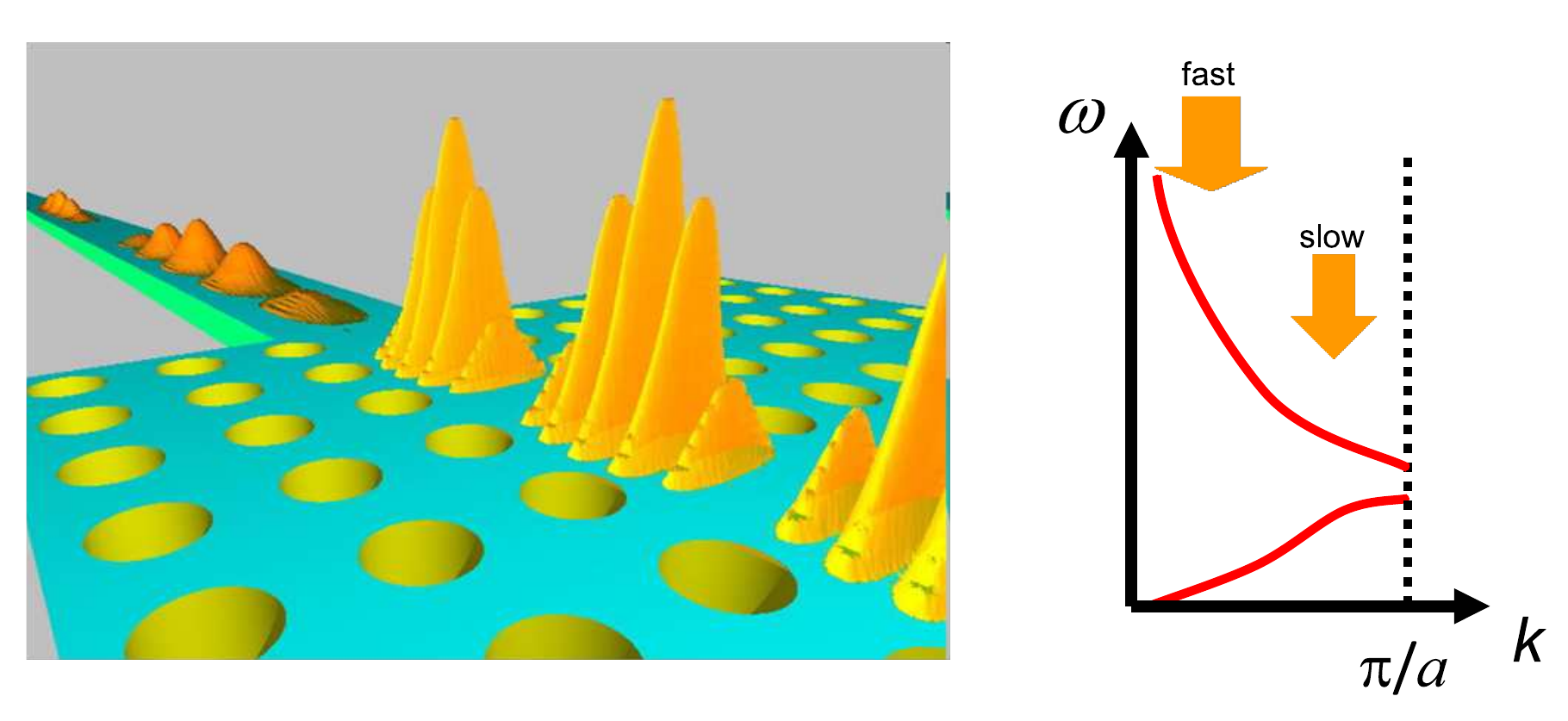}
\caption{Photons entering a wave guide. In the wave guide, the pulses are compressed and their intensity increased. The wave vector is close to the zone boundary and their propagation speed is slowed down (from~\cite{ref20}).}
\label{fig:08}
\end{figure}

\section{Magnonic crystals -- spin waves in periodic potentials}
We will now in more detail discuss the effect of structuring a magnetic film to form a magnonic crystal. That is, only two-dimensional systems are taken into account and beginning with a continuous film, the effects of a periodic structures are considered on an analytical as well as numerical level.
We will show that for the dipolar modes, the dispersion differs for different propagation directions~$k$ relative to the magnetization vector~$M$. Namely, the Damon-Eshbach ($k \perp M$) and the backward volume mode ($k \parallel M$) exist, which are different due to dipolar interactions.
As a consequence, the spin-wave states in a square magnonic crystal have a lower symmetry than the underlying lattice structure. The manipulation of the local magnetization direction or an applied field on a nanometer scale allows active dynamic control of the spin-wave diffraction on the nanometer scale.
Thus, a rotation of the magnetization relative to the lattice changes the energy landscape dramatically.

\subsection{Spin-wave dispersion from nanometer to micron range}\label{sec:spin-wave-ranges}
Before we discuss the propagation of spin waves in a structured medium, in analogy to the preceding chapters on electrons and photons in periodic potentials, we first have to say some words to the equation of motion for spin waves to be solved in the periodic environment. In a continuous, non-periodic system, the magnetization follows the Landau-Lifshitz-Gilbert equation of motion, which reads
\begin{equation}\label{eq:landau-lifshitz-gilbert}
\frac{d\boldsymbol{M}}{dt}=-\gamma\mu_0\boldsymbol{M}\times \boldsymbol{H}_\mathrm{eff}+\frac{\alpha}{M_S}\left(\boldsymbol{M}\times \frac{d\boldsymbol{M}}{dt}\right).
\end{equation}
We will only consider a field applied in the plane of the film. The case of perpendicular magnetization is technologically less favorable since it demands strong fields for saturation, and it is also physically less interesting because the spectrum is isotropic. In the macrospin approximation, all the individual spins are considered to precess in phase and the thin-film solution to Eq.~\eqref{eq:landau-lifshitz-gilbert} is given by the Kittel equation
\begin{equation}\label{eq:kittel}
\left(\frac{\omega_k}{\gamma\mu_0}\right)^2=H_x \left(H_x + M_S - \frac{2K_z}{\mu_0M_S}\right).
\end{equation}
Here, $K_z$ represents an effective anisotropy in the out-of-plane direction. If one allows for solutions to Eq.~\eqref{eq:landau-lifshitz-gilbert} other than uniform precession, it becomes important to distinguish two different interactions coupling magnetic moments or individual spins, respectively. The corresponding solutions will be discussed in the following sections \ref{sec:dipolar-spin-waves} and \ref{sec:exchange-spin-waves}.
\begin{figure}
\centering
\includegraphics[width=227pt]{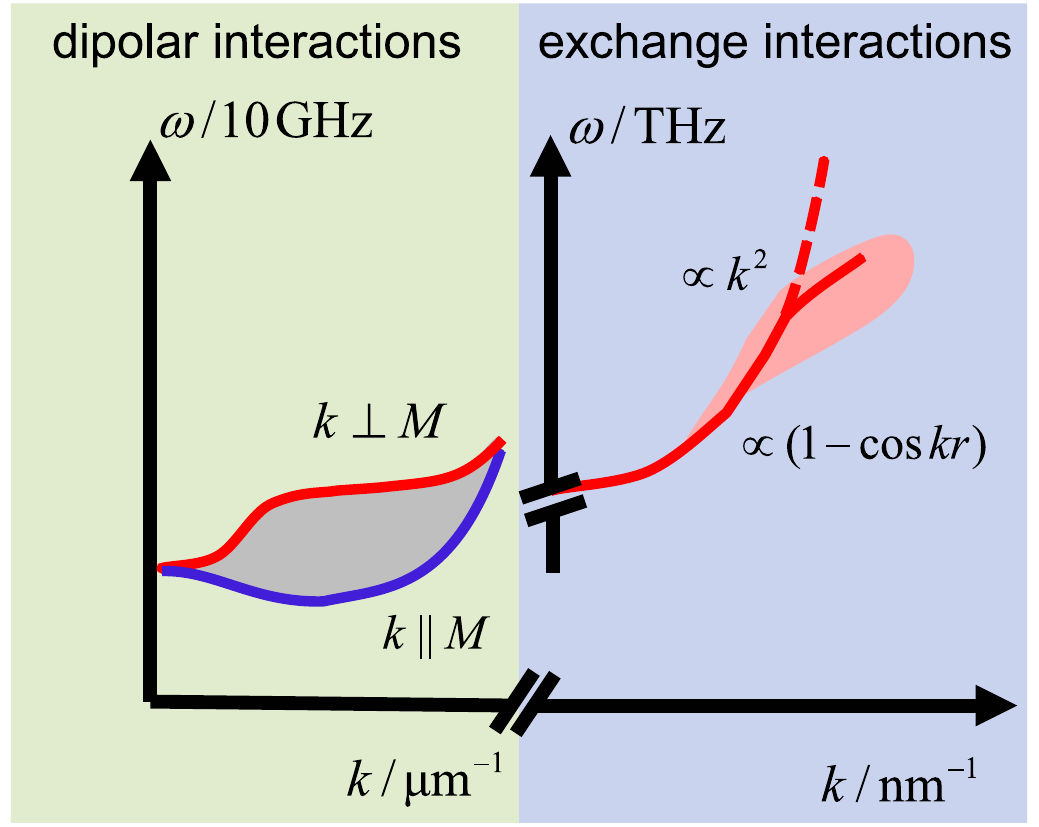}
\caption{Schematic spin-wave dispersion. In the micron wave length (left), dipolar interactions dominate. The Damon-Eshbach and backward volume modes are shown. At an arbitrary angle in between further modes are found (gray shaded area). In the nanometer region (right), the dispersion shows cosine-like behavior. Here, the exchange interaction is dominant and can be approximated by a parabola for small energies. It intersects in a broad region of high-energy spin-waves excitation, where spin waves are heavily damped (red shaded area).}
\label{fig:09}
\end{figure}

\subsubsection{Dipolar spin waves}\label{sec:dipolar-spin-waves}
Solving Eq.~\eqref{eq:landau-lifshitz-gilbert} in the magnetostatic limit yields a manifold of dynamic solutions for a thin magnetic film of thickness~$t$~\cite{kalinikos86,Damon,ref65}.
All can be classified by their propagation direction with respect to the magnetization. Modes whose frequency lies above the one of the uniform precession generally tend to localize at the surface of the film and have a wave vector pointing perpendicular to the magnetization.
The geometry with $k \perp M \parallel H$ is identified as the Damon-Eshbach (DE) geometry~\cite{Damon}. The particular relation between the magnetic field and mode frequency is given by
\begin{equation}\label{eq:damon-eshbach}
\left(\frac{\omega_\mathrm{DE}}{\gamma\mu_0}\right)^2= H_x \left(H_x + M_\mathrm{S} - \frac{2K_z}{\mu_0M_\mathrm{S}}\right)+\frac{M_\mathrm{S}^2}{4}\Big(1-e^{-2 |k_\mathrm{DE}| t}\Big).
\end{equation}
For directions $k \parallel H$, the so-called backward volume waves occur, with a reduced precession frequency compared to the uniform precession. This leads to a negative dispersion, as the waves travel ``backward'' in phase. In between, one finds a manifold of spin waves corresponding to the continuous change of angle from parallel to perpendicular (grey shaded region in Fig.~\ref{fig:09}). This peculiar energy dependence on the angle arises from the dipolar interactions.
For wave lengths below a micron, the energy difference becomes smaller and both dispersions are degenerate in energy. Here, the exchange interaction becomes important, so that this contribution has to be taken into account for mixed dipole and exchange spin waves in an intermediate region of length scales~\cite{kalinikos86}. The steep uprise of the energy for smaller wave length marks the dominance of the exchange interaction.
In a thin film, the confinement perpendicular to its plane is dominated by the exchange interaction, while in lateral directions no such restrictions exist and dipolar magnetostatic spin waves may be formed.

\subsubsection{Exchange spin waves}\label{sec:exchange-spin-waves}
Since spin-wave lengths span several orders of magnitude from tens of microns (even higher for low-damping materials) to below $1\, \mathrm{nm}$, also their frequencies may vary from GHz to THz. In addition, the frequency for a given wave length can be shifted by the magnetic field.
This broad region in length and time scales is one reason that makes spin waves so interesting for high frequency applications. However, also the dominating interaction varies: at wave lengths below $100\,\mathrm{nm}$, the dispersion is dominated by the exchange interaction. The magnetostatic contribution to the energy of the wave can be neglected.
This simple picture is solved in many solid states physics textbooks for a chain of precessing spins, where next neighbors are coupled by the exchange interaction. The solution is a $(1-\cos(kr))$ like behavior. As a consequence of neglecting the anisotropic dipolar contribution, the dispersion in the exchange limit does not change with the magnetization direction.
It only depends on the next neighbor distance~$r$ and the strength of the exchange interaction, and can be calculated in the ``frozen magnon'' picture from the electronic structure. For small $k$, the dispersion can be approximated to a quadratic form. In that region, the energy increases quadratically with momentum~$\hbar k$ similar to the free electron behavior. One can think of realizing ``free electron like'' magnonic materials.
\begin{figure}[!ht]
\centering
\includegraphics[width=400pt]{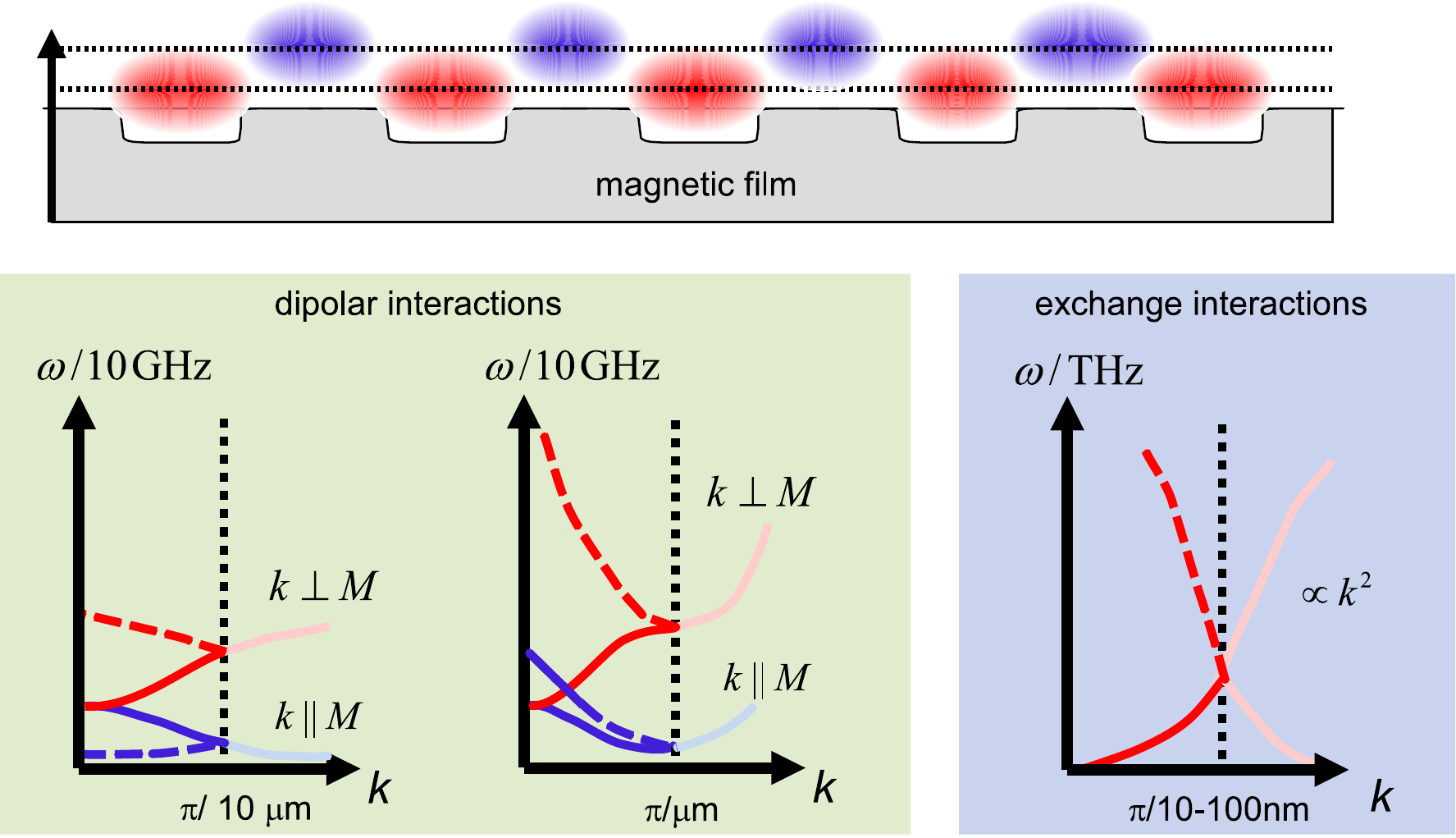}
\caption{Schematic spin-wave dispersion for the different length scales in the periodic zone scheme. For a spin-wave Bloch state, prominent effects in the band structure are expected: $10\,\mathrm{\mu m}$ range (left), $1\,\mathrm{\mu m}$ (middle) and $100\,\mathrm{nm}$ (right).}
\label{fig:10}
\end{figure}

At even higher energies, the spin waves approach the THz regime. Their energy gets comparable to single spin-flip excitations (Stoner excitations) between the bands of different spin character. These high-energy spin-wave modes in the THz range are heavily damped, indicated by the red shaded area.
Dynamic excitations have to be thought of as a superposition of multiple spin excitations propagating through the ferromagnet~\cite{ref21,ref22}. Life times of these high-energy spin waves of nm length scales are a few picoseconds before the decay into other spin-wave excitations of lower energy~\cite{ref23}.
Investigated heavily in the 80s~\cite{ref24}, their physics have drawn new interest. They are thought to be the key to the further understanding of ultrafast demagnetization processes after femtosecond laser excitations. Because of their short wave length and lifetime, at the moment, these are not suited for studies of the formation of magnonic bands.
They propagate only a few nanometers. In the following, we will discuss the formation of magnonic band structures in the dipolar region.

\subsection{Band structures of magnonic crystals}
In this section we will discuss first what are the expectations for a band structure in a magnetic material that is periodically modified. If novel spin-wave states appear due to the periodic modification, a magnonic crystal is formed. Second, we will calculate for one example the respective band structure.
For wave lengths larger than $1\, \mathrm{\mu m}$, the dispersion of the spin waves is dominated by dipolar interactions. Corresponding frequencies are below $20\,\mathrm{GHz}$, depending on the magnetic material and applied field.
A steep quadratic increase is related to the nanometer wave length range. The strong magnetic exchange interaction takes over as the frequency reaches $1\,\mathrm{THz}$. The dispersion has completely different slopes for both regions.
The consequence is shown in Fig.~\ref{fig:10}: in the short wave length range a periodic modification will lead to a ``quasi free electron''-like band structure in the periodic zone scheme (schematically constructed on the left), similar to what is found in micromagnetic simulations~\cite{ref25}. If dipolar interactions dominate, the energy splits for the Damon-Eshbach and backward volume modes: the band structure will be anisotropic with regard to the applied field.
When the structure size is larger than the micrometer range, the negative dispersion of the backward volume mode leads to the surprising result that the second band can be lower in energy than the first band. We will see that this unexpected feature is found as well in our band structure calculations. A very complex spin-wave manifold can develop with many bands having similar energies. The three types of magnonic crystals classified in Fig.~\ref{fig:10} will show a very different behavior: dipolar bands with negative or positive dispersion as well as quadratic exchange dominated bands.

In the following, we will discuss in greater detail how, similar to the Bloch theorem applied to electrons and photons, a magnonic band structure of a periodic ferromagnetic system can be computed. A theory was developed by Puszkarski et al.~\cite{Vasseur,Krawczyk}. A periodic alteration of the saturation magnetization was put into the Landau-Lifshitz Gilbert equation and solved by a plane-wave method.
This theory is based on infinitely extended, three dimensional magnonic crystals. While this ansatz emphasizes the close relationship of photonic and magnonic crystals, it neglects the particular property of inhomogeneous internal magnetic fields by setting $H_\mathrm{int}=H_\mathrm{ext}$.
Note, that only if the regions with an inhomogeneous $H_\mathrm{int}$ are restricted to a vanishing fraction of the magnonic crystal's unit cell, this can be justified.
This is for example the case for a two-dimensional magnonic crystals, whose typical structure size is much smaller than its thickness. We will in the following show, how the plane-wave ansatz can be modified to apply to such a magnonic medium --~exemplarily represented by an antidot lattice.

Spin waves with a wave length~$\lambda$ which is much bigger than the film's thickness~$t$ (external field in the film's plane), have almost uniform mode profiles in the z-direction.
Thus, in the following the lower order modes' profiles will be assumed to be uniform and the so-called uniform-mode-analysis will be applied (see Hurben and Patton for details~\cite{Hurben}).
The basic idea of this approach is to neglect the exchange interaction, valid for sufficiently small $k$ as opposed by the full theory in Ref.~\cite{kalinikos86}, and solving the linearized Landau-Lifshitz equation
\begin{align}
    i\Omega m_y -m_z+\frac{M_S}{H}h_z&=0\nonumber\\
    i\Omega m_z +m_y-\frac{M_S}{H}h_y&=0.\label{gl:LLGuni}
\end{align}
The ansatz for the dynamic magnetization in the framework of the uniform-mode-analysis reads:
\begin{align}
    m_y(x,y)=m_ye^{i\left(k_xx+k_yy\right)}\nonumber\\
    m_z(x,y)=m_ze^{i\left(k_xx+k_yy\right)}.\label{gl:linLL}
\end{align}
In the above equations $m_y$ and $m_z$ are constant across the thickness. From the electromagnetic boundary conditions, it follows that the dynamic magnetic field reads:
\begin{align}
    h_y&=-m_y \frac{k t}{2} e^{i\left(k_xx+k_yy\right)}\sin^2\phi\nonumber\\
    h_z&=\left(-m_z+m_z\frac{k t}{2}\right)e^{i\left(k_xx+k_yy\right)}.\label{gl:huni}
\end{align}
By combining Eq.~\eqref{gl:linLL} and Eq.~\eqref{gl:huni} into Eq.~\eqref{gl:LLGuni}, the following direction-dependent dispersion relation can be derived:
\begin{align}
    \omega=\frac{g\mu_B\mu_0}{\hbar}\sqrt{H_\mathrm{ey}H_\mathrm{ez}},\label{gl:uniform}
\end{align}
In Eq.~\eqref{gl:uniform} $H_\mathrm{ey}=H+M_\mathrm{S}\tfrac{k t}{2}\sin^2\phi$, $H_\mathrm{ez}=H+M_\mathrm{S}-M_\mathrm{S}\tfrac{k t}{2}$, where $M_\mathrm{S}$ is the saturation magnetization, $g$ is the gyromagnetic ratio and $\phi$ is the angle between the external field and the wave vector. As a next step, a periodic modulation of the saturation magnetization is introduced:
\begin{align}
    M_\mathrm{S}( \boldsymbol{r} )=\sum_{\boldsymbol{G}} M_\mathrm{S}( \boldsymbol{G} ) e^{i \boldsymbol{G r} }.\label{gl:msfourier}
\end{align}
Where $\boldsymbol{G} = \left[G_n,\,G_m\right]^T=\left[\tfrac{n2\pi}{a},\,\tfrac{m2\pi}{a'}\right]^T$ is a two-dimensional vector of the reciprocal lattice. Any geometry can now be specified by an analytic expression for the Fourier components $M_\mathrm{S}(\boldsymbol{G} )$.
In order to circumvent additional boundary conditions arising from the air regions in the experimentally interesting geometry of antidot lattices, the following trick can be done: by filling the antidots with some artificial ferromagnet with a very high magnetic moment, spin waves in such a system either exist in the antidot, or in the surrounding matrix. Thus, a periodic energy landscape is constructed, which confines the spin waves.
After solving the eigenvalue problem, one can remove the unphysical solutions, which predominantly dwell in the antidot.
The rather smooth nature of the boundary, as a result of the cutoff by the Fourier expansion~\eqref{gl:msfourier} tends to mix the solutions for the matrix and the antidot. This results in non-vanishing imaginary parts for the frequencies of the modes in the matrix.
However, one would need to use an infinite number of Fourier components to model a sharp transition between matrix and antidot, which would immediately violate the condition $\lambda\gg t$.

In the case of magnetic discs ($M_\mathrm{S}^2$) with radius~$r$, which are periodically arranged on a square lattice with side length~$a=a'$ and embedded in a magnetic matrix ($M_\mathrm{S}^1$), an analytical expression for the Fourier components of the magnetization profile can be found in~\cite{Vasseur}
\begin{align}
    M_\mathrm{S}( \boldsymbol{G} )=\frac{2f\left(M_\mathrm{S}^1-M_\mathrm{S}^2\right)}{P}J_1\left(P \right),\nonumber
\end{align}
where $P=r| \boldsymbol{G} |$, $f=\pi r^2\, a^{-2}$ is the filling fraction and $J_1$ is the Bessel function of first order. Note that a modulation of other material properties like the gyromagnetic ratio or the exchange stiffness are not considered here. The generalization of Eq.~\eqref{gl:linLL} for the periodic material is a Bloch wave expansion:
\begin{eqnarray}
    \boldsymbol{m} ( \boldsymbol{r} )=\sum_{\boldsymbol{G}} \boldsymbol{m_k} (\boldsymbol{G}) e^{i(\boldsymbol{k} + \boldsymbol{G}) \boldsymbol{r}}\label{gl:mdynfourier}
\end{eqnarray}
Using the uniform-mode-analysis, the dynamic magnetic field components are
\begin{align}
    h_y&=\sum_{\boldsymbol{G}} h_{y,\boldsymbol{k}}(\boldsymbol{G})  e^{i(\boldsymbol{k} + \boldsymbol{G}) \boldsymbol{r}}\nonumber\\
    &=\sum_{\boldsymbol{G}} \left(-m_{y,\boldsymbol{k}}(\boldsymbol{G}) \frac{|\boldsymbol{k} + \boldsymbol{G}|\cdot t}{2}\sin^2\phi_{\boldsymbol{G}}\right)e^{i(\boldsymbol{k} + \boldsymbol{G}) \boldsymbol{r}},\nonumber\\
    h_z&=\sum_{\boldsymbol{G}} h_{z,\boldsymbol{k}}(\boldsymbol{G})  e^{i(\boldsymbol{k} + \boldsymbol{G}) \boldsymbol{r}}\nonumber\\
    &=\sum_{\boldsymbol{G}} \left(-m_{z,\boldsymbol{k}}(\boldsymbol{G}) + m_{z,\boldsymbol{k}}(\boldsymbol{G})\frac{|\boldsymbol{k} + \boldsymbol{G}|\cdot t}{2}\right) e^{i(\boldsymbol{k} + \boldsymbol{G}) \boldsymbol{r}},\label{gl:hdynfourier}
\end{align}
where $\phi_{\boldsymbol{G}}$ is the angle between the external field and $\boldsymbol{K}=\boldsymbol{k}+\boldsymbol{G}$. The equations~\eqref{gl:mdynfourier}, \eqref{gl:hdynfourier} and~\eqref{gl:LLGuni} provide a system of equations (finite number of $N$ lattice vectors)
\begin{align}
    \tilde{M}\vec{m}_{\boldsymbol{k}}^j=i\Omega_j \vec{m}_{\boldsymbol{k}}^j,\label{gl:EWP}
\end{align}
the eigen values of which have to be determined. Here, $\vec m_{\boldsymbol{k}}^j = \left[m_{y,\boldsymbol{k}}^j (\boldsymbol{ G}_1),..., m_{y,\boldsymbol{k}}^j(\boldsymbol{G}_N) ,m_{z,\boldsymbol{k}}^j (\boldsymbol{G}_1),...,m_{z,\boldsymbol{k}}^j(\boldsymbol{G}_N) \right]^T$. The eigenvalues $\Omega_i=2\pi{f_i}\left(\gamma\mu_0 H\right)^{-1} $ are proportional to the eigenfrequencies $f_i$ and the mode profiles can be constructed from the eigenvectors $\vec{m}_{\boldsymbol{k}}^j$. Note that the $2N\times 2N$-matrix $\tilde{M}$ has a block-diagonal form:
\begin{align}
    \tilde{M} & =\left\{ \tilde M_{yy}, \tilde M_{yz}; \tilde M_{zy}, \tilde M_{zz} \right\}.\nonumber
\end{align}
In the equation above $\tilde M_{yy}^{ij} = \tilde M_{zz}^{ij} = 0$ and
\begin{align}
    \tilde M_{yz}^{ij}&=\delta_{ij} - \frac{1}{H} \left(-1+\frac{|\boldsymbol{k}+\boldsymbol{ G}^j|d}{2}\right) M_S(\boldsymbol{G}^i-\boldsymbol{G}^j) + \frac{H_\mathrm{dem}((\boldsymbol{G}^i-\boldsymbol{G}^j) )}{H} \nonumber\\
    \tilde M_{zy}^{ij}&=-\delta_{ij} - \frac{1}{H} \frac{|\boldsymbol{k}+\boldsymbol{G}^j|d}{2} \sin^2\phi_{\boldsymbol{G}^j} M_\mathrm{S}(\boldsymbol{G}^i-\boldsymbol{G}^j) - \frac{H_\mathrm{dem}((\boldsymbol{G}^i-\boldsymbol{G}^j))}{H} ,\nonumber
\end{align}
with $i,j=1\ldots N$. In this formula a locally varying static demagnetizing field~$H_\mathrm{dem}$ was taken into account by including a Bloch wave formulation of this field, which is similar to the expression used for the magnetization profile:
\begin{align}
    H_\mathrm{dem}(\boldsymbol{r})&=\sum_{\boldsymbol{G}} H_\mathrm{dem}(\boldsymbol{G})e^{i\boldsymbol{G r}}.\label{gl:hdem}
\end{align}
In practice, the demagnetizing field can be obtained from numerical simulations, or from analytic expressions. Such a field may strongly alter the boundary conditions for the confinement. Since the model for the antidots confines the waves in yet another way, it cannot be included for this approach.
Nevertheless, artificial structures exist~\cite{ref26}, where the full model, taking into account the demagnetizing field and the exchange field, could similarily be applied. For a realistic situation, one problem has to be solved: the accurate description of a potential demands many reciprocal lattice vectors. But in this case the initial assumption $\lambda \gg t$ may be violated.
A good agreement between theory and experiment may then only be expected for the lowest order branches. For the antidot model, we will only use thirteen lattice vectors here to discuss general trends. Note the sketched theory could be very easily altered to describe the situation, where --~as for active devices~-- the spatially varying quantity is the applied field.
\begin{figure}
\centering
\includegraphics[width=0.8\columnwidth]{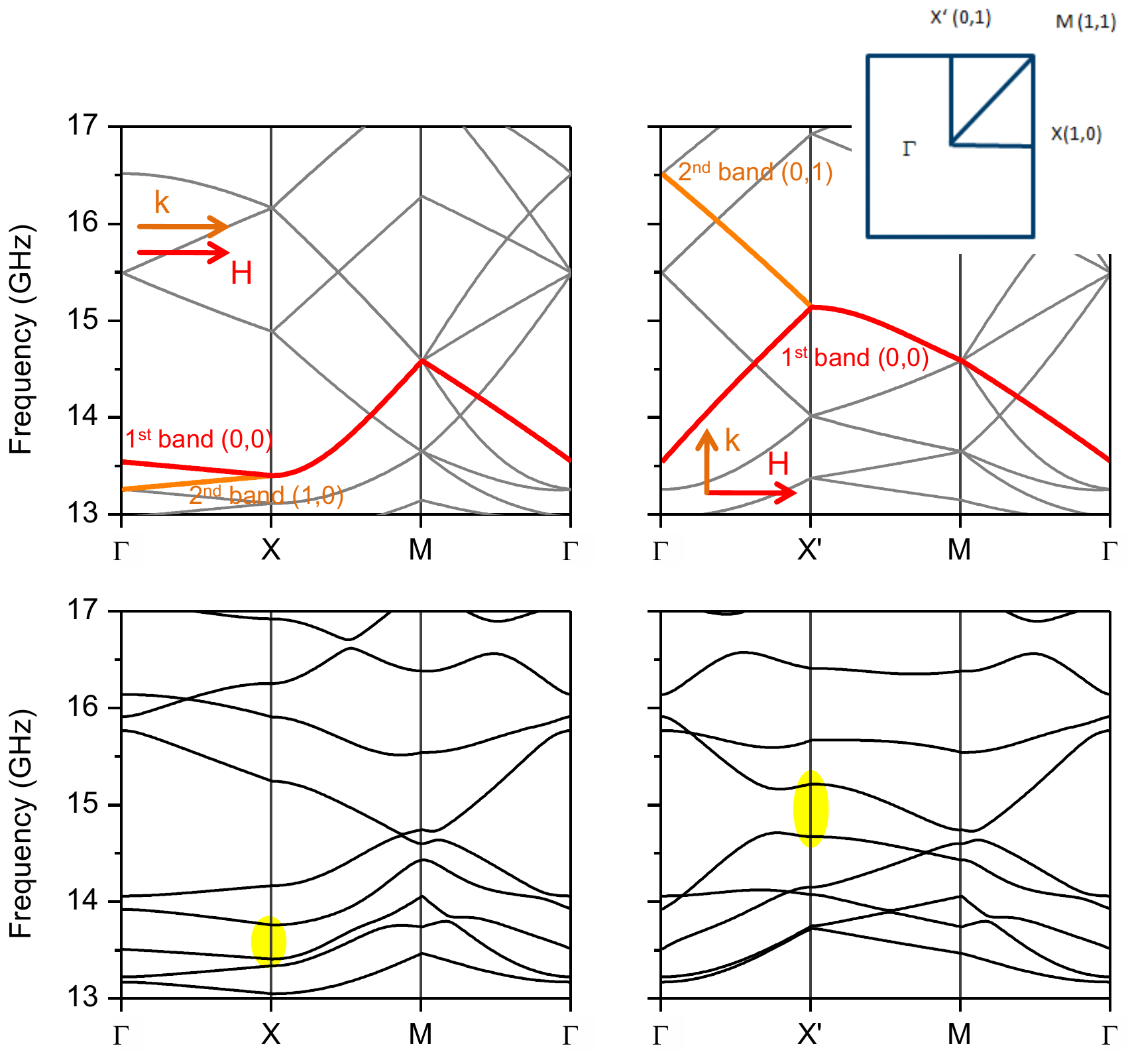}
\caption{Band structure calculation for a CoFeB film for a two dimensional square lattice. Top: Free spin-wave band structure with first and second band marked in red and orange, respectively. Bottom: Solved by a set of Bloch states for a film thickness of $t=50\,\mathrm{nm}$, hole distance $a=3.5\, \mathrm{\mu m}$ and hole diameter $d= 1\, \mathrm{\mu m}$. The splitting at the high symmetry points X, X' at the zone boundary is marked with the shaded yellow area.}
\label{fig:11}
\end{figure}

Band structure calculations are performed for a CoFeB film with a thickness $t=50\, \mathrm{nm}$, a hole distance $a=3.5\, \mathrm{ \mu m}$ and hole diameter $d= 1\, \mathrm{\mu m}$ (Fig.~\ref{fig:11}), and a nickel film with the same geometrical parameters (Fig.~\ref{fig:13}). A field of~$130\, \mathrm{mT}$ is applied in the plane along the (1,0) direction. The discs are filled with an artificial ferromagnet, with a high saturation magnetization of $10\, \mathrm{T}$.
The artificial solutions, which predominantly dwell in the discs, are by virtue of the high magnetic moment shifted above the depicted frequency range. This ansatz separates the solutions into two sets. Only the low frequency which are of interest are given in Fig.~\ref{fig:11}.
\begin{figure}
\centering
\includegraphics[width=0.8\columnwidth]{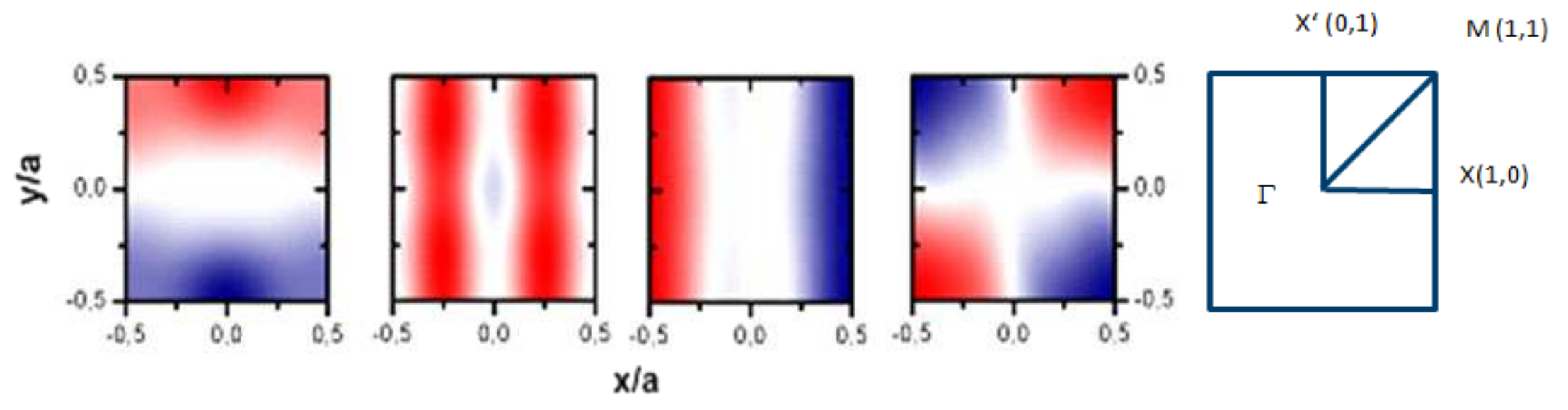}
\caption{Distribution of the spin-wave amplitudes at the point of high symmetry X, X' and M (from~\cite{ref27}). From the left to the right, one can see mode profiles of the first band at X', $\Gamma$, X and M. Different colors denote positive and negative phase.}
\label{fig:12}
\end{figure}
\begin{figure}
\centering
\includegraphics[width=0.75\columnwidth]{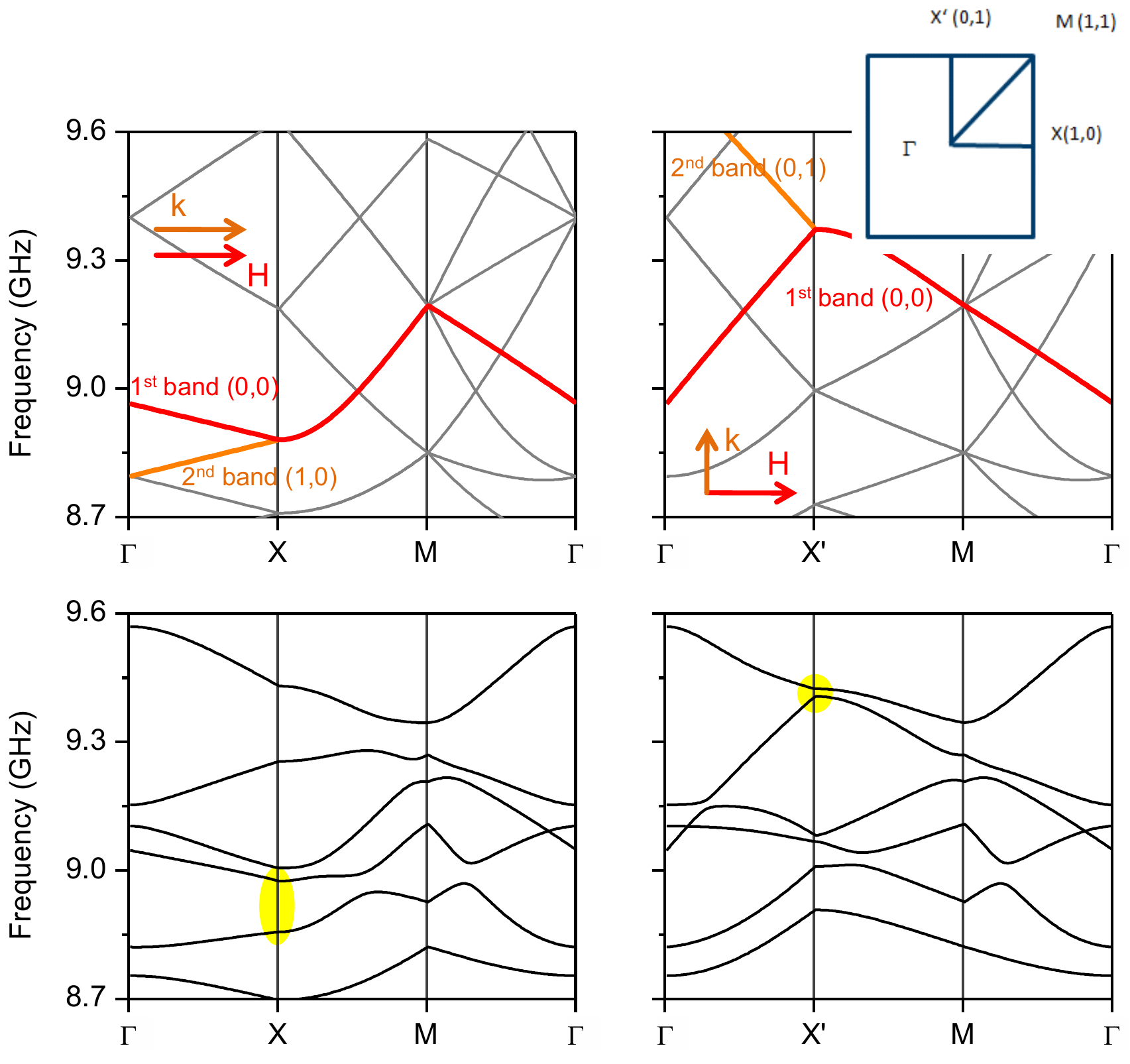}
\caption{Band structure calculation for a Ni film for a two dimensional square lattice. Top: Free spin wave band structure with first and second band marked in red and orange, respectively. Bottom: Solved by a set of Bloch states for a film thickness of $t=50\, \mathrm{nm}$, hole distance $a=3.5\, \mathrm{\mu m}$ and hole diameter $d= 1\, \mathrm{\mu m}$. The splitting at the high symmetry points X, X' at the zone boundary is marked with the shaded yellow area.}
\label{fig:13}
\end{figure}

Because of the larger saturation magnetization~$M_\mathrm{S}$, the bands for CoFeB are much higher in frequency.
In the following we will describe some typical features. For comparison, the free spin-wave band structure is shown, with the first and second band marked. For this purpose, we have taken Eq.~\eqref{gl:uniform} and plotted it in a reduced zone scheme. As expected, no band gaps at the zone boundary are present.
On the left side of Fig.~\ref{fig:11} the propagation direction is parallel to the applied field and thus determined by the backward volume geometry. As expected, with respect to the schematic bands given in Fig.~\ref{fig:10} the second band is below the first band in energy. At the right, for the $\Gamma\to\text{X'}$ direction, the band dispersion is determined by the Damon-Eshbach geometry, showing a steep increase.
The band calculation clearly reveals a different dispersion for the magnetization along the (1,0)-direction. Only for the $\mathrm{M}\to\Gamma$ point the bands are the same.
At the points of high symmetry, a splitting is observed which is of about $0.5\,\mathrm{GHz}$ in frequency.
The uniqueness of the solutions inhibits sections of individual bands, as they appear in the free spin-wave picture. Instead, in the hybridization regions, the modes repel each other and interchange their character.
Detailed theoretical studies of the group around Puszkarski showed the appearance of gaps in the band structure of two-dimensional~\cite{Vasseur} and three-dimensional~\cite{Krawczyk} materials where the spin-wave propagation is forbidden.
They find that the size of the magnonic gap increases if the contrast between the constituents in magnetization or exchange stiffness is increased.

\subsection{Periodic dipolar potential in structured films: micromagnetic simulations}\label{sec:micromag}
Apart from analytical descriptions, numerical simulations provide a convenient tool to investigate magnonic crystals and predict their spectral response, please refer also to a current review especially devoted to this topic~\cite{ref28}. For the purpose of numerical simulation, it is necessary to implement periodicity. The main physical entity to handle in this context is the long-range static demagnetizing field, which is of dipolar origin.
The easiest way to do this is to construct a simulation volume that consists of many unit cells. One can then assume that, at least in the central cell, the surroundings approximate infinite periodicity. Of course, such an approach demands high computational power, especially if one has to respect fundamental length scales arising from the exchange interaction.
The discretization length should not exceed a few nanometers if dynamic or static phenomena on this scale, like exchange dominated spin waves or domain walls, are considered. The result is that simulating many unit cells is only applicable when access to a proper computing facility is given. Implementing periodic boundary conditions for a single unit cell circumvents this problem by using an accordingly modified demagnetization field.
Before the integration of the LLG-equation, a demagnetization tensor is computed from the geometry of the sample~\cite{ref29,ref30}. In a single isolated unit cell, strong divergence of the magnetization occurs at the boundaries.
By assuming an identical magnetization configuration repeated periodically in space, this divergence is removed, and the demagnetization field of a continuous and periodic medium is created. In the strict sense, the imposed exact repetition restricts dynamic computations to spin waves with $k=2n\pi\, a^{-1}$, where $n\in N^0$ and $a$ is the lattice constant.
Only by increasing the number of unit cells within the (periodic) simulation boundary, a finer grid in the reciprocal space can be probed. For the computation of the static equilibrium magnetization configuration or quasi-static simulations of hysteresis loops, the described method provides less ambiguous results.
Note that the approach can also be applied to the dynamical matrix method, which reformulates the linearized LLG-equation as an eigenvalue-problem on a finite grid~\cite{ref31}. Apart from the fact that the dynamic matrix method only applies to the linear regime of magnetization dynamics, it is practically limited to structures not bigger than $100\,\mathrm{nm}$.

\begin{figure}[!ht]
\centering
\includegraphics[width=0.8\columnwidth]{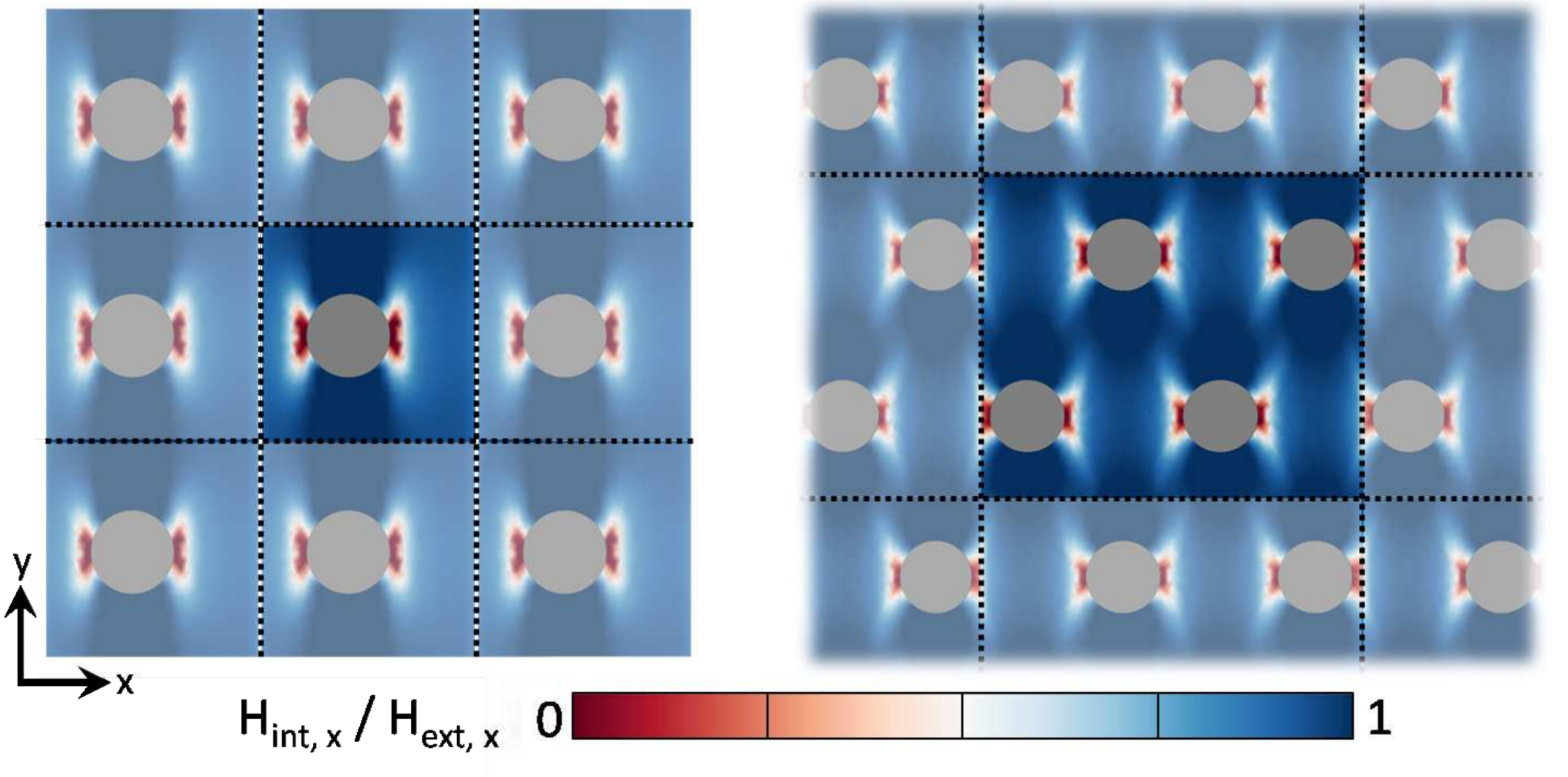}
\caption{Calculation of the total internal field for a square (left) and a hexagonal (right) antidot structure with an applied field in the $x$-direction of $\mu_0 H_\mathrm{ext} = 100\, \mathrm{mT}$. The colors show the minima in the extension of the internal field around each antidot. Please note the slight asymmetry in the square case to rule out unit cell artifacts.}
\label{fig:14}
\end{figure}

Figure~\ref{fig:14} shows the calculation of the internal field ($H_\mathrm{int,x} = H_\mathrm{ext,x} + H_\mathrm{demag,x}$) for a CoFeB sample, where the anisotropy field may be set to zero. If not stated otherwise, the external field was canted by an angle of $30^\circ$ out of the film plane, mirroring the experimental situation to be discussed in a later section.
The pictures show the equilibrium state of a unit cell of a square and an unconventional unit cell of a hexagonal antidot structure. The colors represent the field component parallel to the external field, normalized to the external field amplitude. The cells were computed using the software package Nmag, which can apply periodic boundary conditions.
Calculations of the square lattice show inhomogeneities around the antidots, while in the hexagonal simulation, these regions show higher expansion due to the higher fraction of holes per unit area. In both simulations, a lattice periodicity $a=3.5\,\mathrm{\mu m}$ and an antidot diameter $d=1.36\,\mathrm{\mu m}$ was used, which lead to filling fractions of $11.9\%$ and $13.7\%$ in the case of square and hexagonal lattice, respectively.
Note that concerning low computational power, discrete element sizes in this case are in the range of $60\, \mathrm{nm}$. Therefore, exchange interactions cannot be discussed by this simulation. Significance has been proven by calculations with bigger and smaller elements showing identical results.

\subsection{Micron and nanometer structuring for static and active magnonic crystals}
\begin{figure}[ht!]
\centering
\includegraphics[width=0.75\columnwidth]{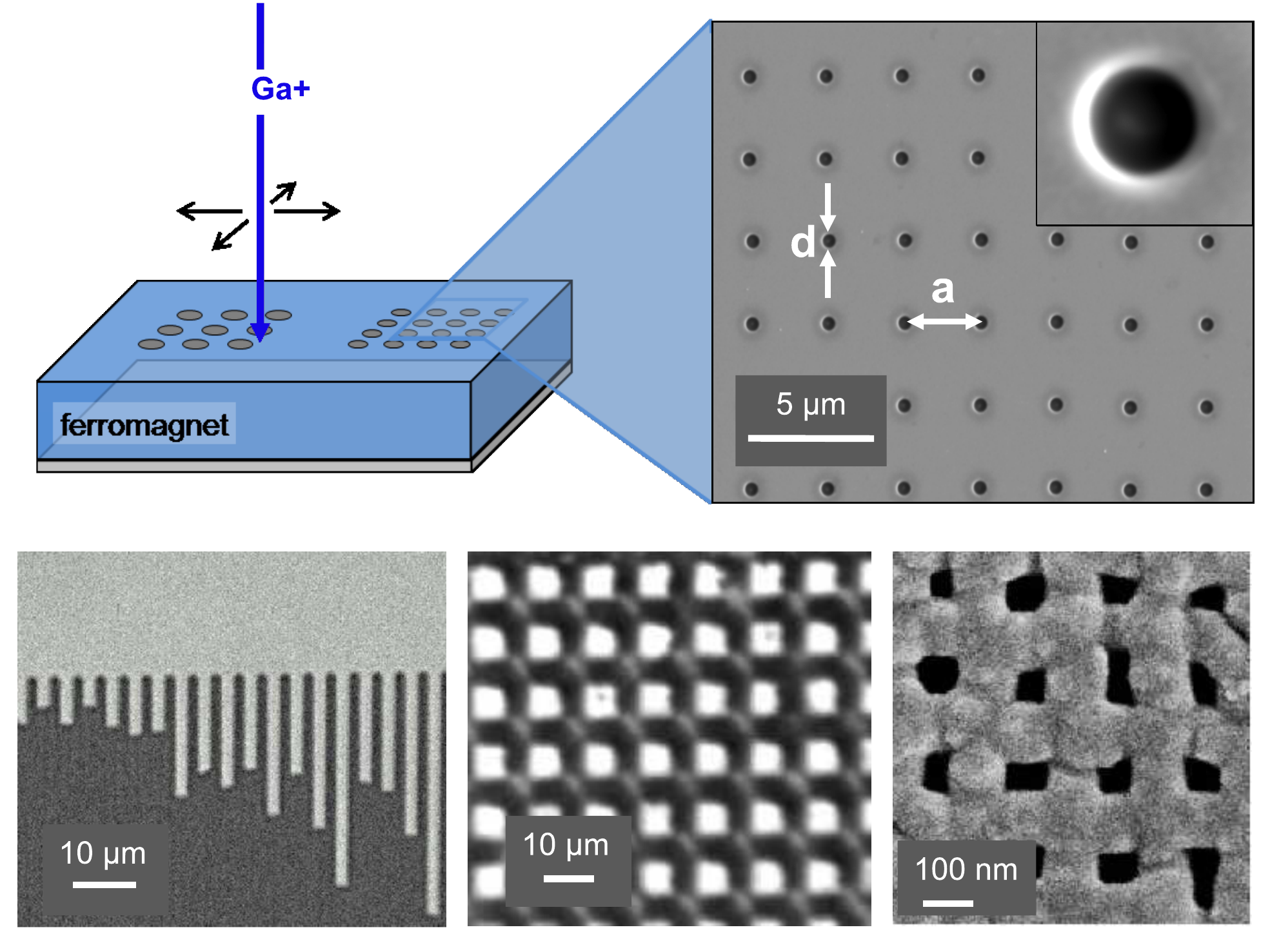}
\caption{Magnetic periodic structures on different length scales and by different methods. Top: Antidot structure prepared by focused ion beam etching. The parameters~$d$ (diameter of the hole) and~$a$ (antidot distance) define the filling fraction. Bottom: other techniques involve ion beam bombardment of exchange biased layers, which modifies the switching field \cite{ref33} (left) or combined soft/hard composite materials~\cite{ref34} (middle), in both cases the magnetic contrast has been imaged. Smallest structures on large areas can be produced by interference lithography or by self-organized or prestructured porous alumina membranes entering into the nm range~\cite{ref35} (right).}
\label{fig:15}
\end{figure}

Thin film technology provides a large variety of techniques to prepare one- and two-dimensional structures.
The relevant length scales may range from nano- to millimeters, depending on the material under investigation. A technique that is becoming increasingly important is lithography using a beam of ions.
Starting from a continuous film, the magnetic material is sputtered off by a focused beam of gallium ions (focus $10\,\mathrm{nm}$). Scanning the beam across the sample provides the flexibility of e-beam lithography without the need for further processing of the specimen. This reduces the efforts and the production of precise large scale nanostructured samples is possible. An example of a two-dimensional structure is given in Fig.~15 showing high precision, long-range order, and in the inset, an enlargement of a well defined constituting antidot. Damage and implantation of Ga ions happens only close to the antidot site.
With this method different lattice geometries can conveniently be achieved and it was used for most of the structures shown in the experimental section. Generally, the damping or magnetization of the material is found to be unchanged. Other standard techniques, such as optical lithography for structures above $1\,\mathrm{\mu m}$ diameters and e-beam lithography combined with lift-off or Ar$^{+}$ ion milling, are also used as methods for structurization. In the bottom row of Fig.~\ref{fig:15} we give some examples of magnetic superstructures.
By ion beam modification, the exchange bias can be modified locally, resulting in a separate switching of different areas of a material~\cite{ref33}. The varied switching field allows an active modification of the superstructure during the experiments. A modified switching field can also be realized by combining soft/ hard materials~\cite{ref34}. An example where dynamic experiments on very small magnonic structures have been realized is shown on the very right of Fig.~\ref{fig:15}: using self-organized or prestructured porous alumina membranes it is possible to enter the range below $100\,\mathrm{nm}$~\cite{ref35}.
Ferrites like yttrium iron garnet (YIG) are particularly well suited for fundamental research. They have a very low Gilbert damping ($\alpha=6\times10^{-5}$) and decay lengths are in the millimeter range. Thus typical structures are modified on larger length scales~\cite{ref36}. In YIG already dynamic magnonic crystals have been realized by using a set of parallel, periodically spaced, current conducting stripes close to the YIG film surface. When the current is turned on the current flow causes a sine-like variation of the film's internal magnetic field~\cite{ref32,Bailleul2010}.

\section{Experiments with magnonic structures}\label{sec:experiments}
Magnonic crystals with different dimensionalities have been experimentally investigated in the past. In one dimension, frequency band filters were demonstrated by the group of Hillebrands (see review on YIG magnonics~\cite{ref36}). For a two-dimensional magnonic crystal, the simplest case is a squared antidot lattice in a thin ferromagnetic film, as discussed by Neusser and Grundler~\cite{ref37}, however the expression `magnonics' had been formed by Kruglyak and Hicken~\cite{ref38}.
In the experiment, the control over the internal field distribution by the filling fraction (area of the holes per total area) and lattice parameter turns out to be crucial. From a broad continuum of spin waves at high energies in low damped CoFeB films (Gilbert damping $\alpha = 0.006$), in which the spin-wave propagation length is more than $100\,\mathrm{\mu m}$, Bloch modes appear only for low filling fractions~\cite{ref39}. The static dipolar fields stemming from a single hole defect have also been studied by combining MFM measurements and micromagnetic simulations~\cite{ref40}.
One can access the confinement distance and determine the interaction of single defects with radius~$r$. It is found empirically that if the separation is smaller than $10r$, collective effects emerge.
This is also true for the dynamic case~\cite{ref41}.
In this experimental section a somewhat larger part will be devoted to the application of femtosecond lasers to study magnonic materials. We will divide the presentation of experimental results into those, where spin-wave localization dominates the collective behavior in the periodically structured material and those, where delocalized Bloch modes predominate.
To give an introduction to this rich field of research, experimental findings from the various magnonic systems obtained with different approaches will be discussed. First a short overview on the experimental techniques to detect and excite spin waves will be given.

\subsection{Techniques}
Experimentally spin-wave modes in periodically structured materials have been studied by different techniques: microwave based techniques (ferromagnetic resonance (FMR), vector-FMR or pulse-inductive microwave magnetometer (PIMM)), in combination with spatial resolved Kerr microscopy; also, optical pump-probe techniques using femtosecond lasers have been applied.
Brillouin light scattering (BLS) takes a special role since it also allows a resolution in $k$-space, and thus, to measure dispersions and band structures directly. The different methods have different advantages. Typical specifications for a study of magnonic materials are that small areas have to be measured for all in common. Structurization of materials in a top-down approach generally allows, as outlined in the previous chapter, only small areas.

\subsubsection{Microwave techniques: resonance and time-resolved experiments}
\begin{figure}[!ht]
\centering
\includegraphics[width=0.75\columnwidth]{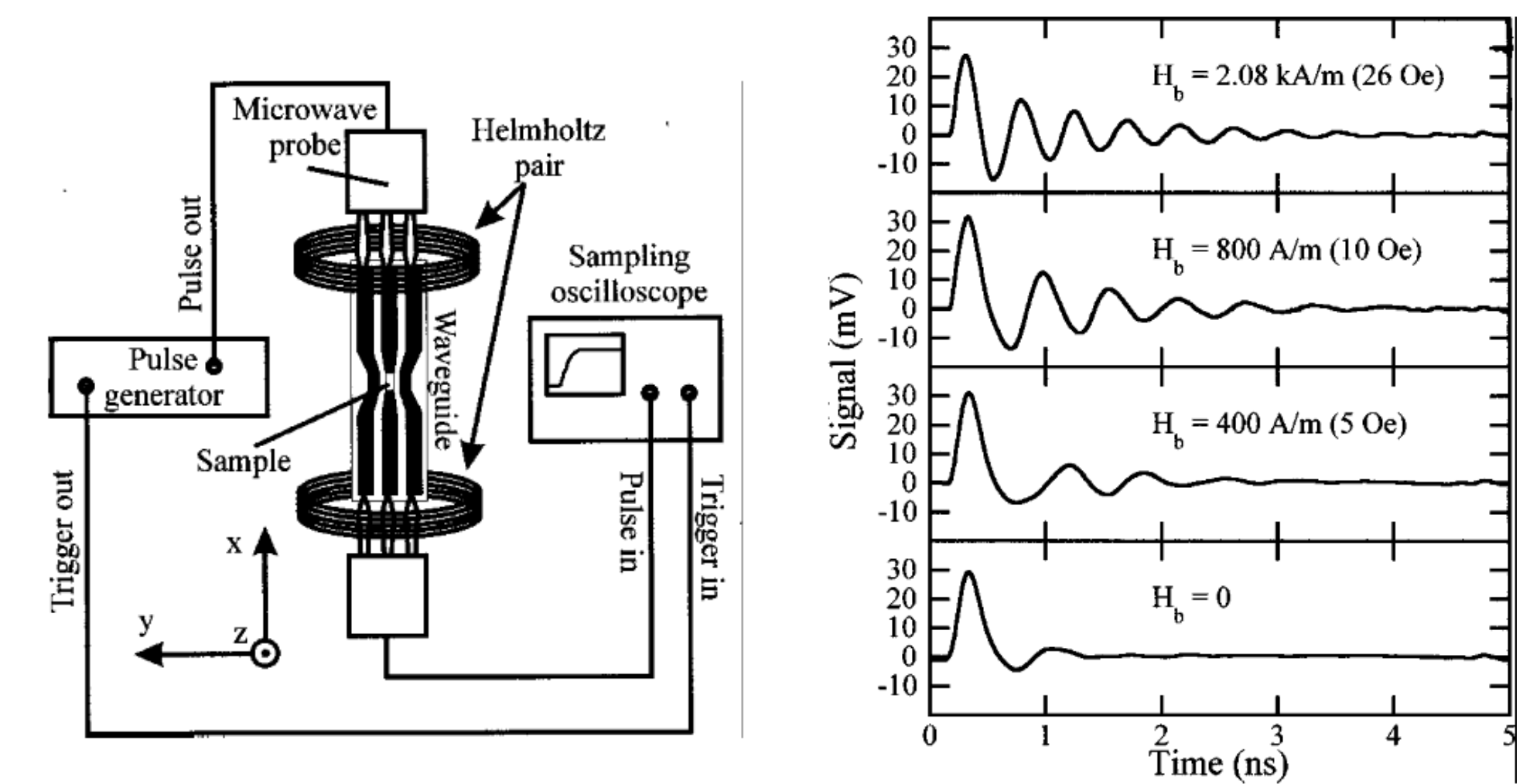}
\caption{Time-resolved wave guide techniques: pulse-inductive microwave magnetometer (PIMM)~\cite{ref43}.}
\label{fig:16}
\end{figure}

Microwave-based techniques owing to high sensitivity on small areas have the strip line geometry in common. Two methods have to be distinguished: ferromagnetic resonance (FMR), or vector network analyzer-FMR, which use a harmonic resonant excitation and methods that are using pulsed activation as pulse-inductive microwave magnetometer (PIMM).
In the first case, a harmonic microwave is guided through the strip line and the transmitted signal is measured~\cite{ref42}, similar to a standard FMR experiment using a microwave hollow conductor measuring the absorption power of the sample in the cavity, however much more sensitive. More complex setups use a local excitation and detection.
That enables guiding the microwave signal through the magnonic structure directly and early studies showed the formation of narrow transmission bands in YIG~\cite{yu}.
Also, this allows studying the transmission and reflection of a magnonic material as a function of the frequency by a vector network analyzer. PIMM uses a pulsed excitation~\cite{ref43}. The ringing down of the magnetic excitations is shown in  Fig.~\ref{fig:16} for different field values.
The signal is the inductance induced voltage into the stripline. To fully exploit the advantage in probing small volumes, the magnetic structure has to be lithographically structured on top of the wave guide.
A local approach is possible by using an inductive magneto-dynamic probe that can be scanned across the sample, in which spin waves are excited by a microwave strip line~\cite{WuPRB2004}.

\subsubsection{Microscopy techniques: imaging of resonant modes}
\begin{figure}[!ht]
\centering
\includegraphics[width=0.8\columnwidth]{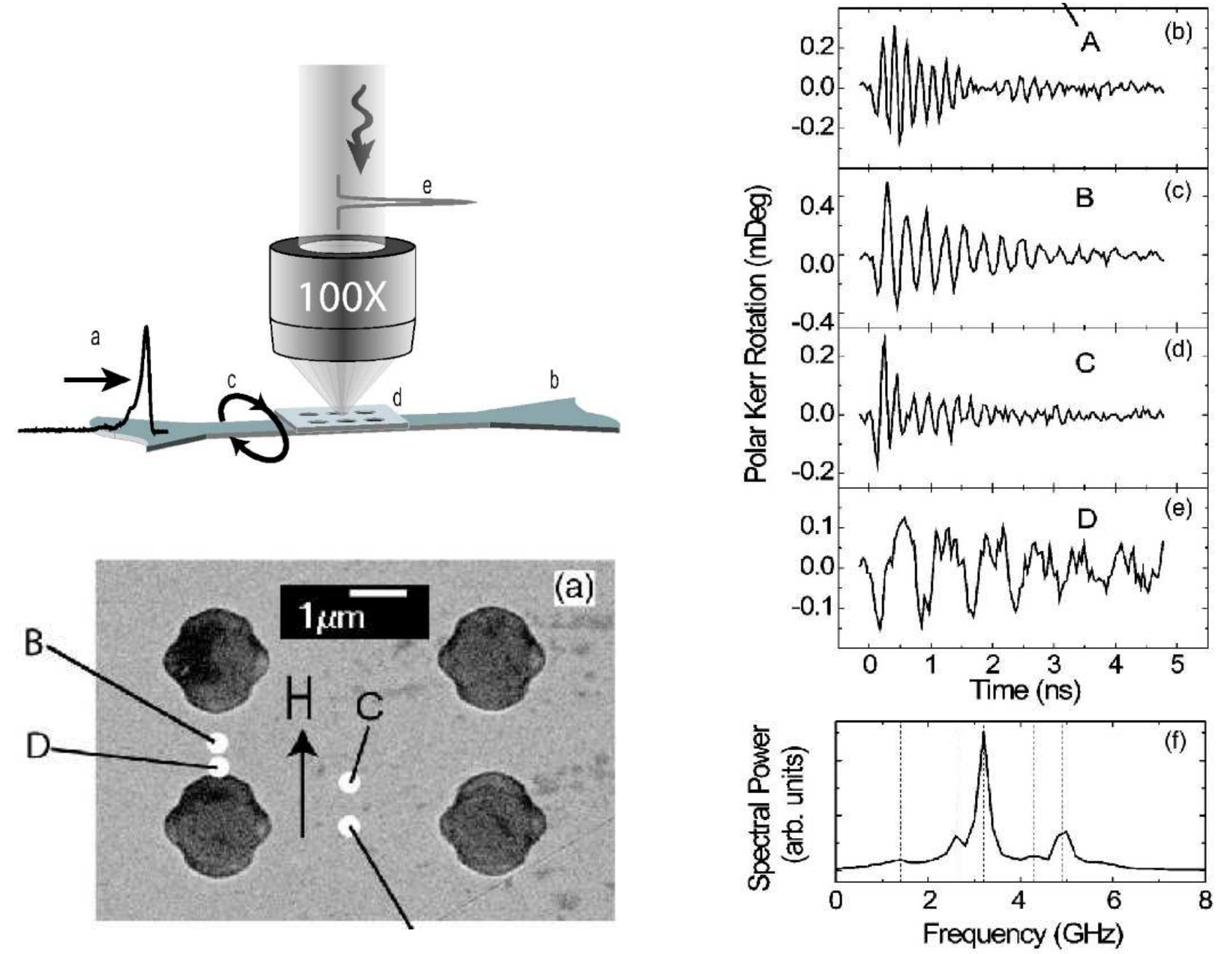}
\caption{Time and spacially resolved MOKE \cite{ref44}.}
\label{fig:17}
\end{figure}

To map magnetic modes with different frequency spatially, microscopy methods have to be used. Experiments on magnonic antidot structures have been performed by Pechan et al~\cite{ref44}. They combined Kerr microscopy with time resolution to map modes at different frequencies.
A current pulse excites the magnetization dynamics, which are probed spatially by rastering the sample and measuring the time resolved Kerr signal. The pump beam and probe beam come from a femtosecond pulsed laser. The response on different spot positions on the sample is reprinted from their work in Fig.~\ref{fig:17}.
To analyze the data, the time-domain data is Fourier transformed to the power spectrum at each spot to a map power distribution, which determines the mode location at the resonance frequencies. A very recent work shows the power of this method to map the localizing of spin-wave modes~\cite{ref45}. Examples will be discussed in section~\ref{sec:excitation-using-lasers}.

\subsubsection{Brillouin light scattering (BLS): micro and nano BLS}
\begin{figure}[!ht]
\centering
\includegraphics[width=0.9\columnwidth]{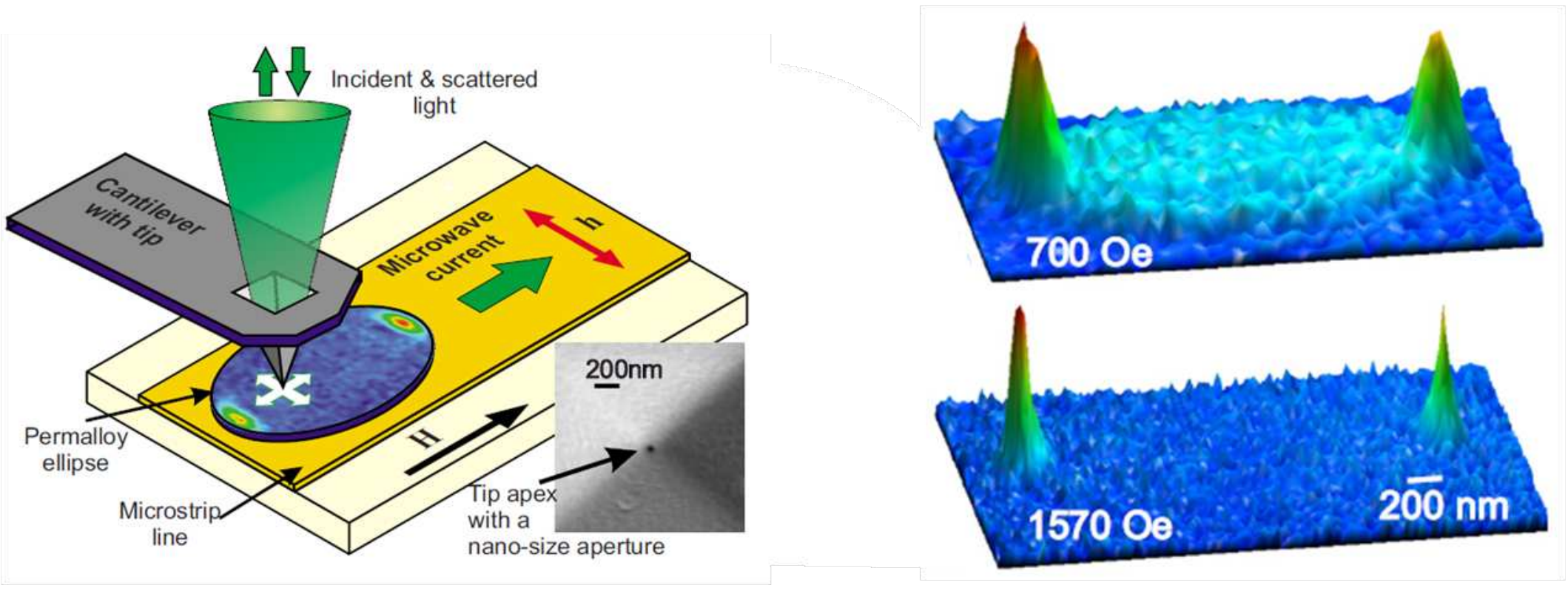}
\caption{Left: Schematic of how to achieve nanometer resolution with a BLS setup (from \cite{ref47}). An AFM tip with a nanosize aperture is used for near-field imaging. Right: Experimental data obtained from an elliptical permalloy disc showing precessional modes localized at the edges. The applied field tunes the spatial extension of the magnetic excitation in a manner proportional to $1/H_\mathrm{ext}$.}
\label{fig:18}
\end{figure}

BLS can access a further degree of information: the measuring of the inelastically scattered light under a certain angle allows the calculation of the $k$-vector and, by measuring the energy shift, the frequency is accessible simultaneously.
This method has been developed and applied to micron sized structures by the Hillebrands group to highest perfection over the last years.
Using methods that allow the determination of the resonance of a magnetic mode alone gives access to the energy of the mode for different field directions but no information on the wave length.
Then, by studying the field dependence, modeling the dispersion and comparison to the experimental data, modes can be identified. Naturally for that reason, BLS is the most powerful method to access the dispersion and spin-wave band structure of a material directly. For example, a wide range of wave vectors can be probed~\cite{Sandweg2010}, and three-magnon process investigated~\cite{PRB.79.144428}.
The method can be combined with microwave excitation to select or populate well defined modes or to investigate the propagation of spin waves along a magnetic wave guide stripe. Two recent developments should be mentioned: by using phase information, phase fronts can be investigated~\cite{vogt2009}.
Also, the resolution of the method has been considerably increased just lately: using high resolution scanning microscopy in combination with BLS, the resolution has been increased, approaching $200\,\mathrm{nm}$ ($\mu$-BLS). This allowed the mapping of modes in permalloy wave guides, shown in Fig.~\ref{fig:guides}.
A further step has been published by the Demokritov group. Using an optical aperture on a cantilever, near field imaging of the BLS intensity allows mapping of the edge modes of a micron sized ellipse, shown in Fig.~\ref{fig:18}, with a resolution of even below~$55\,\mathrm{nm}$. The $\mu$-BLS technique was also used for the first observation of Bose Einstein condensates of magnons~\cite{ref46} under continuous pumping of spin waves at microwave frequency.

\subsubsection{Femtosecond laser techniques: photo-magnonics}
A technique to investigate ultra fast processes involving spin waves is the excitation with femtosecond (fs) laser pulses, a field recently reviewed by Kirilyuk et al.~\cite{ref48}. Such ultra short pulses can be achieved using titanium doped sapphire seed lasers.
Due to the broad emission spectrum ($\text{FWHM}= 200\,\mathrm{nm}$) of the crystal, by mode locking of more than $10^5$ modes fs laser pulses can be generated.
Resulting pulses with a repetition rate of typically $80\, \mathrm{MHz}$ have an average energy of about $6\, \mathrm{nJ}$. At the expense of repetition frequency, amplification of single pulses to $\approx 5\, \mathrm{\mu J}$ then at $250\, \mathrm{kHz}$ is feasible to achieve large degrees of demagnetization upon absorption of the pulse by the sample.
Focusing the pulses to a $60\, \mathrm{\mu m}$ spot (in diameter) onto the sample, one achieves a fluence of $60\, \mathrm{mJ\, cm^{-2}}$.

\begin{figure}
\centering
\includegraphics[width=0.95\columnwidth]{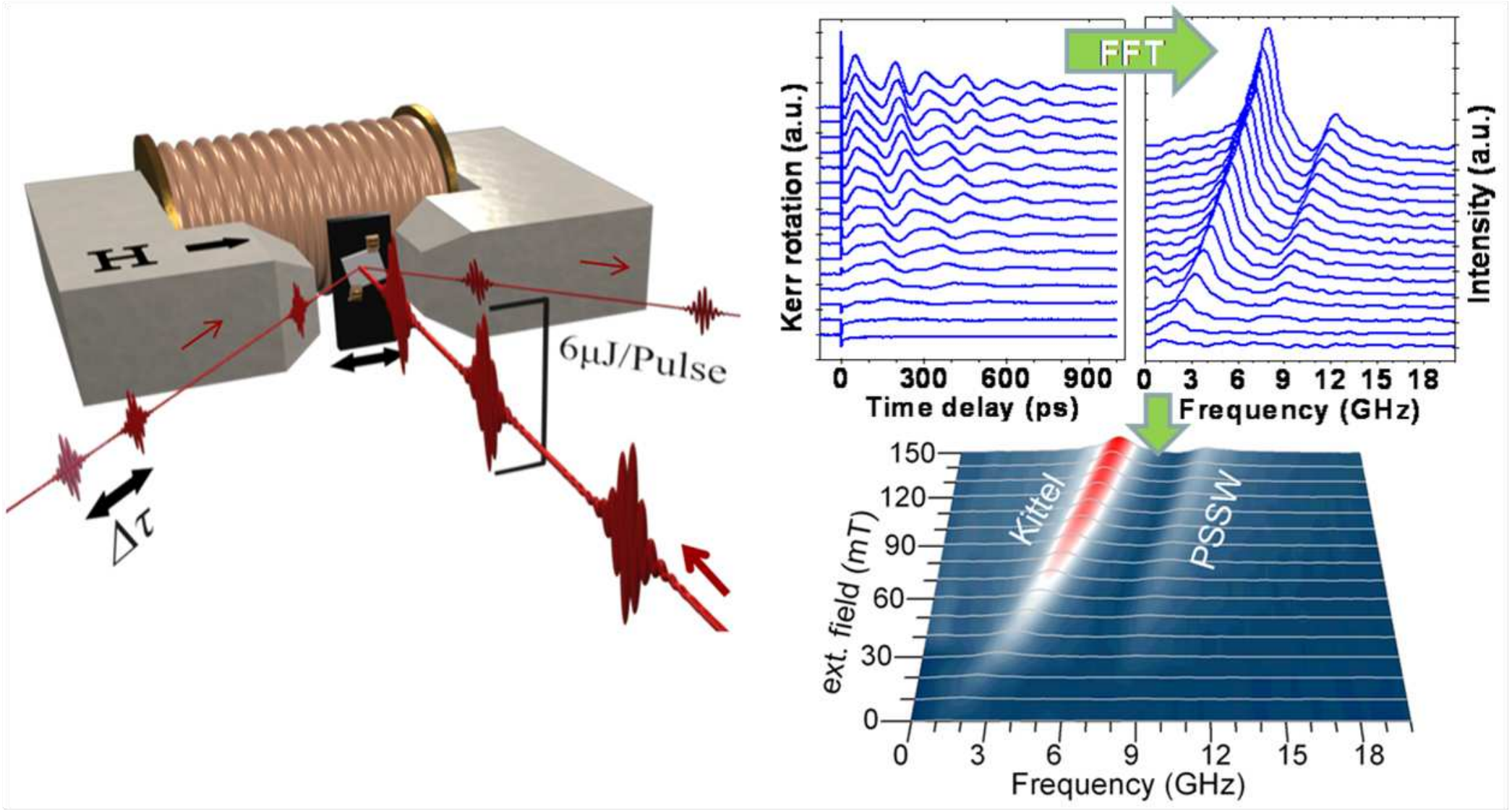}
\caption{Time resolved pump-probe setup (left) showing intense pump pulses for excitation and the detecting probe pulses with tunable time delay $\Delta \tau$. Analysis steps for a $50\,\mathrm{nm}$ Ni film (right): time-resolved trace as a function of the pump-probe delay, Fourier transform and color plot of the Fourier power.}
\label{fig:19}
\end{figure}

However, the use in magnonics is relatively new and we will therefore go into more detail.
For all-optical experiments, the laser beam is split into two pulses. The stronger pump pulse (95\% of the intensity) yields demagnetization in the order of up to 50\% and is used to excite the spin waves in the sample.
The other, less intense, pulse locally probes the magnetization of the sample. Due to the magneto-optical Kerr effect, upon reflection from the surface, the polarization changes depending on the magnetization.
It is remarked that in this technique not the absolute, but only the change of magnetization induced by the pump pulses is recorded. In the experimental setup~\cite{ref49} the pump pulse is guided over a delay stage in order to tune the time delay between both pulses. The magnetic field of up to $\mu_0 H_\mathrm{ext} = 150\, \mathrm{mT}$ is in the longitudinal Kerr-effect configuration, but tilted $30^\circ$ out of the film plane. Due to the very small signal, a double-modulation technique is used with a mechanical chopper operating at $800\, \mathrm{Hz}$ in the pump beam and a photo elastic modulator (PEM) at $50\, \mathrm{kHz}$ in the probe beam.
In the experiments, the time delay between pump and probe can be varied in steps of very few fs up to the maximum range of $1\,\mathrm{ns}$ after excitation. A data set of such a measurement, recorded on a $50\,\mathrm{nm}$ nickel film, is plotted in Fig.~\ref{fig:19} (right). Before the pump pulse arrives, the magnetization is in a state of equilibrium. At $\Delta \tau=0$, the pump pulse hits the sample, leading to the demagnetization of the sample, as shown by the strong change of the magnetization.
As one can already see in the raw data, the magnetization shows two different modes of precession. After analysis (subtraction of the incoherent background and Fourier transformation) these modes can be identified as the uniform Kittel and the perpendicular standing spin-wave modes. Both precession amplitude and frequency increase linearly with the external field, as expected. For a better visualization, all frequency data is also plotted in a color code.

Compared to BLS measurements, the pump-probe technique only probes the Kerr rotation, and thus, is not directly $k$-sensitive. Comparing it to the resonant excitation at a fixed frequency in FMR, significant differences are evident. The excitation mechanism using a pump pulse does not select a certain mode. Instead, high energy spin waves with a broad distribution of wave vectors are excited.
During the first few picoseconds after excitation, these relax to lower energy states~\cite{ref23}, cool down and populate modes given by the sample and magnonic crystal itself. All modes observed are not externally selected (for example, by an alternating microwave field), but are selected by the sample itself.
Hence the results of this technique lead to different information on the magnon density than the ones described previously. This technique combines photo and spin-wave excitation, hence the term photo-magnonics. A recent work by Hamrle showed that both optical techniques, BLS and TRMOKE are comparable in their sensitivity~\cite{ref50}.

\subsubsection{Fundamental mechanisms for spin-wave excitation using lasers}\label{sec:excitation-using-lasers}
The full relaxation path of spin waves after laser excitation is a research field of its own. We will therefore briefly describe the ideas in this separate section.
Since there is no direct coupling between the spin degree of freedom and the light field we want to shed some light on the excitation mechanism specifically used in the examples to follow. How can spin waves be excited and what is the specific mechanism used in photo-magnonics? Just recently two parallel investigations, combined theoretical and experimental proof that it is possible to quantitatively describe femtosecond demagnetization dynamics in its absolute value and characteristic time scales~\cite{ref51,ref52,ref53}.
A slowing down due to intrinsic fluctuations characteristic to the exchange coupled spin system can be mirrored by the models found. It is found for strongest demagnetization or approaching the Curie temperature. Atxitia et al.\ \cite{ref53} used a stochastic equation of motion that is mapped onto a macrospin description, the Landau-Lifshitz-Bloch equation. From spatially resolved micromagnetic simulations, one gets a microscopic insight into the origin of the delayed remagnetization. Short wave length spin waves and higher order spin-wave modes with a high spatial frequency dominate the dynamics within the first ps~\cite{ref23,ref54}.
These excitations are distributed very efficiently from the hot-spin region into the  magnetically not disturbed region. For small excitation spot sizes, the magnetization can be restored very fast by the emission of spin-wave packets and high-energy magnons~\cite{ref55}. High-energy magnons have been studied in detailed experimentally challenging neutron diffraction experiments in ferromagnetic metals in the 80s~\cite{ref24}, investigating the broadening of the spin-wave dispersion curves at high energies as predicted also at that time arising from the strong interaction with the Stoner band, called Landau damping: near the surface of the Brillouin zone, it results in a severe energy dissipation and extremely short lifetime of the high-energy spin waves.
\begin{figure}[!ht]
\includegraphics[width=448pt]{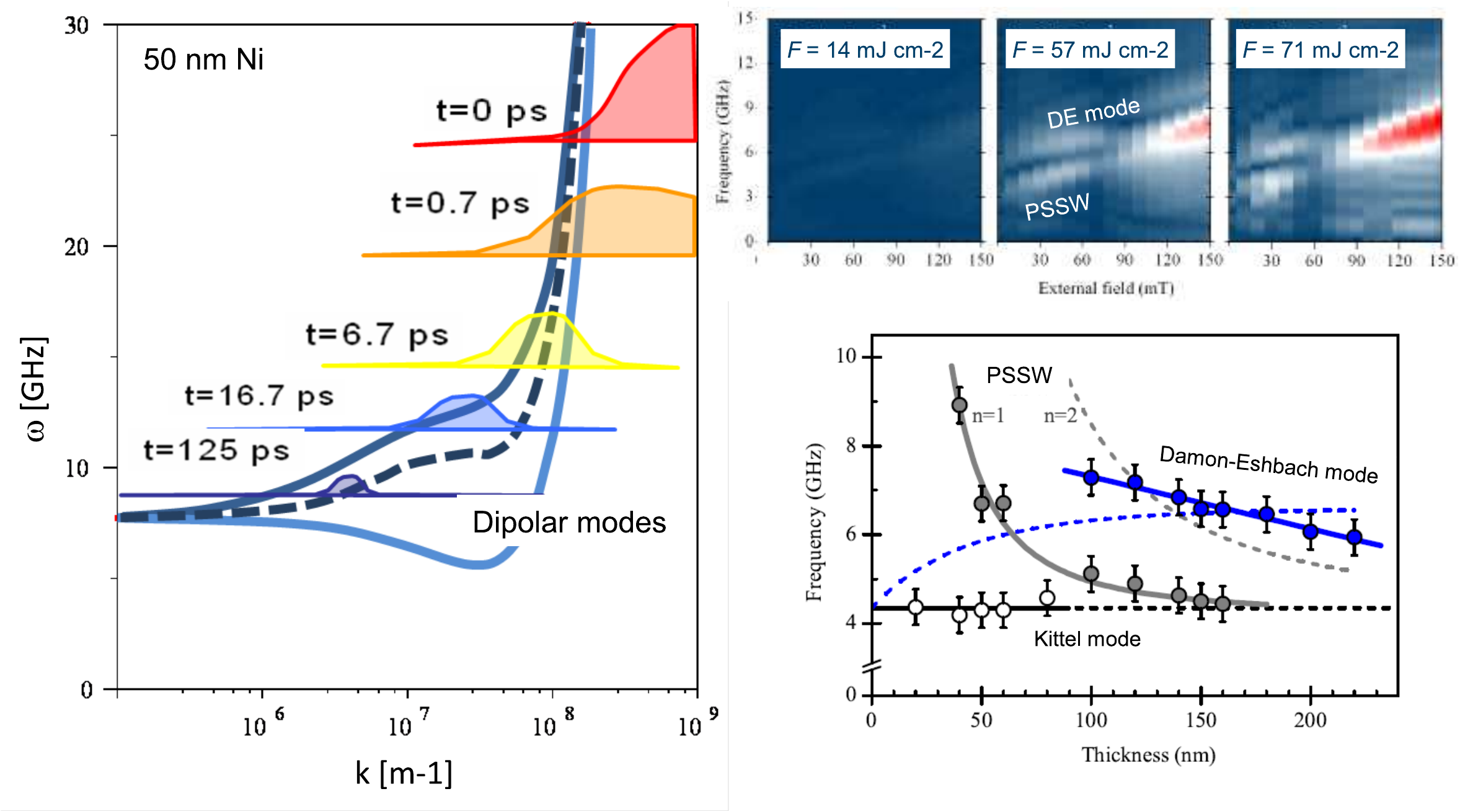}
\caption{Spin-wave relaxation in all-optical experiments after laser excitation. Micromagnetic simulation (left, from \cite{ref23}). Fourier power spectra for different pumping fluencies 14, 57 and $71\, \mathrm{mJ\, cm^{-2}}$, Ni film 150 nm thickness (right). A Damon-Eshbach mode with fixed wave length  $\lambda$ from $0.5$ - $6\, \mathrm{\mu m}$ as determined by the Ni thickness is found. Precession amplitudes can be up to 5$^\circ$. A summary of the mode and wave-length dependence  is given below as a function of the Ni thickness (for a magnetic field of $50\,\mathrm{mT}$) (from \cite{ref65}).}
\label{fig:20}
\end{figure}

In the last years, using spin-resolved electron energy loss spectroscopy~\cite{ref56} and theoretical investigation~\cite{ref21} these excitations have been related to spin-wave packets with lifetimes below $1\,\mathrm{ps}$ and localized within a few nanometers only. The starting point of the relaxation chain is an energy transfer from the high-energy spin-wave excitations resulting from local spin scattering processes to the lower-order spin-wave excitations. The high-energy excitations are overpopulated in an ultrafast demagnetization experiment.
Furthermore they are densely populated (demagnetization typically much larger than 1\%). It seems plausible that the Stoner excitations, that are populated by the elementary spin-flip relaxation processes of the hot electron system (exchange scattering for hot electrons~\cite{ref21,ref22} and Elliott-Yafet spin-flip processes for lower electron energies~\cite{ref57,ref58,ref59,ref60}), decay into short wave length spin excitations and gradually relax to the lower spatial frequency excitations discussed in~\cite{ref23}. A very recent work by Schmidt et al.\ nicely proves these ideas on the femtosecond time scales experimentally~\cite{ref61}.
The energy is transferred from the highest excited mode into modes owning a lower energy following the spin-wave dispersion. From the corresponding Fourier transform, through the magnetization profiles in the micromagnetic model spin-wave excitations are accessible (spatial frequency) given in Fig.~\ref{fig:20}. High spatial frequencies dominate after excitation at $\tau=0$. During the relaxation process, the center of the spectral weight moves towards lower spatial frequencies. For $6.7\,\mathrm{ps}$ the center is at around $0.1\, \mathrm{nm^{-1}}$.
This corresponds to a spatial spin-wave period of $10\,\mathrm{nm}$. The damping results in an overall reduction of the spectral amplitudes, accompanied by a shift of the spectral weight to lower frequencies. It has to be mentioned that the dipolar Damon-Eshbach modes are not within the simulation window because of the film thickness of only $50\,\mathrm{nm}$. Nevertheless simulations are an indication for an energy transfer from high-energy modes to low-energy modes within the spin-wave relaxation. Here we see some similarities to the Bose-Einstein condensation where spin-waves driven by micro-wave fields relax to their lowest state of energy and overpopulate this states so that condensation sets in by permanently driving up the chemical potential~\cite{ref62}. In these experiments the decay from the GHz pump frequency to the frequency of the lowest energy level of the condensate can be followed directly.
The optically induced precession verified by many groups~\cite{ref63} has been successfully used to extract the magnetic Gilbert damping for a variety of materials~\cite{ref64}. The amplitudes can be increased up to few degrees by increasing the pump fluence. In the excitation spectra the mode of the homogenous precession, standing spin waves of up to fifth order for low damping materials~\cite{ref28} and dipolar spin waves with a fixed $k$-vector inversely related to the film thickness have been investigated~\cite{ref65}.
Such a well defined $k$-vector resulting from all optical spin-wave excitation leads to the idea to study the effect of a periodic modification in order to create a magnonic crystal with that periodicity.

\subsection{Localization effects in magnonic crystals}
The idea to excite spin-waves all optically in a magnonic crystal had been put forward by Kruglyak et al.~\cite{ref38}. In the following we will study 2D magnonic crystals. And in fact: when compared to a continuous film as presented in the previous section, magnetization dynamics drastically change after arrays of antidots have been inserted.
\begin{figure}
\centering
\includegraphics[width=414pt]{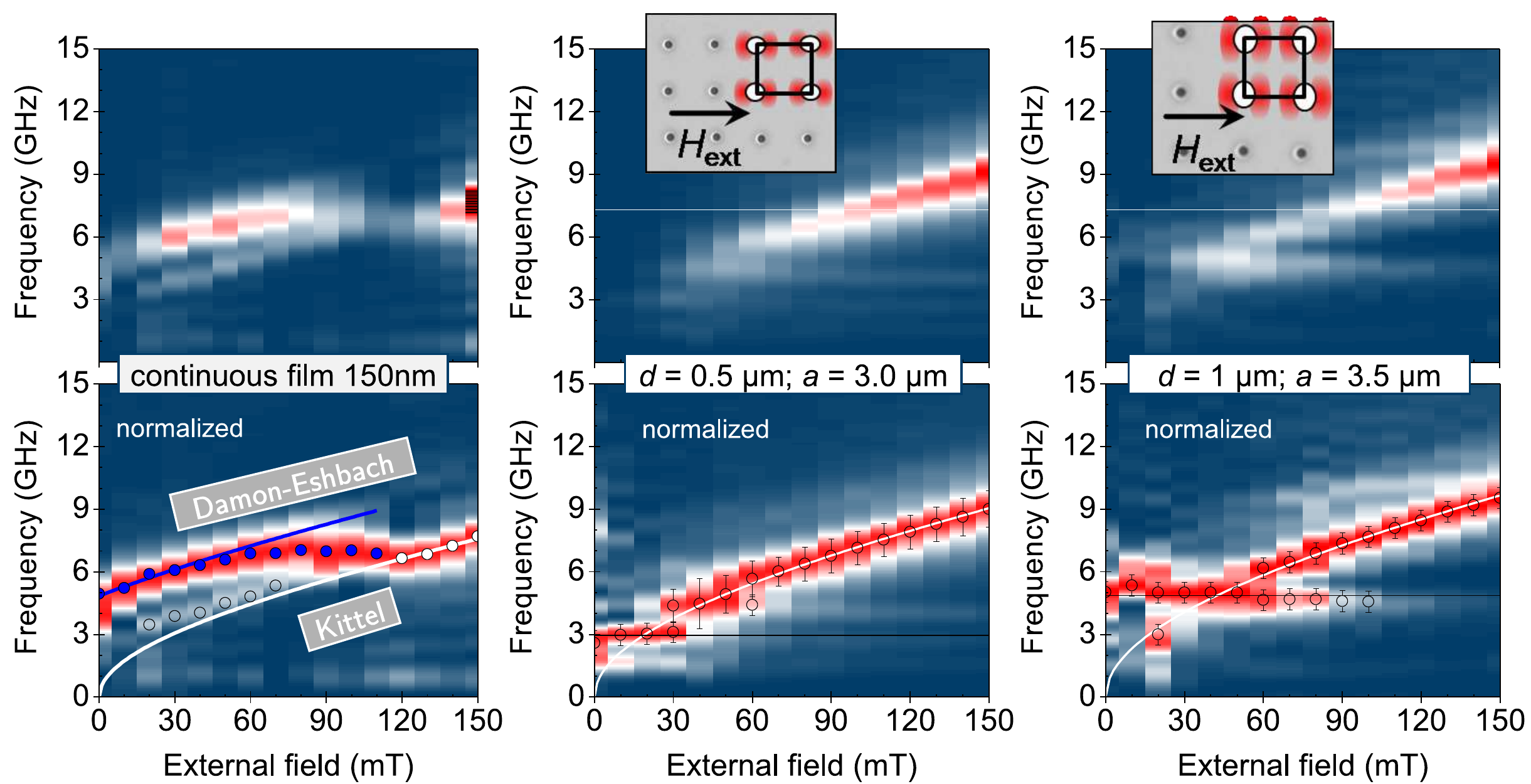}
\caption{Effect of an antidot lattice structure measured by all-optical spin-wave excitation and detection. Left: continuous nickel film of 150 nm. Middle: antidot diameter $0.5\, \mathrm{\mu m}$. Right: antidot diameter $1\, \mathrm{\mu m}$. With the antidot diameter the mode intensity showing weak field dependence and low frequencies is found to increase.}
\label{fig:21}
\end{figure}
The two dimensional periodic modification of the internal magnetic field has been realized by focused ion beam. Shown in Fig.~\ref{fig:21} are three data sets, recorded on a continuous $150\,\mathrm{nm}$ nickel film, as well as on two different antidot lattices of the respective sample.
On the structured areas (lattice parameter $3\,\mathrm{\mu m}$ ($3.5\,\mathrm{\mu m}$), antidot diameter $0.5\,\mathrm{\mu m}$ ($1\,\mathrm{\mu m}$), respectively) the Damon-Eshbach mode as previously excited at field values below $100\,\mathrm{mT}$ is not observed anymore, instead a new magnetic mode is observed. Its frequency does not significantly change with the external field.
This is a hint that a major role is played by the internal magnetic field which is significantly reduced next to the antidots, compared to the applied field. These regions of internal fields, however, become smaller with increasing field and finally disappear. The interaction between different modes at the antidot sites decreases and they are not coupled anymore.
This is in accordance with the disappearance of these weakly field dependent modes at a certain field value which is found to be different for different antidot diameters.
\begin{figure}
\centering
\includegraphics[width=430pt]{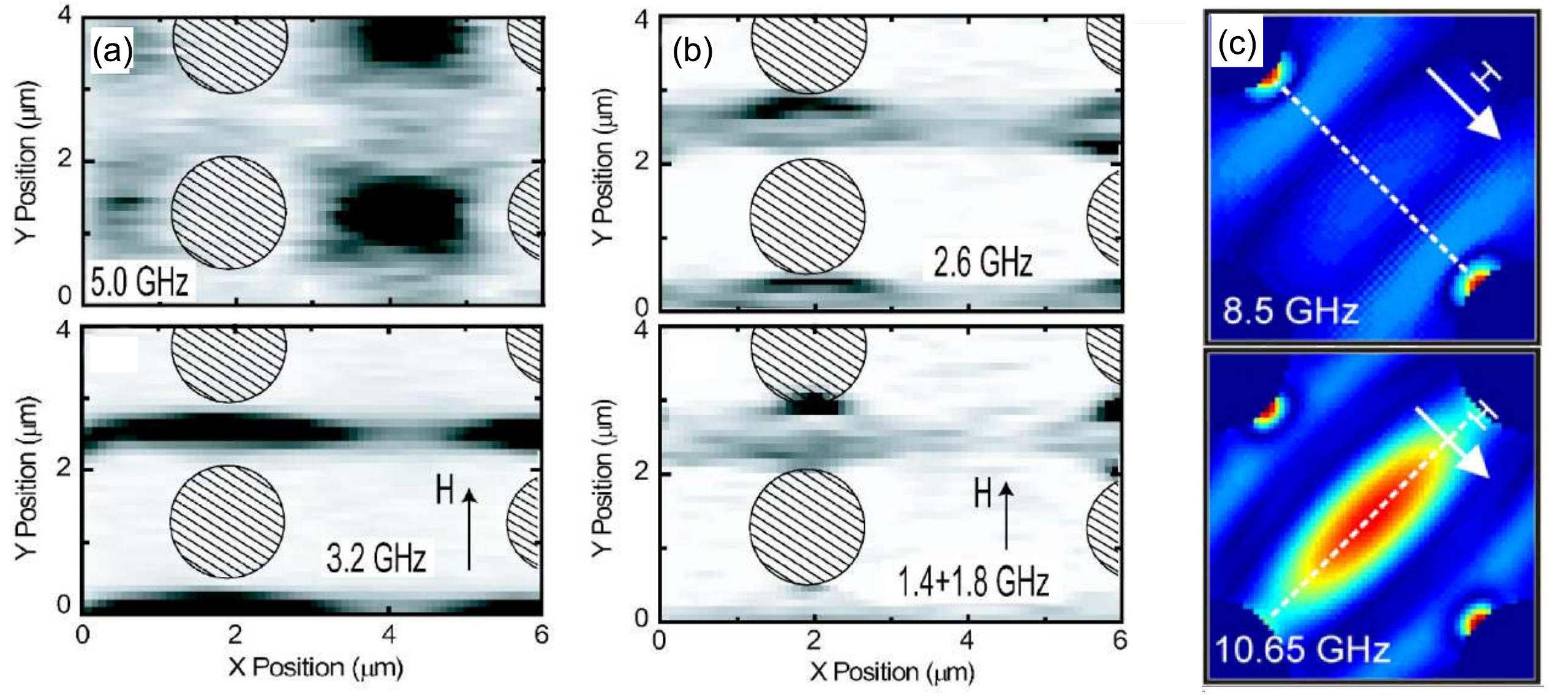}
\caption{(a), (b) Modes measured by time-resolved Kerr microscopy (from~\cite{ref44}): the Fourier power at a resonance frequency is mapped. High frequency modes are found between the antidot structures~(a), low frequency modes localized in the low effective field regions around the antidots~(b) depict the areas of a reduced internal field. In~(c), numerical results from~\cite{ref77} are reprinted that indicate an analogue situation.}
\label{fig:22}
\end{figure}

From different previous works it is well known, that spin-wave modes in antidot lattices can have localized or extended character, referring to the spatial behavior of the mode amplitude. This was shown in micromagnetic simulations for antidot lattices~\cite{ref77} and in earlier experiments by Pechan~\cite{ref44}.
The results of the latter using time-resolved Kerr microscopy is given in Fig.~\ref{fig:22}. In 2005, they had already mapped two kinds of modes: the spin-wave amplitudes of the high frequency modes have dominantly maxima in between the antidots, whereas the low frequency modes have strong intensity around the antidot, showing a typical structure that we know already from section~\ref{sec:micromag}: it mirrors the areas of a reduced internal field. Recently, this was investigated in a combined study by vector network analyzer FMR, time-resolved Kerr microscopy and BLS~\cite{ref67}.
The appearance of delocalized, extended modes strongly depends on the propagation length of the spin wave excited. In the case of Fig.~\ref{fig:21}, where nickel is the ferromagnet under investigation, the spin-wave propagation length is on the order of ten microns due to the rather large intrinsic Gilbert damping ($\alpha=0.02$). This distance is similar to a few unit cells of the antidot lattice.
The data in Fig.~\ref{fig:21} can thus be explained: arising from the demagnetization field, at the antidot edges are potential wells for spin waves with a rapidly changing effective field. At these sites localized modes are excited while in the regions of constant internal field the uniform Kittel mode is present.
A schematic of the internal field has already been depicted in Fig.~\ref{fig:14} for the overlapping inhomogeneities in periodic lattices. In a time-resolved MOKE experiment, also modes extending across several unit cells, hence `feeling' the periodic potential are induced by the structure. These will be presented in detail in section~\ref{sec:bloch-modes} (Bloch modes in magnonic crystals).
The parts (middle) and (right) of  Fig.~\ref{fig:21} display two major differences. The difference of the two antidot arrays being the filling fraction~$f$ of 2.2\% and 6.4\%, respectively, the relative amplitude of the localized modes changes when compared to the uniform precession. Resulting from the larger perturbance by the lattice in Fig.~\ref{fig:21} (right), the localized mode dominates the spectrum over a wider field range than in Fig.~\ref{fig:21} (middle). The area of inhomogeneous internal field grows with the filling fraction so that the observed increase in amplitude further supports the localized interpretation of the non-dispersive modes.
\begin{figure}
\centering
\includegraphics[width=0.95\textwidth]{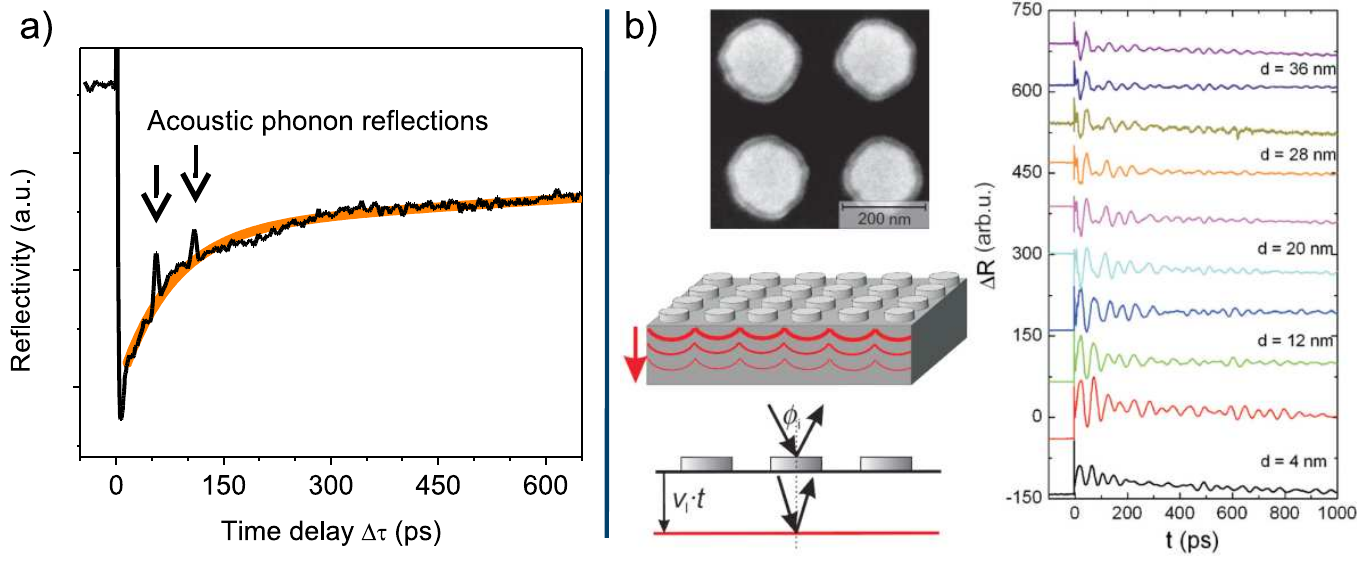}
\caption{Effect of acoustic phonons. (a) Ultrafast heating of the surface of an antidot-structured nickel film generates a stress wave which is reflected at the substrate, as seen at a delay of 70 and $140\,\mathrm{ps}$ as a sharp peak. This film was structured with $a=3.5\,\mathrm{\mu m}$ and $d=1\,\mathrm{\mu m}$. Because of the low filling fraction the lateral heating is homogeneous and no acoustic phonons are excited parallel to the surface which interfere with the magneto-optical measurements. (b)~In the case of nickel dots on a Si substrate the traced reflectivity changes show strong oscillations that can be attributed to a standing acoustic wave governed by the periodicity of the structures (reproduced from~\cite{ref69}).}
\label{fig:23}
\end{figure}

Another possible explanation for a dominantly field-independent mode is to attribute this to a non-magnetic effect, the phonon surface wave of the thin metal film. Corresponding measurements of the time-resolved reflectivity are shown in Fig.~\ref{fig:23} and reveal the propagation and interfacial reflection of heat-induced stress waves normal to the film plane described, which can interfere with the measured transient Kerr rotation~\cite{ref68}.
Though small oscillations are visible on longer timescales, they do not account for the large amplitudes observed and described earlier and are at a different frequency. Additionally, with the velocity of sound of the respective materials, the frequency of possibly excited standing phonons can be estimated:
Assuming the basic mode of phonons with a wave length of $\lambda_\mathrm{phonon}=2a=7\,\mathrm{\mu m}$ and using $v_\mathrm{Ni}=4900\,\mathrm{m\, s^{-1}}$ as the sound velocity in nickel, one finds an expected phonon frequency of $0.7\,\mathrm{GHz}$. Lateral acoustic phonons in the silicon substrate ($v_\mathrm{Si}=8433\,\mathrm{m\, s^{-1}}$) have an expected frequency of $1.2\,\mathrm{GHz}$.
As a consequence, acoustic phonons as an origin for the field-independent modes can be ruled out.
The peaks as seen in Fig.~\ref{fig:23} (a) arise from the stress waves reflected at the film-substrate interface and clearly do not appear in the Kerr spectra, which is a sign that there is no crosstalk of the signal here.
A reason is that the filling fraction is very low, and thus, the whole surface is heated equally. The different possible contributions to the transient signal have been discussed by M\"{u}ller et al.~\cite{ref69} and reproduced in Fig.~\ref{fig:23} (b). In their case, a metallic dot array produced by optical interference lithography was investigated all-optically; the small dots heat up very fast, while the Si substrate keeps cold. The large difference in expansion creates surface acoustic waves with large amplitudes that can be matched to the inter-dot distance, which serves as a periodic lattice for the surface acoustic waves. This poses a clear difference to antidot experiments.
\begin{figure}
\centering
\includegraphics[width=425pt]{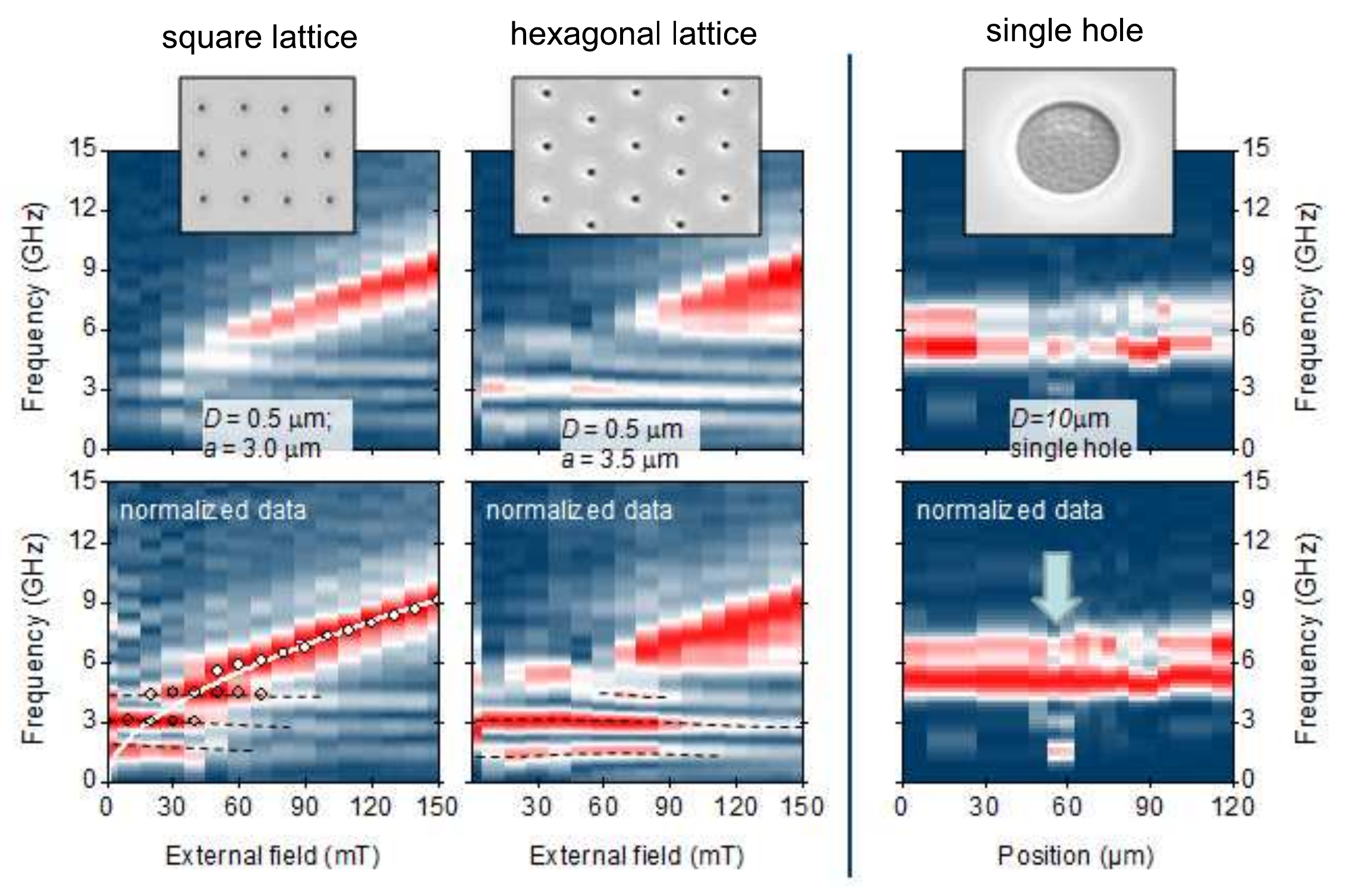}
\caption{Effect of the antidot lattice geometry measured by all-optical spin-wave excitation and detection. Left: square lattice. Middle: hexagonal lattice. Right: mode spectrum around a single hole structure for $\mu_0 H_\mathrm{ext}=30\, \mathrm{mT}$. By scanning the laser spot across a single antidot also here a mode at low frequencies is found that solely exists around the antidot.}
\label{fig:24}
\end{figure}

It is also possible to examine other, more complex symmetries than the simple square geometry; then the spin-wave modes in the structured media should mirror the lattice symmetries if they are not only determined by the potential around a single antidot, but interacting with the next neighbor antidot mode.
Static measurements of hysteretic properties and electronic transport have been performed and resemble the respective lattice symmetries~\cite{wang2006}.
According dynamic experiments are shown in Fig.~\ref{fig:24}. They reveal that localized modes are also observed on hexagonal lattices (Fig.~\ref{fig:24} (middle)).
The mode distance in frequency of about $1.5\,\mathrm{GHz}$ seems very similar for the square and the hexagonal lattice. It is solely determined by the antidot diameter. The reader may note the slightly increased periodicity in the case of the hexagonal lattice in order to maintain the filling fraction of approximately 2\%. Also shown Fig.~\ref{fig:24} (right) are measurements determined on a single antidot.
For a fixed external field, the Fourier power is color-coded as a function of position and frequency. At the position of the single antidot (marked by an arrow), additional modes can be seen.
\begin{figure}
\centering
\includegraphics[width=350pt]{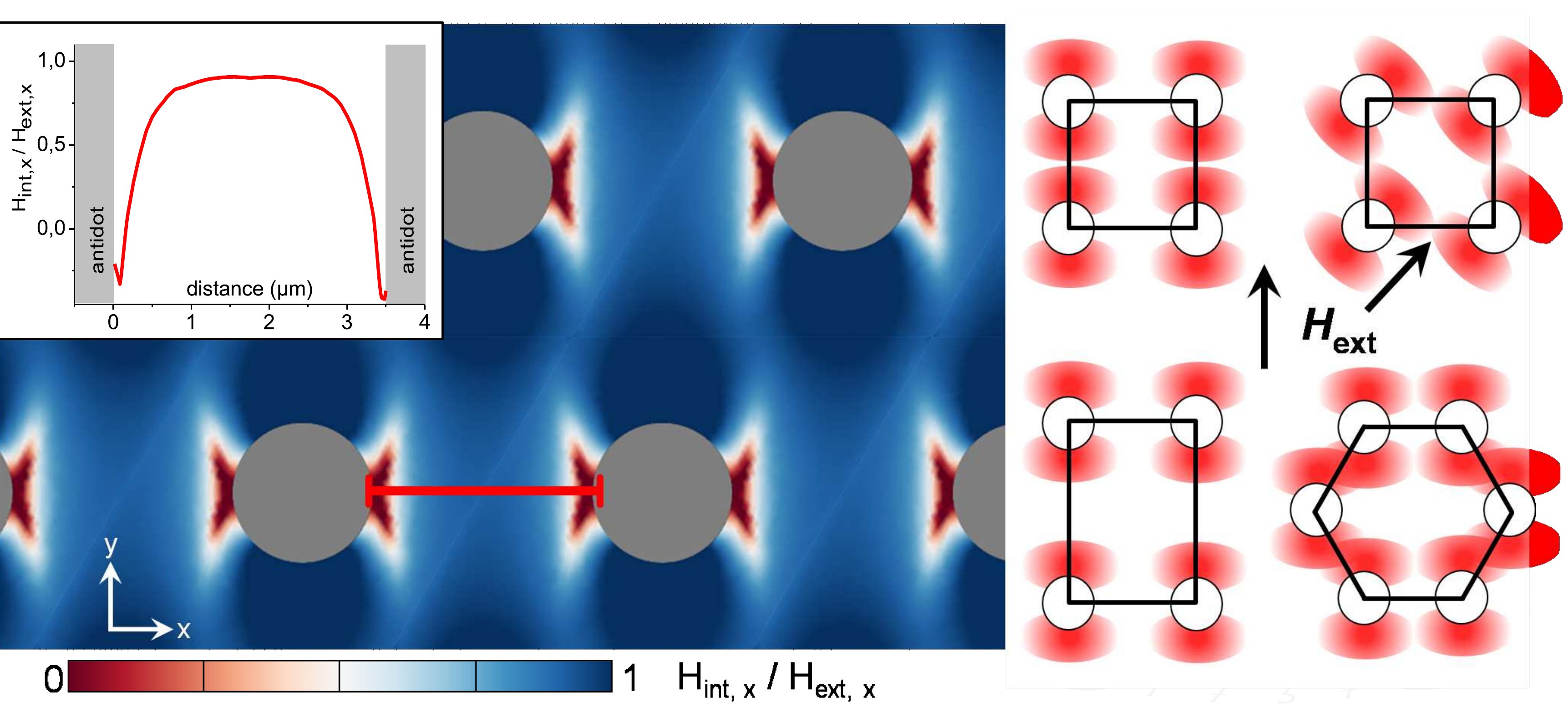}
\caption{Calculation of the total internal field (left) in a hexagonal structure at $\mu_0 H_\mathrm{ext,x}=90\, \mathrm{mT}$ for lattice periodicity $a=3.5\,\mathrm{\mu m}$ and antidot diameter $d=1.36\,\mathrm{\mu m}$. The inset shows the values for a cut along the red line. Schematics of the internal field distribution (right) for square lattice with field along (0,1), (1,1), tetragonally distorted and for a honeycomb lattice.}
\label{fig:25}
\end{figure}
This supports the picture that these low frequency magnetic modes are localized at single antidot edges. To get a better understanding, we will first discuss the internal field distribution around an antidot site in these lattice geometries and then show more detailed angular-dependent experiments. In Fig.~\ref{fig:25} a calculation of the total internal field in a hexagonal structure at $\mu_0 H_\mathrm{ext,x}=90\, \mathrm{mT}$ for lattice periodicity $a=3.5\,\mathrm{\mu m}$ is shown.
The inset reveals that strong changes of the internal field up to distances of $0.5\,\mathrm{\mu m}$ from the dot are found. In the same figure on the right, schematics of the internal field minima for different lattices and magnetic field direction are drawn to illustrate a possible interaction of modes located in these minima.

\begin{figure}[!ht]
\centering
\includegraphics[width=\columnwidth]{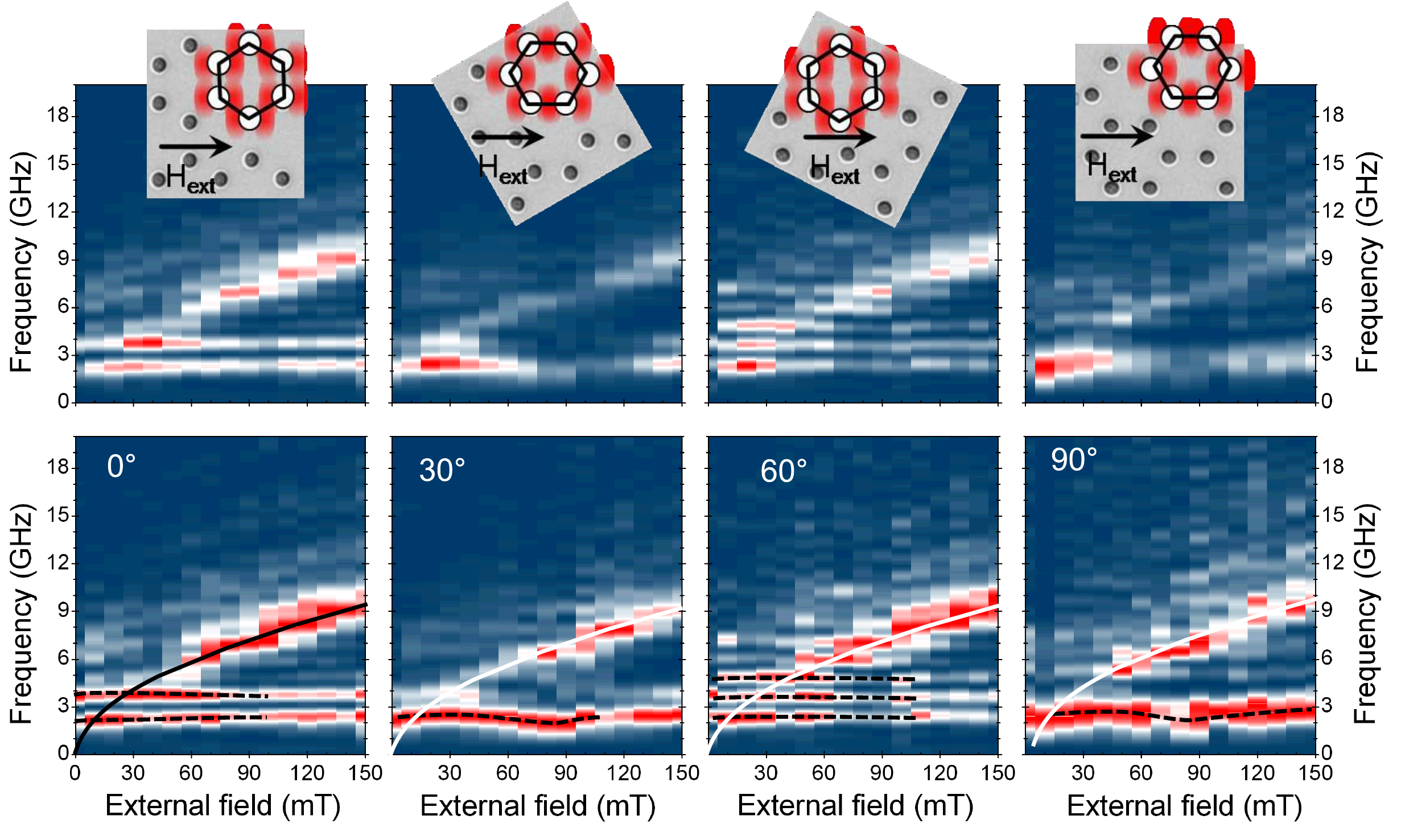}
\caption{Effect of the antidot lattice geometry measured by all-optical spin-wave excitation and detection. The honeycomb lattice is rotated: for $0^\circ$ and $60^\circ$, the mode splitting in energy is low. A weak localization is found. For $30^\circ$ and $90^\circ$, only one mode is found and the mode splitting in energy is high. In the schematic picture, a blocking of the Damon-Eshbach wave propagating perpendicular to $H$ is observed (strong overlap of the red areas, which represents a drop of the internal field).}
\label{fig:27}
\end{figure}

For honeycomb lattices, one expects to find the 6-fold rotational symmetry, which coincides with a 60$^\circ$ repetition of the spin-wave pattern. Here, when compared to hexagonal lattices, regions of rather homogeneous internal field are coexistent with closely packed antidots.
Again, when rotating the sample around the film normal, by tilting the structure with respect to the applied field, the behavior of the localized modes can be controlled. Respective measurements are presented in Fig.~\ref{fig:27} and display the expected symmetry. As expected the data for the angles 0$^\circ$ and 60$^\circ$, as well as 30$^\circ$ and 90$^\circ$ show the same features: opposite to the first two 0$^\circ$ and 60$^\circ$, for 30$^\circ$ and 90$^\circ$ only one dominant mode is found.
The SEM pictures in the insets illustrate the drop in the internal field around the antidots. By rotation of the sample, the overlap between neighboring antidots changes. In the schematic inset, it can be seen that as the overlap in the 30$^\circ$ and 90$^\circ$ is larger, the areas of localization come closer. This results in the observation of a higher Fourier power intensity of the localized, non-dispersive modes and goes with a reduction of the Kittel amplitude for both spectra at angles 30$^\circ$ and 90$^\circ$, respectively.
Only one dominating mode is observed all up to $\mu_0 H_\mathrm{ext,x}=150\, \mathrm{mT}$. Saturation fields of the films are not changed by structuring; they still remain at around $\mu_0 H_\mathrm{ext,x} =10-15\, \mathrm{mT}$. Simulations for a Py honeycomb lattice have been presented~\cite{ref70}.
These reveal the effect of the strong interaction between next neighbor dots that is responsible for the formation of spin wave channels through the structure. The dynamic modes map the inhomogeneities of the internal field -- which means that they have a dominantly localized nature.

\begin{figure}
\centering
\includegraphics[width=468pt]{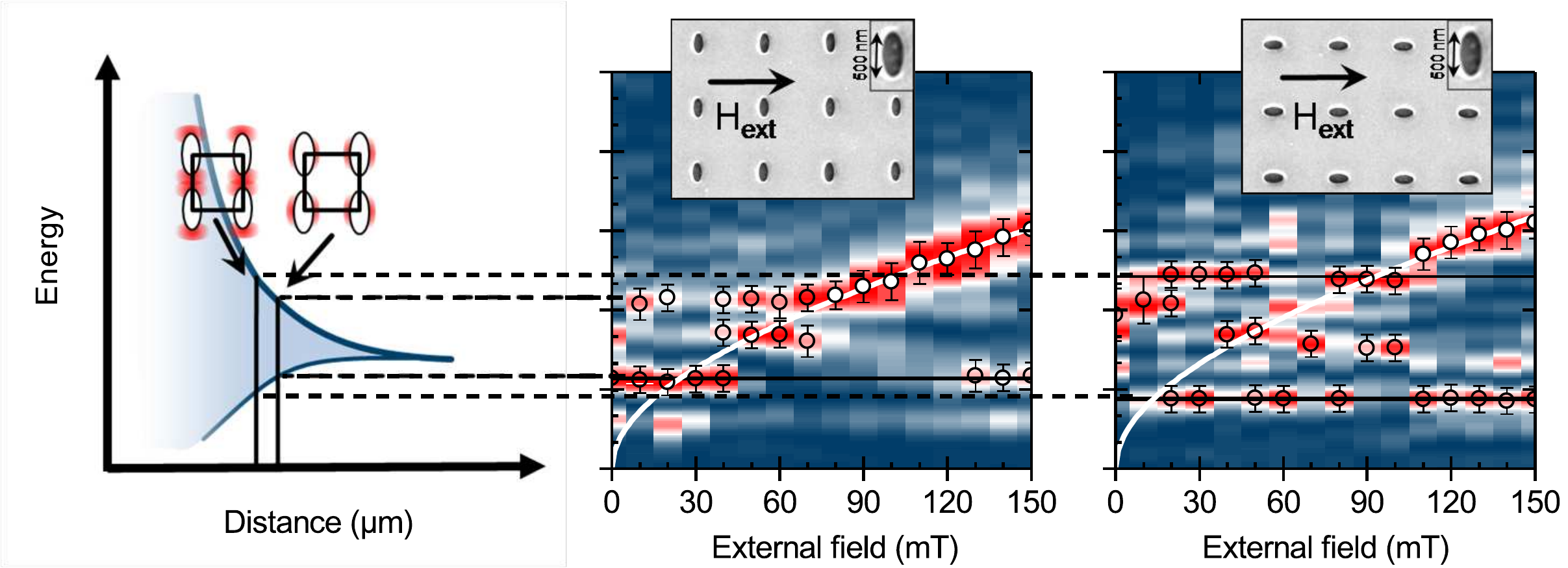}
\caption{Effect of the form factor in elliptical hole structures in a square lattice. Tuning of the interplay between magnetic modes localized at adjacent elliptical antidots can be achieved by making use of the structure's anisotropy. Aligning the larger semi-axis with the external field increases the inter-antidot coupling which in turn yields a larger localization. The frequency width is found to increase. This effect is schematically drawn on the left. Semi-axes are $250\,\mathrm{nm}$ and $500\,\mathrm{nm}$, the square lattice has a periodicity of $3.5\,\mathrm{\mu m}$.}
\label{fig:29}
\end{figure}

The effects of a changing overlap and crystal symmetry can also be controlled through use of elliptical holes instead of circular antidots. The ellipses have an anisotropic dipolar field extending into the film for fields applied along the short axis as compared to fields applied along the long axis.
For the modes localized at the field inhomogeneities around the antidots this means a different extension to the next unit cells, then `feeling' the periodic potential and symmetry created by the structure. This overlap is responsible for the considerable changes in the relative Fourier power of the localized and uniform modes in Fig.~\ref{fig:29}, depending on the orientation of the external field.

The structure consists of elliptical holes with semiaxes of $250\,\mathrm{nm}$ and $500\,\mathrm{nm}$, respectively, arranged on a square lattice with a periodicity of $3.5\,\mathrm{\mu m}$.
First, one observes a change in relative amplitude of the Kittel mode with the angle between external field and lattice. Second, two branches of constant frequency occur in Fig.~\ref{fig:29}. They are separated by several GHz and can be shifted by the angle of the applied field. The separation of the modes is large for the magnetic field applied along the long axis of the ellipse. Bearing in mind the localized character, this behavior can be understood:
By tilting the lattice with respect to the external field the distance between to localization sites is reduced and the splitting of modes can be controlled. This is in close analogy to collective modes as observed in arrays of nanomagnets by Kruglyak et al.~\cite{ref71}.
The frequency shift controlled by field direction and lattice geometry allows to deduce that also the localized modes are more than the single mode at one antidot and thus of magnetic origin.
Further studies therefore need to include a local mode mapping around the antidots to to investigate the nature of these modes unambiguously.

\subsection{Bloch modes in magnonic crystals}\label{sec:bloch-modes}
\begin{figure}[!ht]
\centering
\includegraphics[width=420pt]{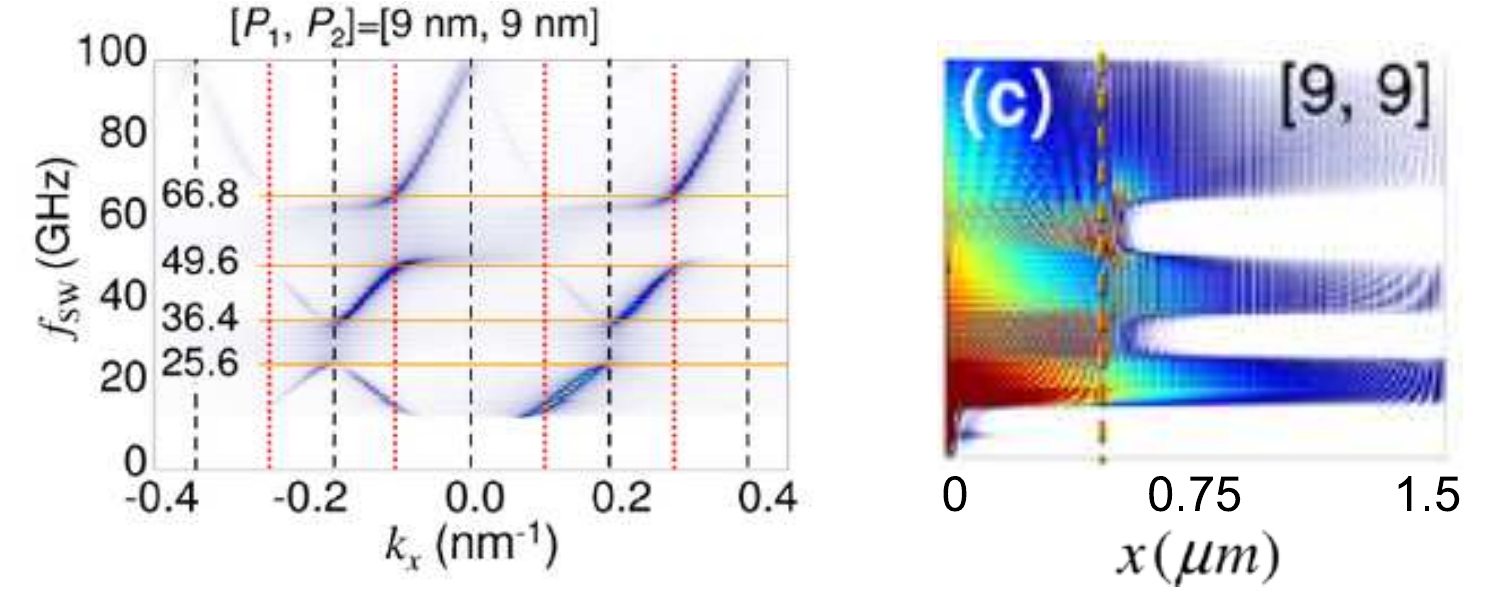}
\caption{Micromagnetic simulation of a one dimensional magnonic crystal: Py stripe with alternating width. Forbidden region (magnonic gaps) in frequency result in a frequency filter effect for transmitted waves. The transmission into the structured region is not allowed and the intensity in the gap region drops down very rapidly (adapted from \cite{ref75}).}
\label{fig:30}
\end{figure}

\begin{figure}[!ht]
\centering
\includegraphics[width=350pt]{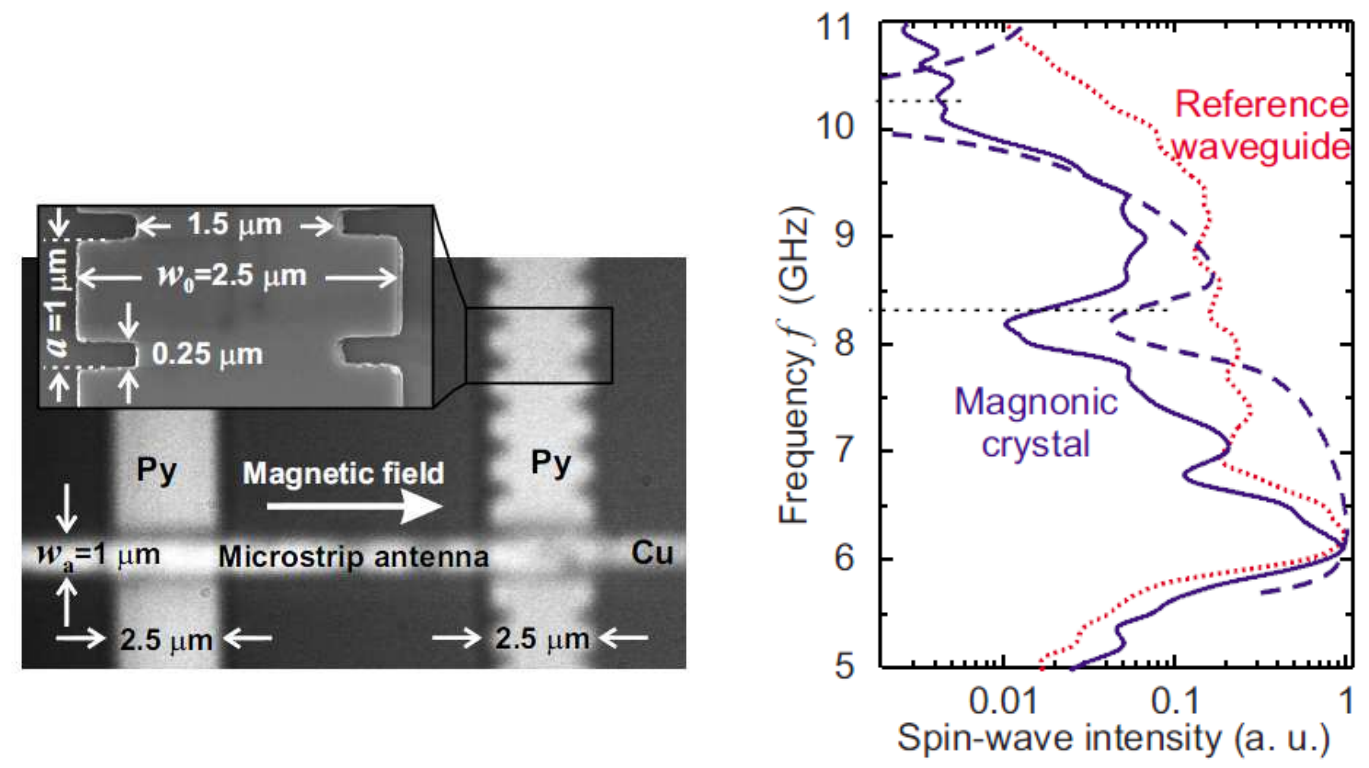}
\caption{Experimentally measured transmission of a one dimensional magnonic crystal: Py stripe with alternating width. A breakdown of the transmitted intensity by more than a factor of 10 is observed at $8.4\,\mathrm{GHz}$ (adapted from~\cite{ref76}).}
\label{fig:31}
\end{figure}

If one wants to study spin-wave materials closer to the ``free electron'' case two preconditions must be fulfilled: first, the propagation length of the spin waves must extend across multiple unit cells. In terms of a resonance this leads to the formation of a narrow band of states in $k$-space.
Gubbiotti et al.\ demonstrated in their recent work a magnonic band gap spectrum in 2D magnonic crystals with submicrometer periods~\cite{ref72}. In general, the formation of a band depends on the spin-wave propagation. For Ni with a Gilbert damping parameter of $\alpha = 0.02$ the propagation length from phase velocity can be estimated to approximately $10\,\mathrm{\mu m}$.
CoFeB has a low damping with $\alpha = 0.006$. Thus the spin-wave propagation length is larger than $100\,\mathrm{\mu m}$. For YIG with $\alpha =6\times10^{-5 }$ even millimeters can be reached.
The second precondition concerns the scattering potential, which must be weak. Already a filling fraction of $0.1$ means that the holes in the film make 10\% of the material. However the distortion in the internal field is much larger.
It will extend a factor two in radius; this variation of the internal field being different from the homogeneous case results in an effective filling fraction, which is much larger. Very nice examples can be found for spin-waves propagating in one dimensional stripes: strong rejection bands are formed. The influence of a zig-zag configuration of the magnetization results in a periodically alternating magnetization. Topp et al.\ showed that spin-wave confinements are also found~\cite{ref73}.
Magnonic gaps in the band structure have been impressively demonstrated in a one-dimensional magnonic material by Chumak et al.~\cite{ref74}. Periodic grooves were mechanically drilled into a ferrite-based wave guide. The transmission shows certain frequency bands determined by their periodicity, while the groove depth in the material determines the transmitted power leading to rejection efficiencies of $30\,\mathrm{dB}$ (contrast of 1/1000). Even in a standard ferromagnet such as permalloy, by simply increasing the width of the wave guide, thus changing the contrast in the periodic energy landscape, a gradual opening of a magnonic gap was realized.
In the case of 1-dimensional systems the filtering effects in width-modulated stripes have been simulated by Kim~\cite{ref75} and subsequently measured by Chumak~\cite{ref76}. Both are reprinted in Fig.~\ref{fig:30}. The distance is in the nanometer range. Clearly, this falls into the exchange dominated spin-wave dispersion. It can be nicely seen, that the bands calculated in the micromagnetic mode, show a quadratic dispersion, which is very similar to a free electron dispersion.
Due to the periodic modification of the width, the gaps opening up span multiple GHz. The transmission into the structured region is not allowed. In Fig.~\ref{fig:31} experimental results from Chumak et al. are displayed for a permalloy wave guide. Clearly, a drop in transmission not as high as in YIG, but also by a factor of 1/10, is observed.

\begin{figure}[!ht]
\centering
\includegraphics[width=286pt]{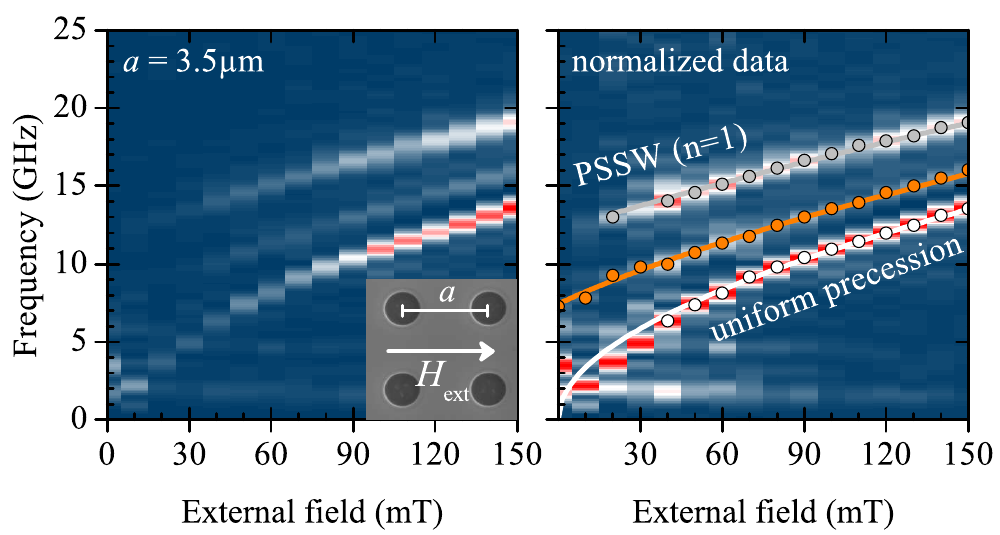}
\caption{Bloch modes in a CoFeB antidot lattice. In a continuous film (not shown) only the mode of  uniform precession and a first order standing spin wave (PSSW) can be found. In a structured film an additional mode appears (orange line and points), whose wavelength is determined by the wave vector at the Brillouin zone boundary $k = \pi / a$.}
\label{fig:32}
\end{figure}

\begin{figure}[!ht]
\centering
\includegraphics[width=474pt]{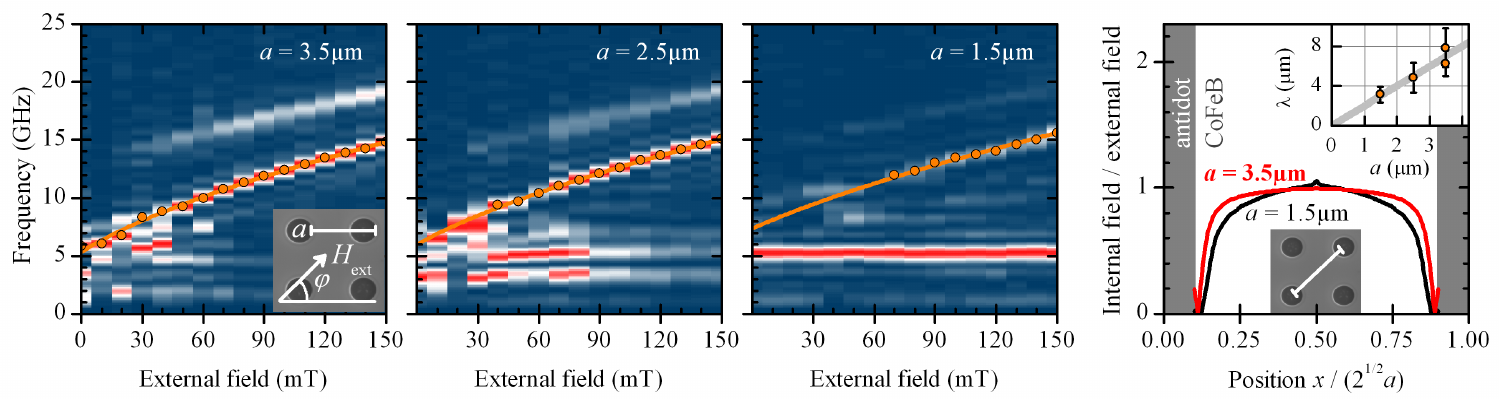}
\caption{For a magnetic field applied under 45$^\circ$,  Bloch modes in a CoFeB antidot lattice by the wave vector at the Brillouin zone boundary $k=\pi / a$ dominate the spectrum. The antidot lattice is varied as a function of the lattice distance for the same filling fraction. Upon reduction of the lattice constant, a mode appears that hardly depends on the magnetic field.}
\label{fig:33}
\end{figure}

The effect of two dimensional structures on propagating modes was studied by Neusser by vector network analyzer-FMR~\cite{ref77}. We will present a study here as well using all-optical spin-wave excitation and detection in a low damped CoFeB film. Figure~\ref{fig:32} shows spectra of a structured CoFeB film (thickness $t=50\,\mathrm{nm}$), measured by means of TRMOKE.
In an unstructured film, with the same thickness, one can identify two modes, which are the uniform precession mode and the first order standing spin wave. The structure (square antidot lattice with lattice constant $a=3.5\,\mu$m, field applied along lattice side) alters the dispersion, and introduces band gaps at the zone boundary, which is at $k=\pm\pi / a$.
Due to a diminished slope of the dispersion at this point, one here finds an increased density of states. The fingerprint of this effect can be seen in the measured spectrum in Fig.~\ref{fig:32}, where a new mode appears, which obeys the dispersion of a Damon-Eshbach-mode with $k=\pm\pi / a$. Since for $t \ll a$ the mode profiles for both directions ($\pm 90^\circ$ with respect to the field) are essentially constant and the dispersion is degenerate with regard to this directions, a superposition of both spin waves yielding a standing wave is most likely.
When the field is applied in an angle of $45^\circ$ with respect to the antidot lattice (see Fig.~\ref{fig:33}), the appropriate dispersion for the same propagation direction with respect to lattice yields again $k=\pm\pi / a$, and is now fourfold degenerate ($\pm 45^\circ$, $\pm 135^\circ$ with respect to the field).
The degeneracy accompanied by an increase of the density of states results in the observation, that only the Damon-Eshbach modes are populated after optical excitation.

These modes are examples for delocalized, extended Bloch-modes, whose excitation and detection in non-frequency and non-$k$-selective experiments become possible due the artificially altered band structure.

\subsection{Magnonic wave guides}
\begin{figure}%
\centering%
\includegraphics[width=426pt]{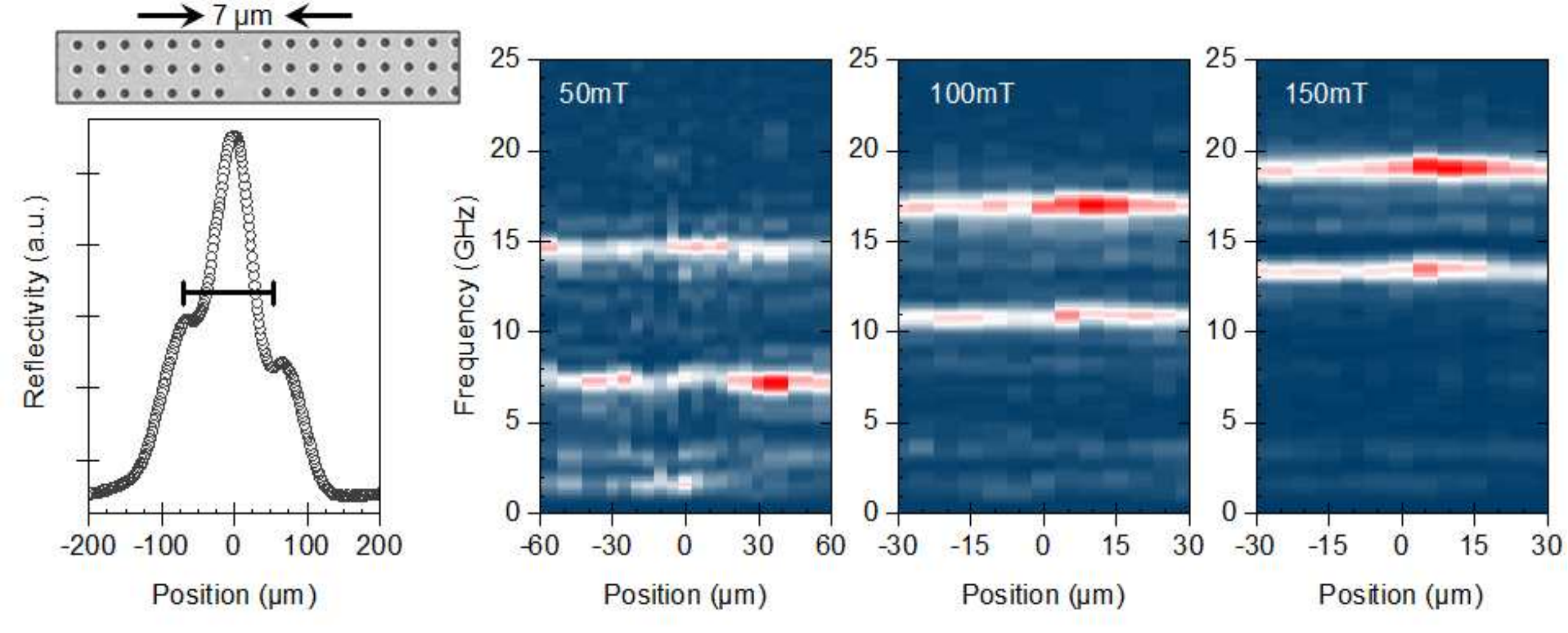}%
\caption{Magnonic spin-wave guide prepared by FIB: a ``line defect''. Below: position of the wave guide scanning with the pump beam in the reflectivity data. Right: magnetic spectra for different field values scanning the position of the probe beam.}%
\label{fig:34}%
\end{figure}
\begin{figure}
\centering
\includegraphics[width=0.8\textwidth]{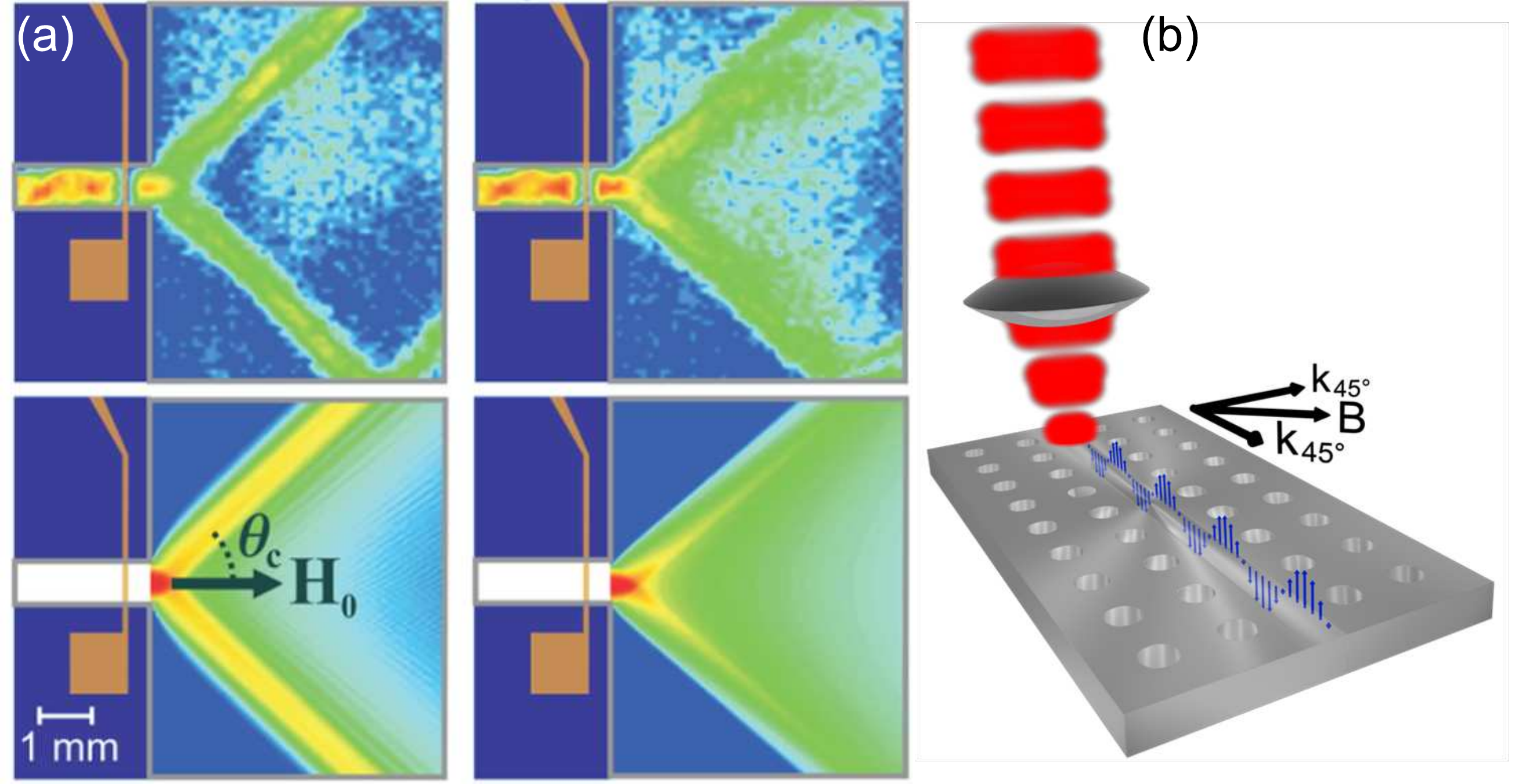}
\caption{(a) Spin-wave caustics observed with BLS show beam-like propagation (from~\cite{SchneiderPRL2010}). (b)~An artist's view of a magnonic spin-wave guide: femtosecond laser pulses allow to seed spin waves freely by moving the laser spot, hence ``photo-magnonics''. Depicted is the situation for the external field applied under an angle of $45^\circ$ with respect to the antidot lattice.}
\label{fig:35}
\end{figure}

%
We have discussed the possibility of completely different behavior of photons in a photonic wave guide. So called ``slow photons'' may emerge that propagate in the way of the religious procession of Echternach, i.e.\ three steps forward and two steps backwards, similar to pictures drawn in the Krauss article~\cite{ref78}. Effectively, the propagation is slowed down, which ultimately is a signature of the periodic potential. Other features like band gap formation allow for the achievement of photonic resonators with highest $Q$ factors.
Moreover, wave guides embedded in a structured material allow to facilitate the guidance of photons around corners or the splitting of wave packages into two. In the introduction we have already discussed the differences between optical and magnonic wave guides: in the latter, the internal field may result in a localization of the spin-wave modes. This can be very different for magnonic wave guides.
Mapping the ideas from photonics to magnonics is not straightforward. Strong differences are found in the anisotropy of the band structure. A Bloch state with the period of the lattice will be a standing wave at the zone boundary. However, the mode perpendicular to it will not feel the same confinement.
By removing one row of the holes, the localized state may propagate along this defect, while for the perpendicular state the situation remains unchanged. A solution is to apply the field under $45^\circ$ with respect to the defect, as depicted in an artist's view in Fig.~\ref{fig:35}~(b). Then the states perpendicular and parallel to the channel will have the same dispersion and thus show the same standing-wave behavior of modes with $k$ at the zone boundary (in close analogy to Fig.~\ref{fig:33}).
Simpler wave guides have been discussed in the beginning (Fig.~\ref{fig:guides}) and even beam-like propagation can be realized by spin-wave caustics as reproduced in Fig.~\ref{fig:35}~(a)~\cite{SchneiderPRL2010, DemidovPRL2009}.
We have performed experiments on a ``line defect'', which means one missing line in the magnonic crystal. First experiments of this kind are shown in Fig.~\ref{fig:34}. Scanning the probe beam across the defect shows magnetic modes inside the magnonic wave guide that are otherwise lower in intensity in the structured material.
While this reveals the feasibility of the experiments, future approaches probing the modes propagating in the wave guides need to take into account the anisotropic magnon dispersion with respect to the external field. Especially, the mode velocity and propagation length as well as the scattering between modes with $k$ parallel or antiparallel to $M_\mathrm{S}$ and $H_\mathrm{ext}$ will be a focus aim.
An absolute advantage of photo-excitation and detection of spin waves becomes evident here: excitation and detection can be freely moved in the structure by moving the two laser spots (in the experiment $60\, \mathrm{\mu m}$ for the pump pulses and $14\, \mathrm{\mu m}$ for the detection were used here).

\section{Outlook: controlled spin-wave localization}
We have described in our review the possibilities of spin waves in artificially structured media: thereby, we started with novel concepts of spin-wave computing; we described the peculiarities of spin-wave guides compared to optical wave guides; and then, we switched to the effect of structuring materials -- what are the conditions and solution of forming novel states.
To develop grounds for different types of spin-wave materials, we repeated basic concepts of solid state physics of band formation of electrons, and clearly distinguished between the case of a small perturbation -- starting from a freely propagating wave -- and the tight binding case. We show that similar concepts can be applied to magnonic crystals.
Experimentally, one finds both localization effects and wave vectors that correspond to a Bloch mode at the zone boundary originating from the periodicity of the antidot lattice.

What kind of development may be projected into the future? One major difference compared to photonics is that the peculiar shape of the magnetic potential forming at each antidot has a rich complexity.
While for the photonic case, the index of refraction is given by the alternating material itself and can be changed only in small fractions, the filling fraction in the magnetic case extends the inter-material border since it is given by the dipolar interaction distorting the internal fields around the antidot, which even can be changed by rotating the applied magnetic field.
The minima in internal fields where spin waves localize can be designed by the antidot shape, as we have demonstrated for elliptical antidots. By changing the applied field, the potential landscape can be rotated: the overlap of the localized states is changed.
One can compare this situation with the tight binding approach. Each ``atom'' is defined by the shape of the dipolar fields, which allows certain solutions of the spin-wave function. These ``atomic'' spin functions can then be used as a basis set to model the interactions of the emerging localized spin-wave crystal, which may have interesting features, to model correlation effects in correlated materials. These ideas have been successfully applied to lattices of single atoms in atomic physics. It would be interesting to see if such concepts could apply here, as well.
For example, the external applied field could gradually change the overlap, e.g.\ ``hopping'' between the localized spin-wave orbital wave functions. However, such ideas are more farsighted. Current experiments will have to show that low damping materials will improve the quality of spin-wave crystals and formation of Bloch states.
Furthermore, different material combinations need to be explored to go beyond the antidot lattice case. Spin-wave computation and spin-wave data transmission has to be developed to go beyond the proof of principle and to show that indeed a high data throughput may be an advantage of spin-wave based computation in the future.

\end{document}